\documentclass[a4paper,12pt]{article}
\usepackage[top=3cm, bottom=3cm, left=3cm, right=3cm]{geometry} 
\usepackage{color}
\usepackage{rotating}
\usepackage{graphicx}
\usepackage{amsmath}
\usepackage{amssymb}
\usepackage{float}
\usepackage{subfig}
\usepackage[countmax]{subfloat}
\usepackage{hyperref}
\def\displayandname#1{\rlap{$\displaystyle\csname #1\endcsname$}%
                      \qquad \texttt{\char92 #1}}

\begin{document}
\title{\bf{Simulations Study of Muon Response in the Peripheral Regions of
the Iron Calorimeter Detector at the India-based Neutrino Observatory}}

\author{R. Kanishka\thanks{email:
kanishka@puhep.res.in}~, Meghna K. K.$^\dagger$, \\
Vipin Bhatnagar$^*$, D. Indumathi$^\dagger$, Nita Sinha$^\dagger$ \\
{\it $^*$Physics Department, Panjab University, 
Sector 14, Chandigarh 160 014, India} \\
{\it $^\dagger$The Institute of Mathematical Sciences,
CIT Campus, Chennai 600 113, India} \\
}
\maketitle

\begin{abstract}
{The magnetized Iron CALorimeter detector (ICAL) which is proposed to be
built in the India-based Neutrino Observatory (INO) laboratory, aims
to study atmospheric neutrino oscillations primarily through charged
current interactions of muon neutrinos and anti-neutrinos with the
detector. The response of muons and charge identification efficiency, angle and energy resolution as a function of muon momentum and direction are studied
from GEANT4-based simulations in the peripheral regions of the
detector. This completes the characterisation of ICAL with respect to
muons over the entire detector and has implications for the
sensitivity of ICAL to the oscillation parameters and mass hierarchy
compared to the studies where only the resolutions and efficiencies of
the central region of ICAL were assumed for the entire detector.

Selection criteria for track reconstruction in the peripheral region of
the detector were determined from the detector response. On applying
these, for the 1--20 GeV energy region of interest for mass hierarchy
studies, an average angle-dependent momentum resolution of 15--24\%,
reconstruction efficiency of about 60--70\% and a correct charge
identification of about 97\% of the reconstructed muons were obtained. In
addition, muon response at higher energies upto 50 GeV was studied as
relevant for understanding the response to so-called rock muons and cosmic
ray muons. An angular resolution of better than a degree
for muon energies greater than 4 GeV was obtained in the peripheral regions,
which is the same as that in the central region.}

\end{abstract}

\newpage

\section{Introduction}

The India-based Neutrino Observatory (INO) \cite{Athar:2006yb} is the
proposed underground facility that will house a magnetized Iron
CALorimeter detector (ICAL) designed to study neutrino oscillations
with atmospheric muon neutrinos. In particular, ICAL will focus on
measuring precisely the neutrino oscillation parameters including the
sign of the 2--3 mass-squared difference $\Delta
m_{32}^{2}$ ($= m_3^2 - m_2^2$) and hence help to determine the neutrino
mass hierarchy through matter effects. Oscillation signatures
for neutrinos and anti-neutrinos are different in the presence of
matter effects which become important in the few GeV energy region.
These parameters are sensitive to the momentum $P$ and the zenith angle
$\cos\theta$ (through path length travelled) of neutrinos. Reconstruction
of these parameters further depends on the energy and direction of muons
and hadrons \cite{hadronresponse} produced in charge-current interactions of the neutrinos
in the detector; hence studies of muon energy and direction resolutions
are crucial.

Since ICAL is designed to be mostly sensitive to muons, the main
physics issues that ICAL will probe will be through studies of charged
current (CC) muon neutrino (anti-neutrino) interactions with muons
(anti-muons) produced in the final state. Two types of interactions
are relevant: in the first, the neutrino enters the detector,
interacting with (dominantly) nucleons in the iron nucleus.  These
events are identified by vertices which are inside the detector
(tracks begin inside the detector). In the second type of interaction,
the neutrino interacts with rock around the detector and the produced muons lose energy while propagating through the rock (these are the so-called rock muons or upward-going muons). These events have vertices outside the detector with  their tracks starting  at an edge of the detector. The first type of interaction dominantly gives low energy (few GeV) muons due to the rapidly falling atmospheric neutrino flux. While in the case of rock muons, most of the lower energy muons are stopped in the rock itself, so that the fraction of higher energy ($>$ 10 GeV) muons is higher in this sample. In addition, cosmic ray muons also enter the detector from above.

In an earlier paper \cite{central}, the response of ICAL to few-GeV
muons with respect to both momentum magnitude and direction
reconstruction was studied through simulations for muons generated in
the central region of ICAL where the magnetic field is largely uniform
both in direction and magnitude. Here we present for the first time a
GEANT4-based simulations study of the muon response in the peripheral
region of ICAL, where the magnetic field is not only highly
non-uniform in both magnitude and direction, but there are significant
edge effects as most of the tracks will be partially contained. 
Note that a substantial fraction of rock muon events
 traverses such regions in the detector; hence it is important to
understand the muon response in these regions for such physics studies.

The inclusion of muon response in the peripheral region of the detector can significantly change the oscillation and mass hierarchy results since it contains resolutions and efficiencies in the energy range 1--50 GeV. The first set of simulations results for precision measurement of neutrino oscillation parameters and the mass hierarchy were produced using only the central region resolutions of about 9--14\% and efficiency of about 80\% (see Ref.~\cite{physics}, \cite{physics1} for more details) in the energy range 1--20 GeV.

The paper is organised as follows: in Section 2 we briefly discuss some
relevant details about the GEANT4-based simulation of the detector
geometry and magnetic field, as well as the methodology of hit and
track generation and track reconstruction. In Section 3, we discuss
the general features of muon propagation in the different regions
of ICAL. In Section 4, we discuss the selection criteria used. We
then present in Section 5 the results, with these selection criteria,
for the muon efficiencies and resolutions in the peripheral region of
ICAL. A comparison of the response in all the regions of ICAL (central and peripheral) is given in Section 6. We conclude in Section 7,
with discussions.

\section{The ICAL Detector Simulation Framework}

Details of the ICAL detector can be found elsewhere \cite{central}. Here
we briefly review the relevant simulation inputs for the geometry and
magnetic field. The ICAL detector geometry is simulated \cite{central}
using GEANT4 software \cite{geant}. It comprises of three identical
modules of dimension 16 m $\times$ 16 m $\times$ 14.45 m. The direction along which the modules are placed is chosen to be the $x$-direction (and is
labelled with the azimuthal angle $\phi=0$) and the direction perpendicular to the $x$-axis in the horizontal plane is the $y$-direction. The vertical direction is the $z$-direction with the $z$-axis pointing upwards (so the polar angle is also the zenith angle). Each module comprises a stack of 151 layers of 5.6 cm thick magnetized iron, separated by a 4 cm air gap in which the active detector elements, the RPCs are placed. The $y$-direction is in the plane of the iron plates, parallel to the slots of the magnet coil, as shown in Fig.~\ref{fig:magfield}, with the origin of the coordinate system being the centre of the central module.

Apart from coil slots (thin 8 m long slots centred around $y=0$ at $x=
x_0 \pm 4$ m, where $x_0$ is the central $x$-coordinate of each module),
there are vertical steel support structures which are placed at every 2
m along the whole detector in both $x$ and $y$ directions, to maintain
the air gap between plates. These are dead spaces that affect the muon
reconstruction. In addition, the magnetic field is also not uniform
everywhere and so the quality of reconstruction depends on the region
where the event is located.

\subsection{The Magnetic Field}

\begin{figure}[htp]
\renewcommand{\figurename}{Fig.}
\begin{center}\includegraphics[width=0.55\textwidth]{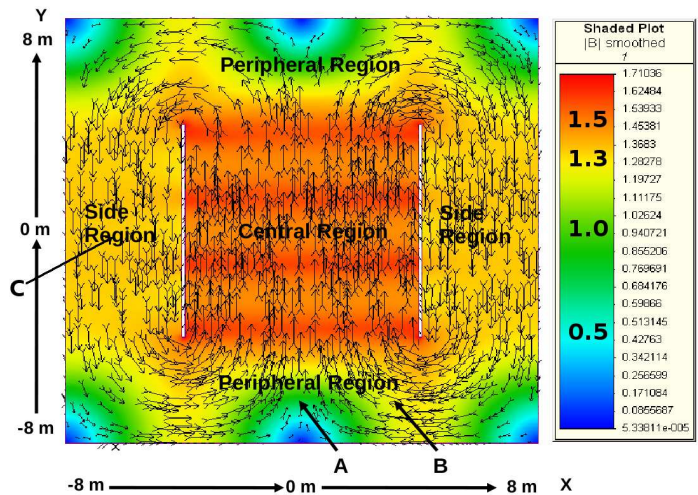}
\end{center}
\caption{Magnetic field map as generated by the MAGNET6 \cite{magnetcode}
software in the central iron layer of the central module. Points A =
$(0, -650, 0)$ cm, B = $(300, -650, 0)$ cm, C = $(-2270, 0, 0)$ cm (in
the $1^{st}$ module of ICAL), are defined for later use. Notice that C
is actually in the left-most module of the detector and is simply marked
here for convenience.}
\label{fig:magfield}
\end{figure}

The magnetic field has been simulated in a single iron plate using the
MAGNET6 software \cite{magnetcode}. The magnetic field map is generated
at the centre ($z=0$) of the plate and is assumed uniform over the entire thickness of the plate. The magnetic field lines in a single iron plate
in the central module are shown in Fig.~\ref{fig:magfield}. The field is generated by current circulating in copper coils that pass through slots in the magnetized plates. The slots can be seen as thin white vertical lines at $x=\pm 4$ m in the figure. The direction and length of arrows
denote the direction and magnitude of the magnetic field.

The ``central region'' is defined as the volume contained within
the region $\vert x \vert \le 4$ m, $\vert y \vert \le 4$ m with $z$
unconstrained, that is, the region within the coils slots in each
module. The central region has the highest magnetic field that is
uniform in both magnitude and direction ($B_y$) while the region labelled
``peripheral region'' (outside the central region in the $y$ direction,
$\vert y \vert > 4$ m) has the maximally changing magnetic field in both
direction and magnitude, with the field falling to zero at the corners
of the module.

In an earlier paper \cite{central}, the muon response was studied in
the central region; here we study the peripheral region where, apart
from the changing magnetic field, the reconstruction is also affected by
edge effects. In addition, there are two small regions outside the coil
slots in the $x$ direction ($\vert x \vert > 4$ m), labelled as the
``side region'' in Fig.~\ref{fig:magfield} where the magnetic field
is about 15\% smaller than in the central region and in the opposite
direction. The side regions in the central module are contiguous with the
side regions in the adjacent modules and the quality of reconstruction
here is expected to be similar to that in the central region. However
the left (right) side region of the left (right)-most module will suffer
from edge effects and we shall consider them separately in the study.

\subsection{Event Reconstruction}

In each analysis, 10,000 muons are propagated in the detector with
fixed momenta and direction and with the starting point of the muons
 in either the peripheral or side regions. In contrast to the
earlier study in the central region \cite{central}, here the muon
momenta vary from 1--50 GeV/c, with the higher energies being of
interest for rock muon studies. The muon tracks are reconstructed
based on a Kalman filter \cite{kalman} algorithm. The RPCs that signal
the passage of muons through them have a position sensitivity of about
$\pm 1$ cm in the $x$- and $y$-directions (the RPC strip width is 1.96
cm) and about $\pm 1$ mm in the $z$-direction (the gas gap in the RPCs
being 2 mm). In addition, ``hits'' or signals can be generated in
adjoining RPC strips as well, and further that the RPC efficiency and time resolution are about 95 \% and 1 ns respectively \cite{rpc_char}. For more details regarding the generation of hits, tracks and their reconstruction, see
Ref.~\cite{central}.

\section{General Feature of Muon Response in the Peripheral and Side Regions}

We first discuss the general expectations, based on the detector geometry
and the orientation of the magnetic field.

\begin{figure}[htp]
\renewcommand{\figurename}{Fig.}
\begin{center}\includegraphics[width=0.55 \textwidth]{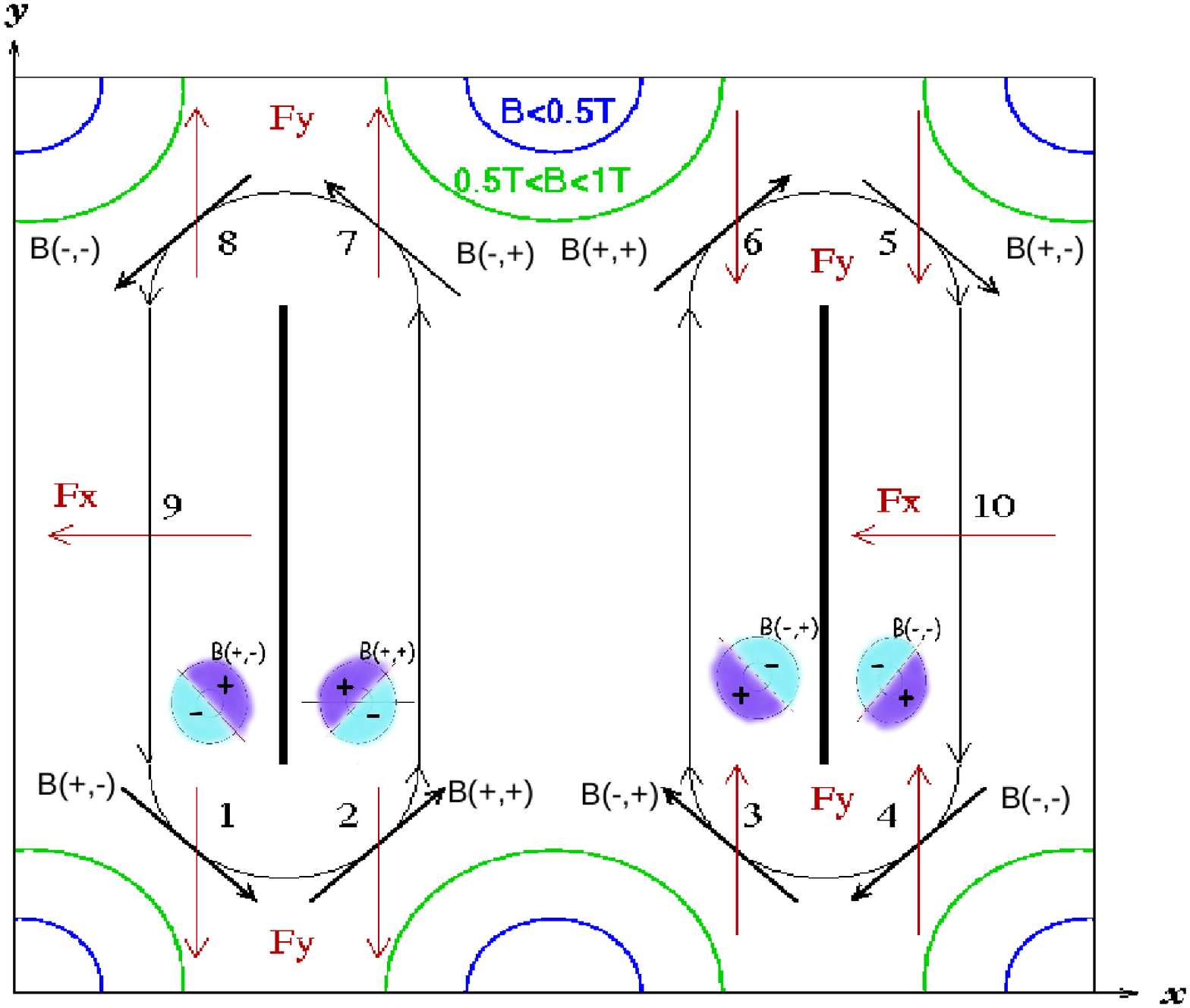}
\end{center}
\caption{Magnetic field map with the net force directions in the
peripheral and side region. The thick black arrows indicate the direction
of the magnetic field with labels $B(i,j), i=+,-,$ that denote the sign
of the $B_x$, $B_y$ components in each region. The brown arrows indicate
the direction of $F_x$ or $F_y$ force components that will act on a
negatively charged upward-going muon. The small coloured circles
indicate the direction of $F_z$ in each region, with purple (cyan)
denoting $F_z > 0 (< 0)$. (Note that side regions 9 and 10
are in the $1^{st}$ and $3^{rd}$ modules of the detector respectively
and are shown together in the same module for convenience.)}
\label{fig:map}
\end{figure}

The Lorentz force equations are $\vec{F} = q (\vec{v} \times \vec{B})$,
where $\vec{B}$ is the magnetic field that is confined to the $x$-$y$
plane, $q$ is the charge of the particle with momentum $\vec{P}$
and energy $E$ so that its velocity $\vec{v}$ is directed along the
momentum with magnitude $v = Pc^2/E$. Since $q = -1$ for $\mu^{-}$,
the components of force in the peripheral region for an upward-going muon,
momentarily ignoring energy loss, are,

\begin{eqnarray} \nonumber
F_{x} & = & v_{z} B_{y}~; \nonumber \\
F_{y} & = & - v_{z} B_{x}~; \nonumber \\
F_{z} & = & v_{y} B_{x} - v_{x} B_{y}~,
\label{eq:periforce}
\end{eqnarray}
whereas the analogous components of force in the side region are given as:
\begin{eqnarray} \nonumber
F_{x} & = & - v_{z} B_{y}~; \nonumber \\
F_{y} & = & 0~; \nonumber \\
F_{z} & = & v_{x} B_{y}~. 
\label{eq:sideforce}
\end{eqnarray}
It is seen that in both regions, $F_{x}$ and $F_{y}$ are independent
of $\phi$ (that is, independent of the momentum components in the plane
of the field) and depend on $P_{z}$ (i.e., on the zenith angle $\theta$
alone) while $F_z$ depends on both $\theta$ and $\phi$. Depending on the
components of magnetic field $B_{x}$ and $B_{y}$, Eqs.~\ref{eq:periforce}
and \ref{eq:sideforce} give the net force in the different regions of
ICAL. Consider the in-plane ($x$, $y$ components) forces in the regions
denoted as 1--10 in Fig.~\ref{fig:map}. It can be seen that in regions
$1,2,7,8$, $F_{y}$ is such that it causes an upward-going muon
($\cos\theta>0$) to be bent in a direction going
out of the detector, thus lowering the reconstruction efficiency. The
effect is just the opposite in regions $3,4,5,6$. If $F_{z} > 0$ as well,
the already upward-going muon traverses more iron layers as
discussed in Ref.~\cite{central} and hence gives good resolution; hence,
$F_y$ affects the reconstruction efficiency while $F_z$ determines
the quality of reconstruction. Since $F_z$ depends on both $(\theta,
\phi)$ as well as the magnetic field, the sign of $F_z$ is shown in
Fig.~\ref{fig:map} inside a circle of $\phi$ in each of regions $1,2,3,4$
(for negatively charged upward-going muons), for $\vert B_x \vert \sim
\vert B_y \vert$ with purple (cyan) regions denoting $F_z > (<) 0$.

\begin{figure}[btp]
\renewcommand{\figurename}{Fig.}
  \centering
\includegraphics[width=0.35\textwidth]{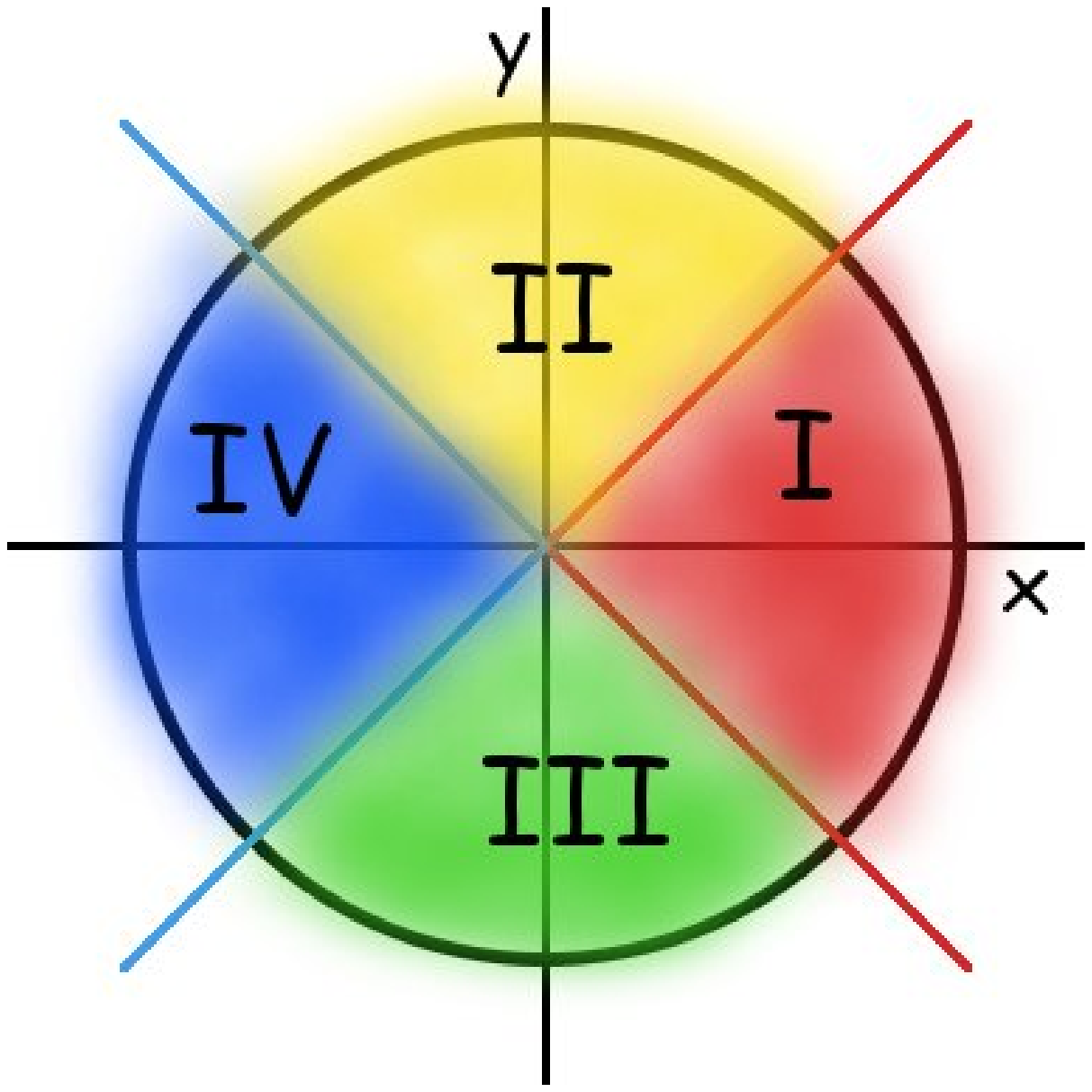}
\hspace{0.5cm}
\includegraphics[width=0.35\textwidth]{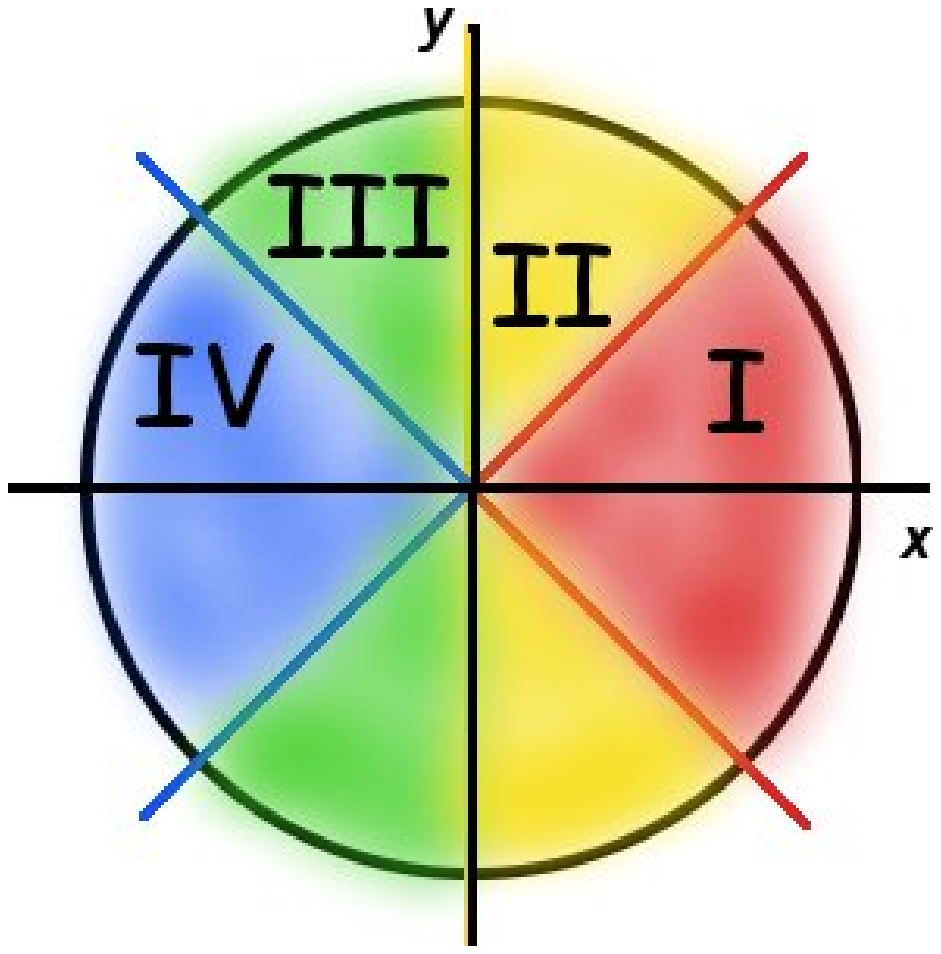}
\caption{The choice of $\phi$ bins in the peripheral (left) and side (right) regions.}
\label{fig:phi_choice}
\end{figure}

Hence the magnetic field that determines the quality of reconstruction
breaks the azimuthal symmetry, so muons in different $\phi$ directions
(for the same momenta and polar angle $\cos\theta$) have different
detector response. This was discussed in detail in Ref.~\cite{central}. Going by the force equations, we therefore analyse the muon response in
the peripheral region in four different set of $\phi$ bins as shown in
Fig.~\ref{fig:phi_choice}: bin I: $\vert \phi \vert \le \pi/4$, bin II:
$\pi/4 \le \phi < 3\pi/4$, bin III: $-3\pi/4 \le \phi < -\pi/4$, and
bin IV: $3\pi/4 < \vert \phi \vert \le \pi$.

For muons with starting point in the negative $y$ peripheral region (regions marked
$1,2,3,4$ in Fig.~\ref{fig:map}), this implies that most of the muons
with momenta such that $\phi$ lies in bin III (but otherwise having
same magnitude and $\cos\theta$) are prone to exit the detector from
the side; however, there will be marked differences in the quality of
reconstruction between regions $(1,2)$ and $(3,4)$ due to the different
$F_y$ force that turns the track back into the detector in regions
$3,4$. Hence the average detector response in this region is an average
over these two different behaviours. In addition, $F_z > 0~(< 0)$ for
bin II muons in regions $1,2$ $(3, 4)$ and this helps to improve the
reconstruction in regions $1,2$, so that bin II muons can be expected to
have the best quality of reconstruction of the regions 1--4.

A similar analysis can be done for muons which begin in the positive $y$ peripheral region (regions marked $5,6,7,8$). Of course, tracks at the edge of a
region or of high energy muons may move from one region to another where
a different magnetic field applies; however, we simply bin the events
according to the region in which the muon originates.

In the side region 9, $F_{x}$ causes the particle to go out of the
detector but $F_{z} > 0$ and so this region has good resolution but worse
efficiency. The results are opposite in side region 10 since $B_{y}$
$<$ 0. We therefore define the same $\phi$ bins for the side region as
were used for the central region, viz., bin I: $\vert \phi \vert \le
\pi/4$, bin II: $\pi/4 < \vert \phi \vert \le \pi/2$, bin III: $\pi/2 <
\vert \phi \vert \le 3\pi/4$, and bin IV: $3\pi/4 < \vert \phi \vert \le
\pi$. The difference is in the definition of the second and third bins,
see Fig.~\ref{fig:phi_choice}, and is more appropriate from the point of
view of the geometrical configuration in this region. Region 9 (10)
will have the worst reconstruction in $\phi$ bin IV (I). However, in
region 10, the direction of the $F_x$ force is expected to improve the
results just as in the case of regions $3,4$. 

The results will be the same for downward-coming $\mu^+$ (with $\cos\theta <
0$) while that for downward-coming muons or upward-going anti-muons can be
obtained by symmetry. In our analysis, therefore, we study the muon
response in the peripheral regions 1--4 and the side region 9.

It is important to keep in mind that the support structures and coil gaps
also break this azimuthal symmetry in a non-trivial way and the effects of
the geometry may alter the trends of the distributions as discussed above.

The net effect of the nature of the detector geometry and magnetic field
can be seen in the muon resolutions and efficiencies and we will now
discuss these.

As discussed above, we study the response in the following peripheral
and side regions.

\paragraph{In the Peripheral Region}:
Here, 10,000 muons ($\mu^-$) were propagated with fixed input momenta
$P_{\rm in}$ and direction $\cos\theta$ (and smeared over the entire
azimuthal angle $\phi$), with their starting point uniformly smeared over the region
centred at $(0, -600, 0)$ cm and extending upto $\pm$ $(2400, 200,
720)$ cm from it; this comprises the whole peripheral region along the
three modules of the detector in the {\em negative $y$ region} where
the magnetic field is non-uniform.

\paragraph{In the Side Region}: In the side region, muons ($\mu^-$) were
propagated with the same procedure as above but with point of origin smeared in
a region centred around $(-2200, 0, 0)$ cm and $(2200, 0, 0)$ cm (which
are in the $1^{st}$ and $3^{rd}$ modules of the detector respectively)
and smeared uniformly in $\pm$ $(200, 400, 720)$ cm around these.

\section{Selection Criteria Used}

All tracks which satisfy the loose selection criterion $\chi^2/\hbox{ndf}
\le 10$, are used in the analysis, where $\chi^2$ is the standard
$\chi$-squared of the fit and ndf are the number of degrees of freedom, 
$\hbox{ndf} = 2 \times N_{hits} - 5$, where $N_{hits}$ are the
number of hits in the event, $N_{hits} \ge 5$ and the Kalman filter involves the fitting of 5 parameters.

Further selection criteria are used to get reasonable fits and hence
resolutions. Two major constraints have been applied in both the
peripheral and side regions to remove low energy tails. The first is
similar to that applied when analysing tracks in the central region
\cite{central}: the Kalman filter algorithm may generate more than one
track. While this may be correct and useful in the case of genuine
neutrino CC interactions, where one or more hadrons  accompany the muon,
this is a problem for single muon analysis and arises because of
detector dead spaces (for instance, two portions of a track on either
side of a support structure may be reconstructed as two different tracks). This problem will be mitigated by the identification of a vertex
in a genuine neutrino interaction, here, we place a constraint and
analyse only those events for which exactly one track is reconstructed,
leading to a consequent loss in reconstruction efficiency. The second
selection criterion is specific to the peripheral and side regions and is
described below.

Initially, events were generated at fixed points of origin to understand the
effect of the magnetic field. In the peripheral region, the starting point was
chosen to be either at point A (in a region of nearly zero magnetic
field) or B (large magnetic field with both $x$- and $y$-components
non-zero), while in the side region, a generic point C was studied (see
Fig.~\ref{fig:magfield}). Results of this study clearly indicated that a
large fraction of events whose tracks were truncated because the particle
exited the detector (so-called partially contained events) were relatively
poorly reconstructed. These could not be eliminated by tightening the
constraint on $\chi^2$ of the fits; however, they could largely be
removed by demanding that the track contain a minimum number of hits,
such that either $N_{hits} > n_0$ or $N_{hits}/\cos\theta > n_0$ (note that
there may be multiple hits per layer), where $n_0$ needed to be carefully
optimised. It was found that for a given momentum and direction of
the muon, $n_0$ needed to be larger (smaller) in regions where the
magnetic field strength is small (large). Where the muon does not leave
the detector, so that the entire track is contained in the detector
(fully contained events), no constraint on $N_{hits}$ is needed. With
this understanding, the generic peripheral and side region response was studied.

\subsection{Effect of Selection Criteria}

\begin{figure}[htp]
\renewcommand{\figurename}{Fig.}
  \centering
\includegraphics[width=0.48\textwidth]{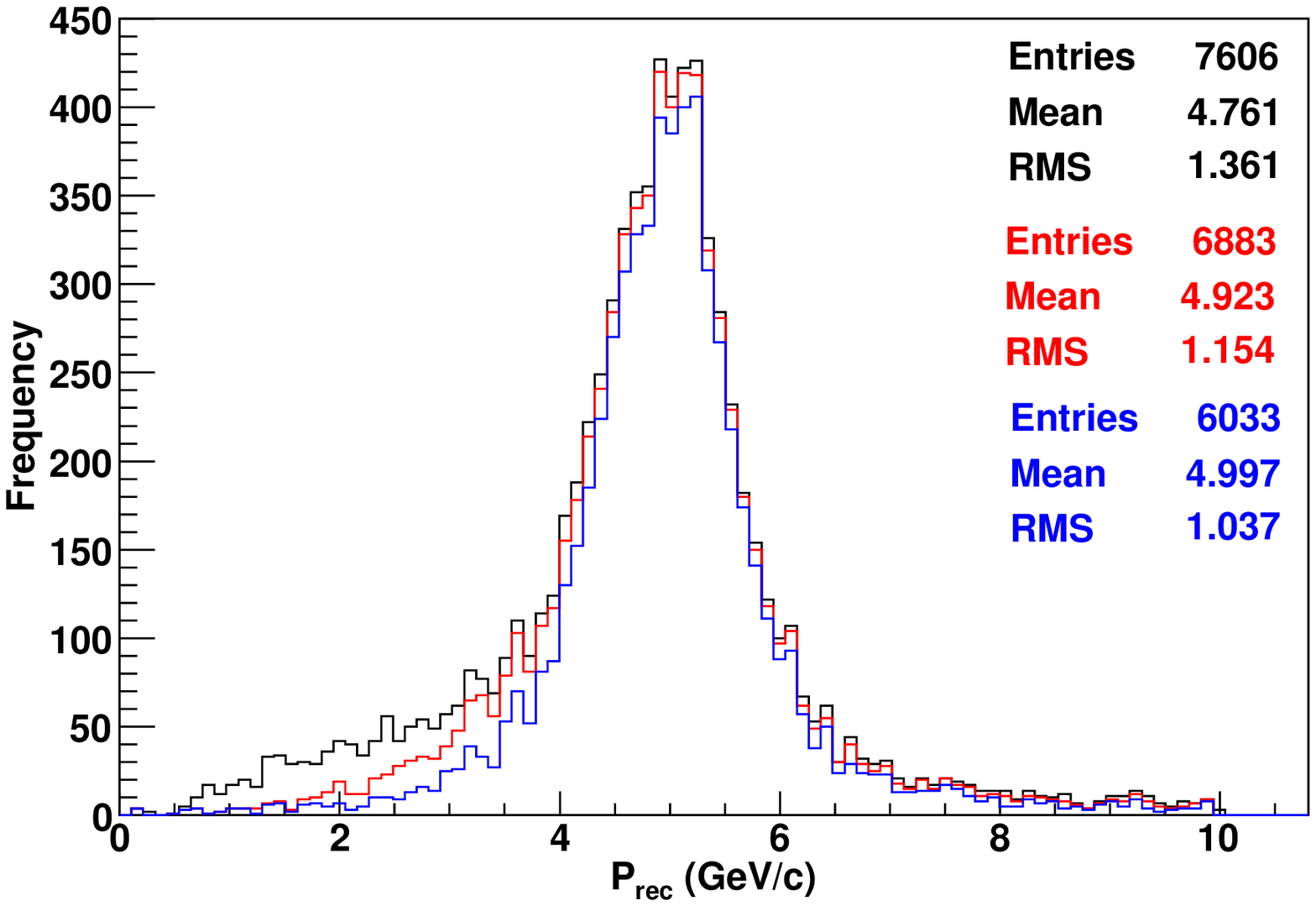}
\includegraphics[width=0.48\textwidth]{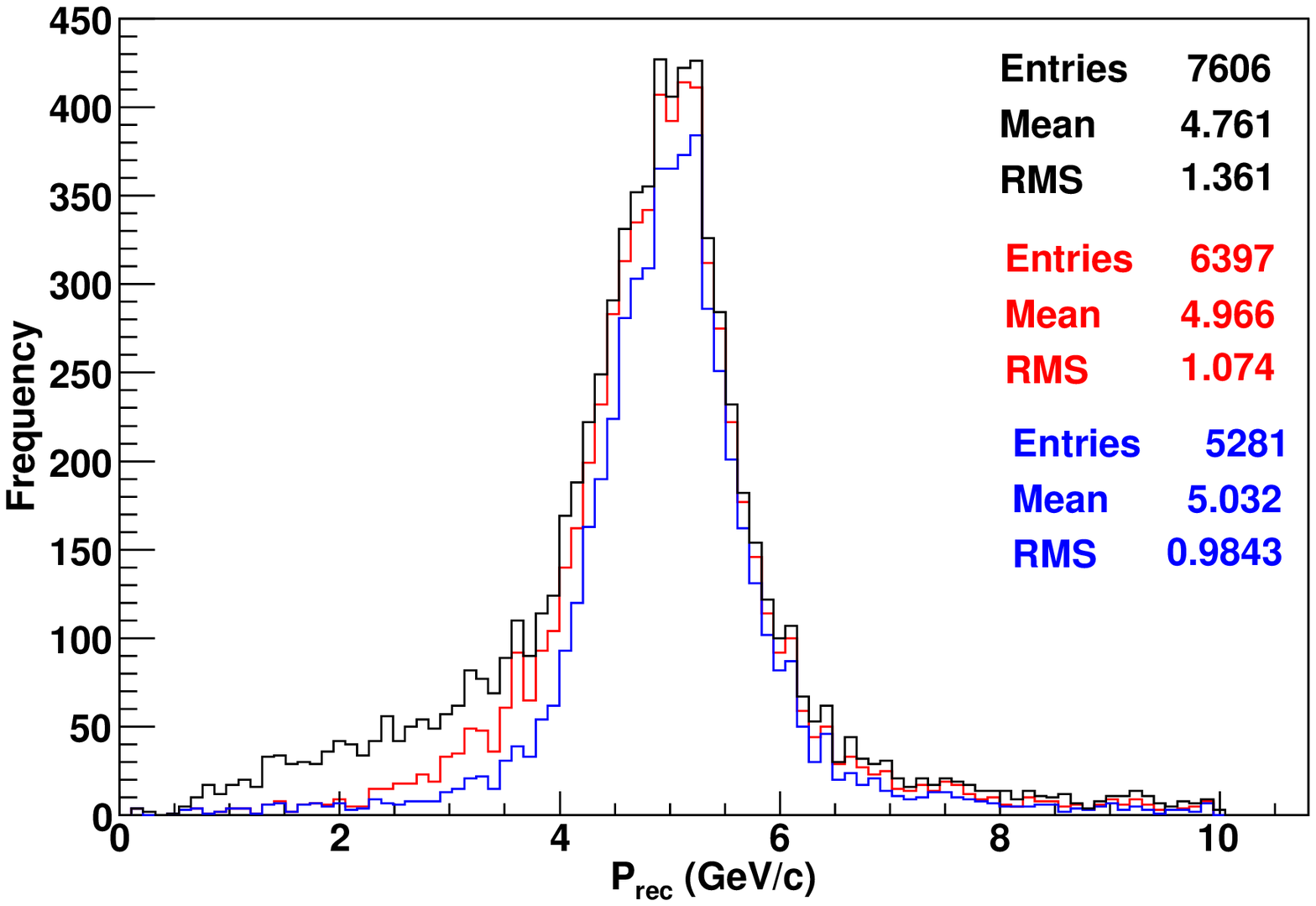}\\
\includegraphics[width=0.48\textwidth]{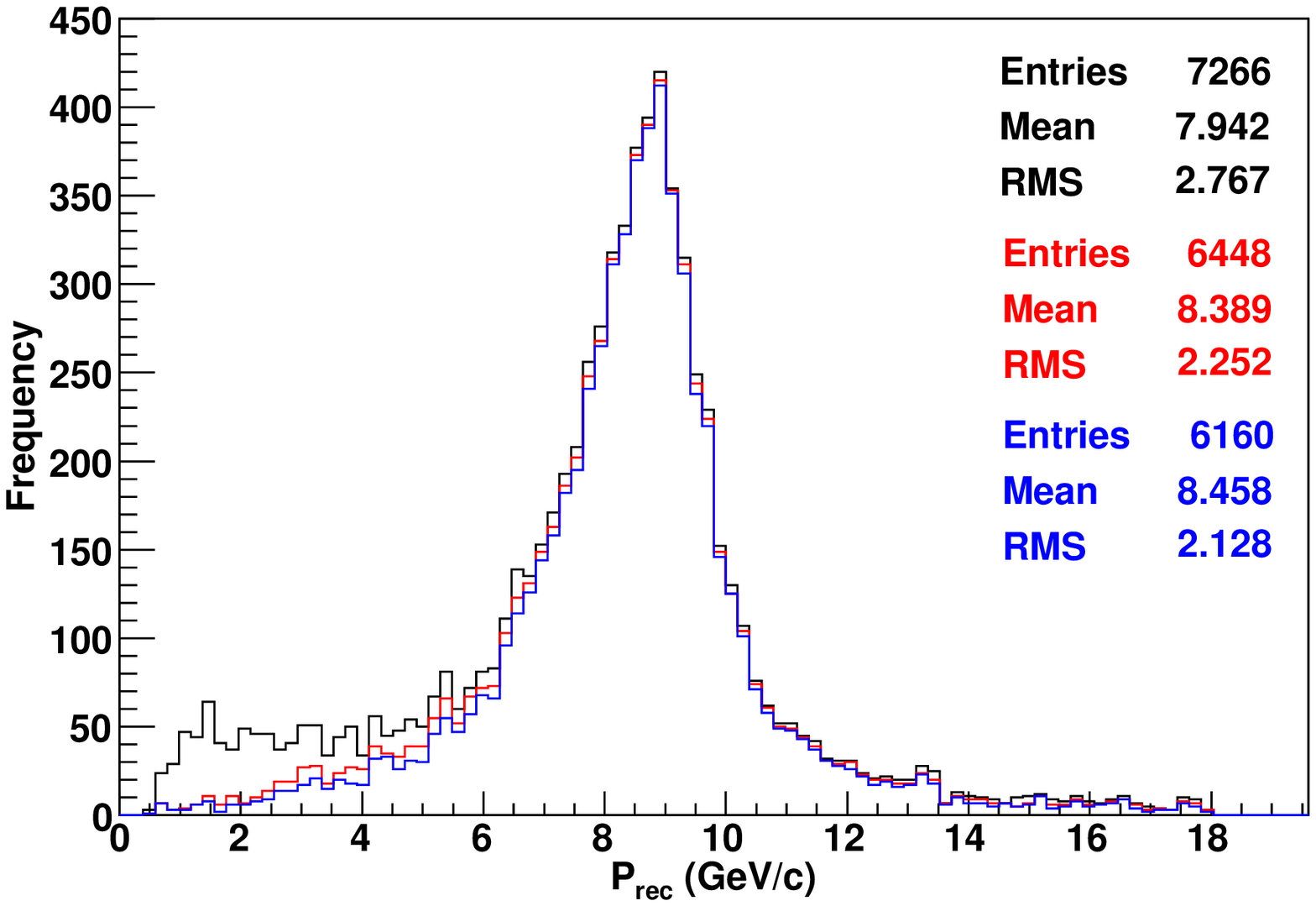}
\includegraphics[width=0.48\textwidth]{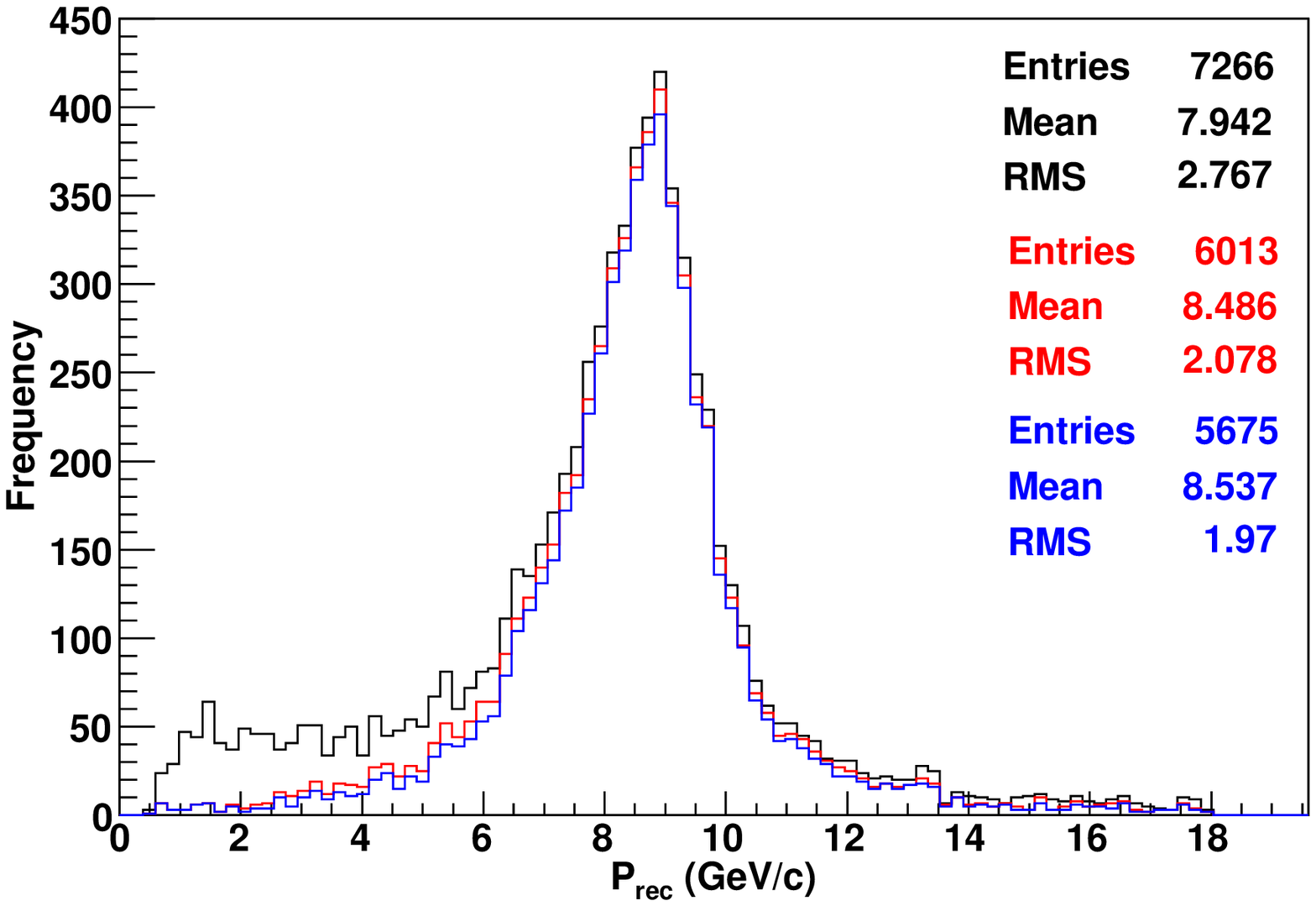}
\caption{Top (bottom) figures show the reconstructed momenta
$P_{rec}$ using selection criteria $N_{hits}>n_0$ for partially contained events
in the peripheral region with ($P_{in}$, $\cos\theta$) = (5 GeV/c, 0.65)
(top) and (9 GeV/c, 0.85) (bottom) with $n_0 = 15~(20)$ in the left
(right) figure. Fully contained events have no $N_{hits}$ constraint. In
each figure, the black curve is without constraints on $N_{hits}$, red is
with $N_{hits}/\cos\theta > n_0$ and blue is for $N_{hits} > n_0$.}
\label{fig:fixedcuts-per}
\end{figure}

\begin{figure}[htp]
\renewcommand{\figurename}{Fig.}
  \centering
\includegraphics[width=0.48\textwidth]{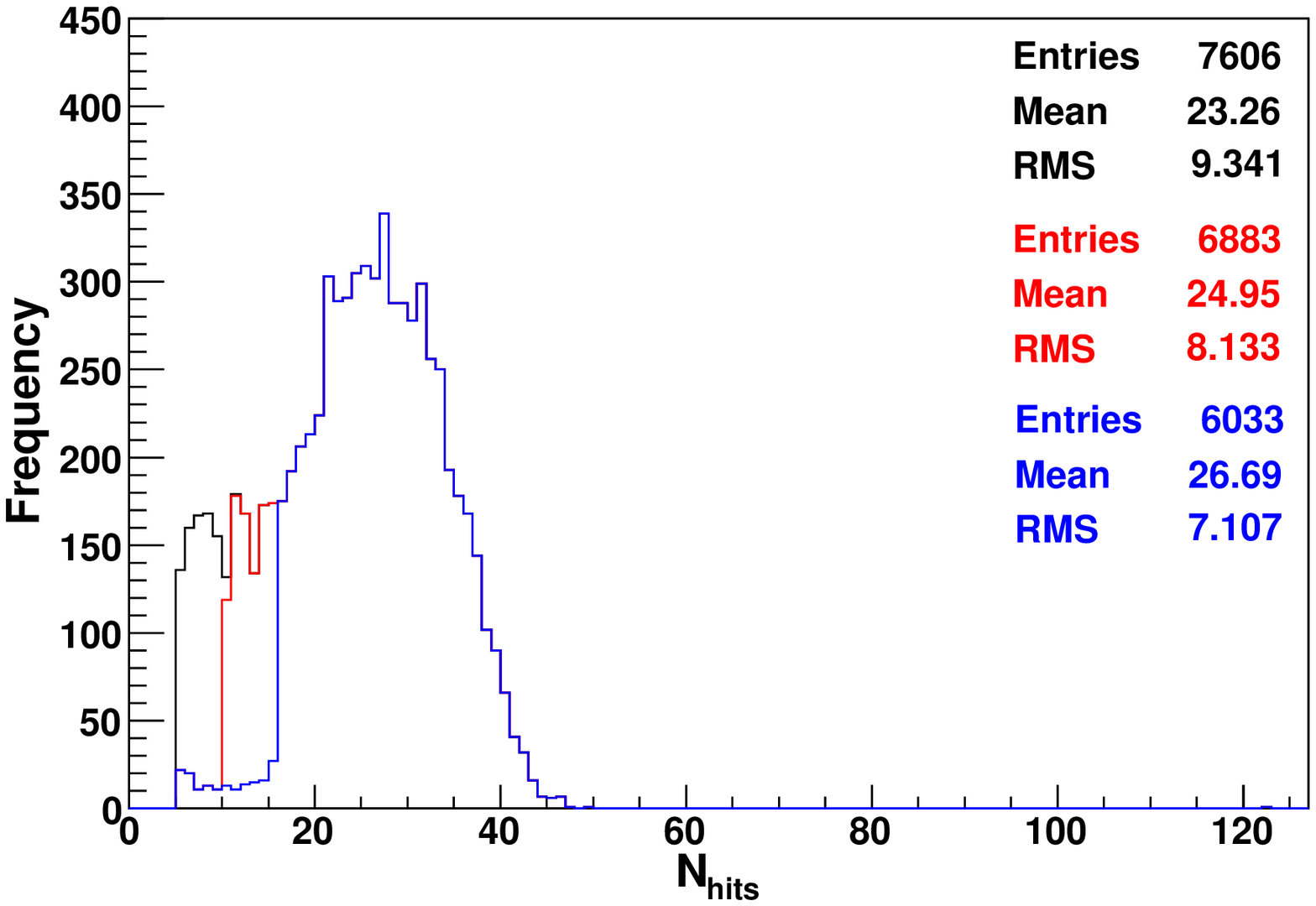}
\includegraphics[width=0.48\textwidth]{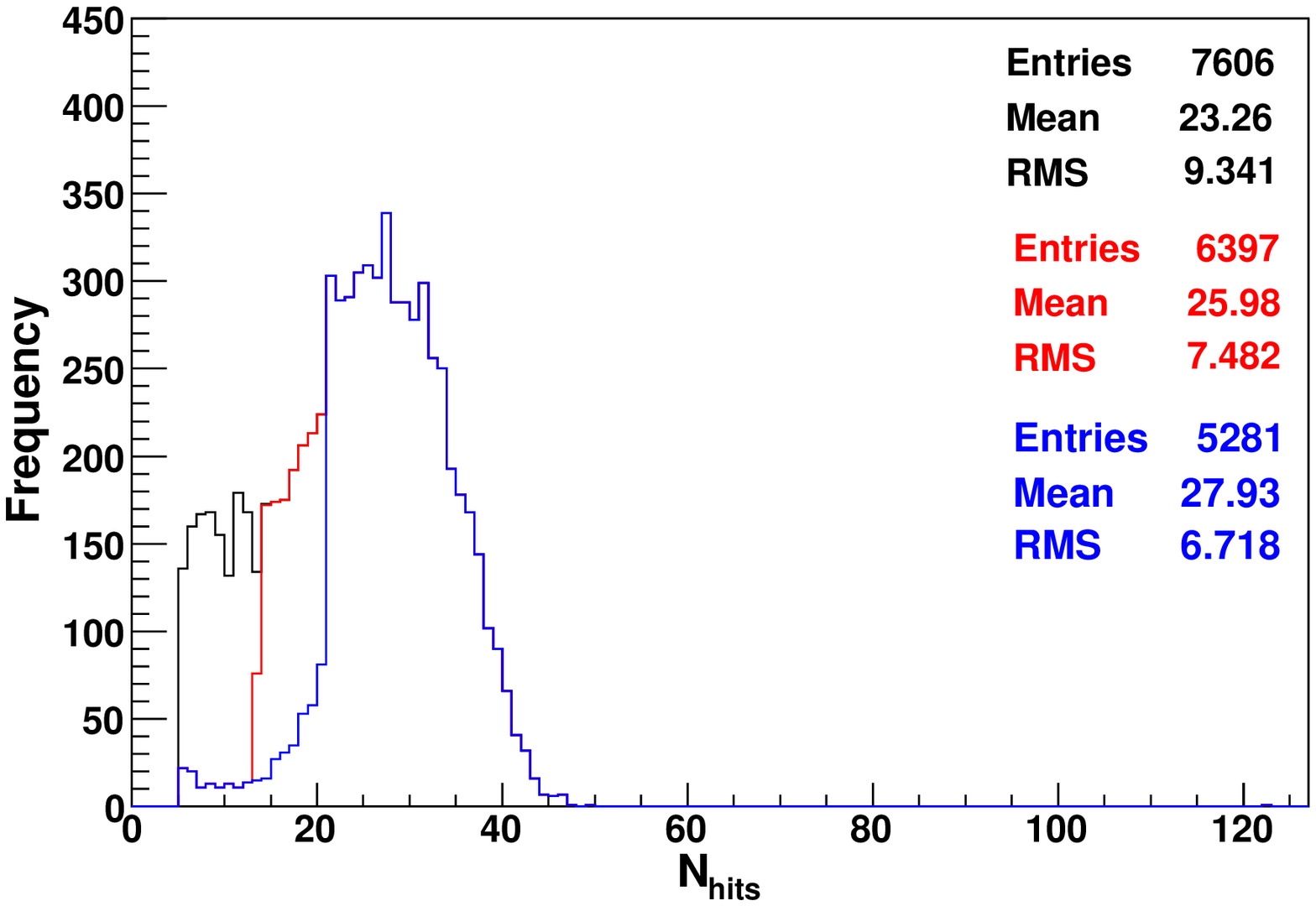}\\
\includegraphics[width=0.48\textwidth]{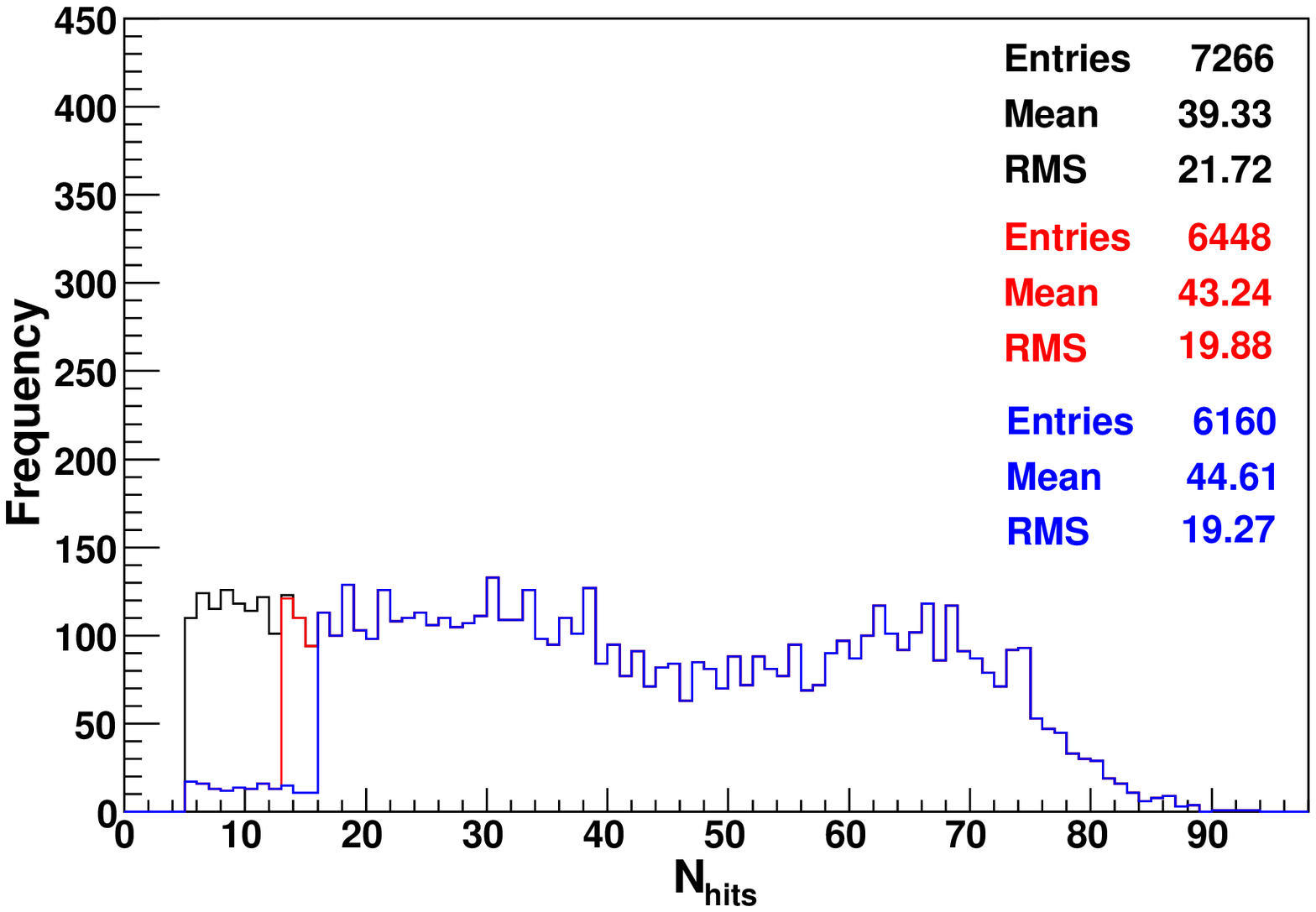}
\includegraphics[width=0.48\textwidth]{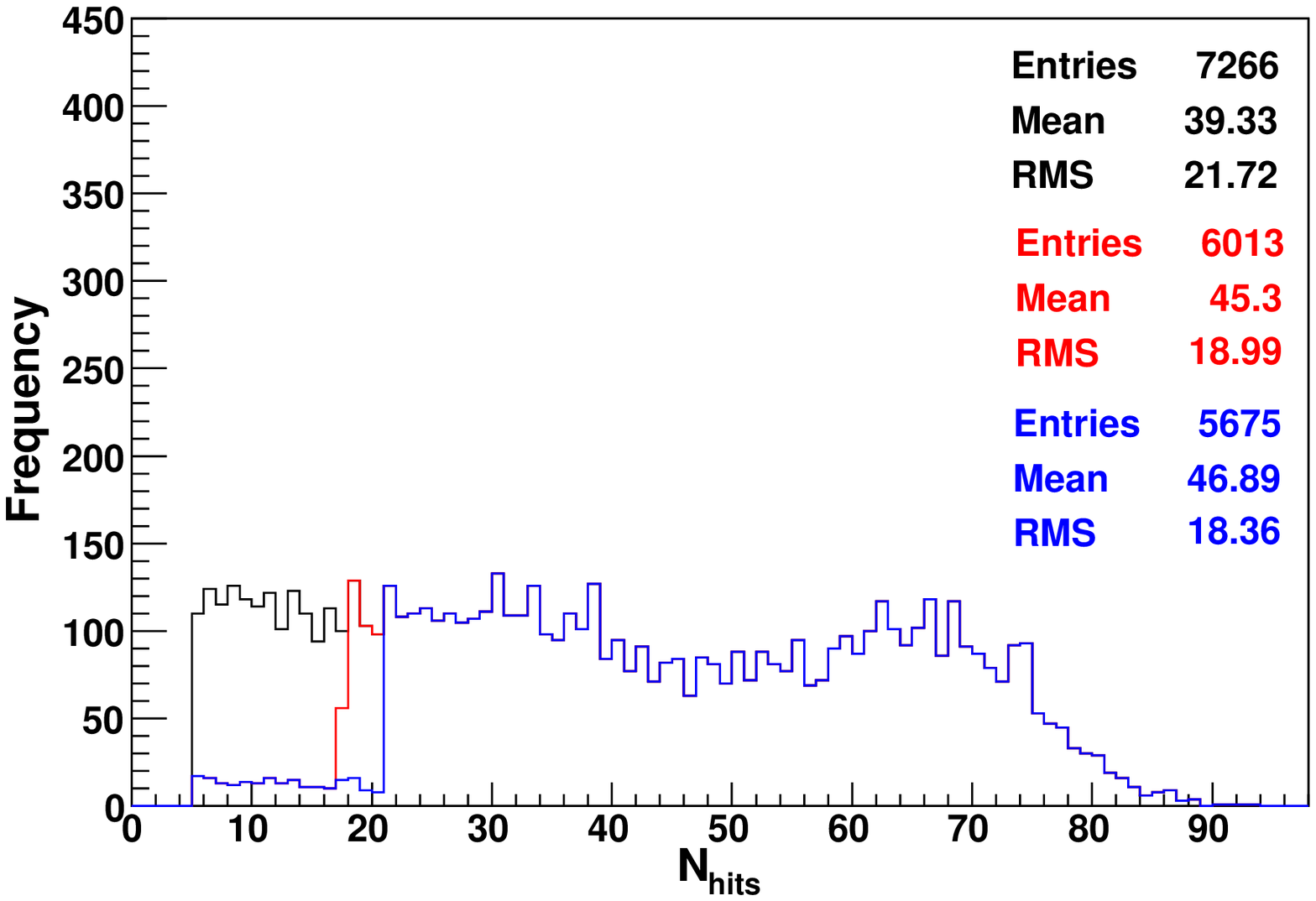}
\caption{Top (bottom) figures show the $N_{hits}$ distributions using selection criteria $N_{hits}>n_0$ for partially contained events
in the peripheral region with ($P_{in}$, $\cos\theta$) = (5 GeV/c, 0.65)
(top) and (9 GeV/c, 0.85) (bottom) with $n_0 = 15~(20)$ in the left
(right) figure. Fully contained events have no $N_{hits}$ constraint. In
each figure, the black curve is without constraints on $N_{hits}$, red is
with $N_{hits}/\cos\theta > n_0$ and blue is for $N_{hits} > n_0$.}
\label{fig:nhitsdist-per}
\end{figure}

The effect of $N_{hits} > n_0$ or $N_{hits}/\cos\theta > n_0$ can be seen from
Fig.~\ref{fig:fixedcuts-per} for the peripheral region. If the event is
fully contained, there is no constraint on $N_{hits}$; the effect of $n_0
= 15$ is shown in the left-hand side of Fig.~\ref{fig:fixedcuts-per} for
$(P_{in}, \cos\theta) = (5 \hbox{ GeV/c}, 0.65)$ and $(9 \hbox{ GeV/c},
0.85)$ where the histogram in the magnitude of the reconstructed momentum
$P_{rec}$ is plotted. For $P_{in} = 5$ GeV/c, it is noticed that the
$N_{hits}$ constraint does not affect the $P_{rec}$ momentum distribution
much, as most of the events are fully contained. But it gives a better
(more symmetrical) shape to the distribution by reducing the low-energy
tail. On the other hand, the effect for $P_{in} = 9$ GeV/c is stronger, with the hump at lower $P_{rec}$ being eliminated with the
$N_{hits}$ selection criteria.
Fig.~\ref{fig:nhitsdist-per} shows the effect of $N_{hits} >
n_0$ or $N_{hits}/\cos\theta > n_0$ on $N_{hits}$ distributions in
the peripheral region. In all cases, the few events surviving below the
constraint are from totally contained events, on which no constraint is
placed. These cause the histograms to remain non-zero in the region $N_{hits} \le n_{0}$ or $N_{hits}/\cos\theta \le n_{0}$ as seen in Fig.~\ref{fig:nhitsdist-per}. However, these events are relatively few in number, being less than 2\% (3\%) of the total reconstructed events for $n_{0}$ = 15 (20) in Fig.~\ref{fig:nhitsdist-per}. The constraint $N_{hits}/\cos\theta > 15$ is the most conservative one, with a loss of only about 10\% of the reconstructed events and is found to be more optimal than $N_{hits} > n_0$.

Different choices of $n_0$ can be used. We have shown the effect of
(a) no constraint, (b) $N_{hits} > 15$, and (c) $N_{hits}/\cos\theta > 15$
in the left-hand side of Fig.~\ref{fig:fixedcuts-per}. The last choice
is motivated by the fact that a slant-moving muon (in the absence of
magnetic field) would move a distance $d/\cos\theta$ in comparison to a
vertically upward-going muon of the same momentum that would traverse
a distance $d$. Similar figures on the right use the choice $n_0 =
20$, with the more stringent requirement giving distributions with
correspondingly smaller root-mean-square or square root of the variance
(RMS widths) by about 7--8\%, but showing a decrease in the total number
of reconstructed events by 10--15\%. Note also that increasing $n_0$
eventually leads to removal of well-reconstructed events, as visible
from the loss of events in the peak apart from just trimming the tails
for the lower momentum $P_{in} = 5$ GeV/c for $n_0 = 20$.

\begin{figure}[htp]
\renewcommand{\figurename}{Fig.}
  \centering
\includegraphics[width=0.48\textwidth]{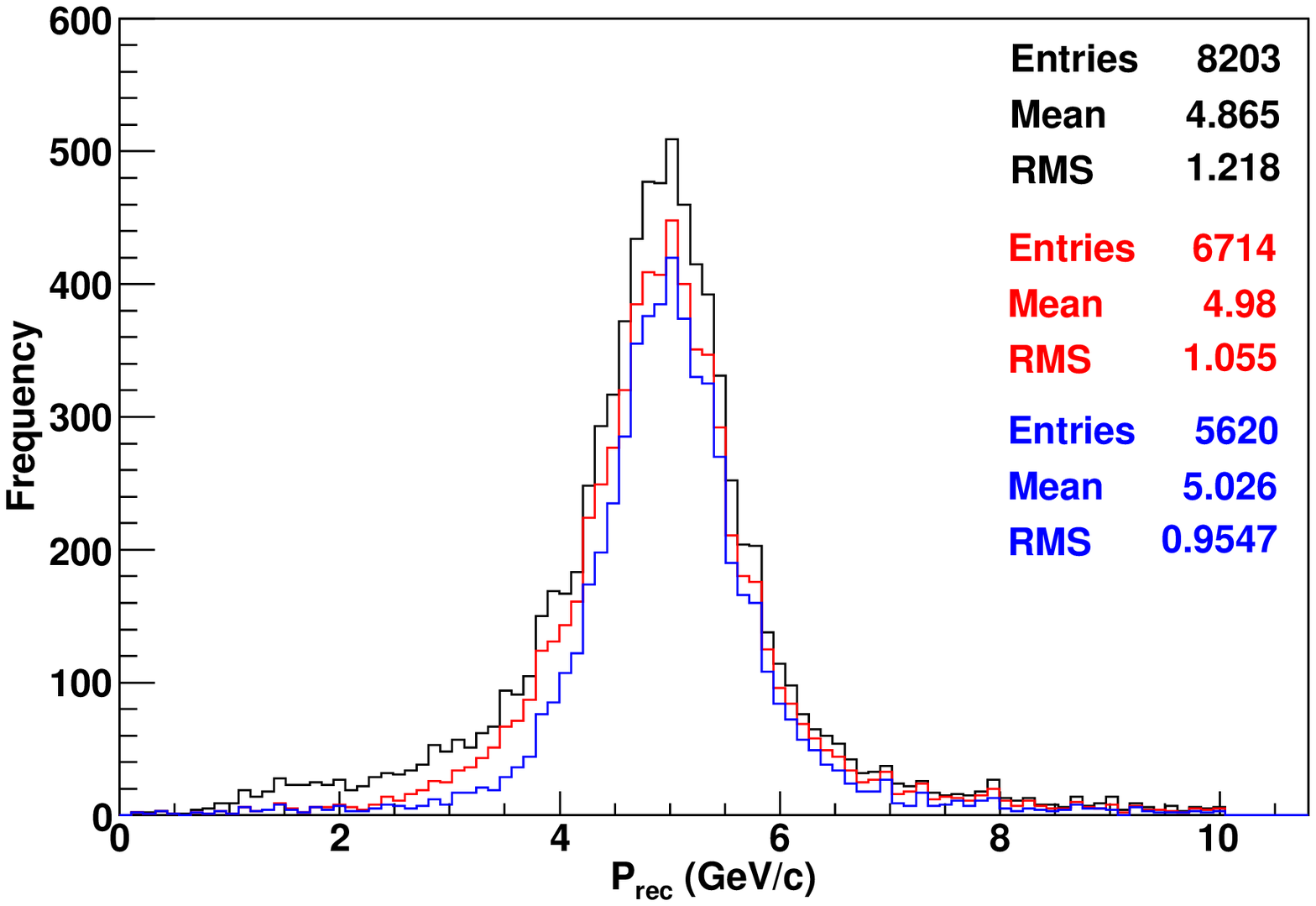}
\includegraphics[width=0.48\textwidth]{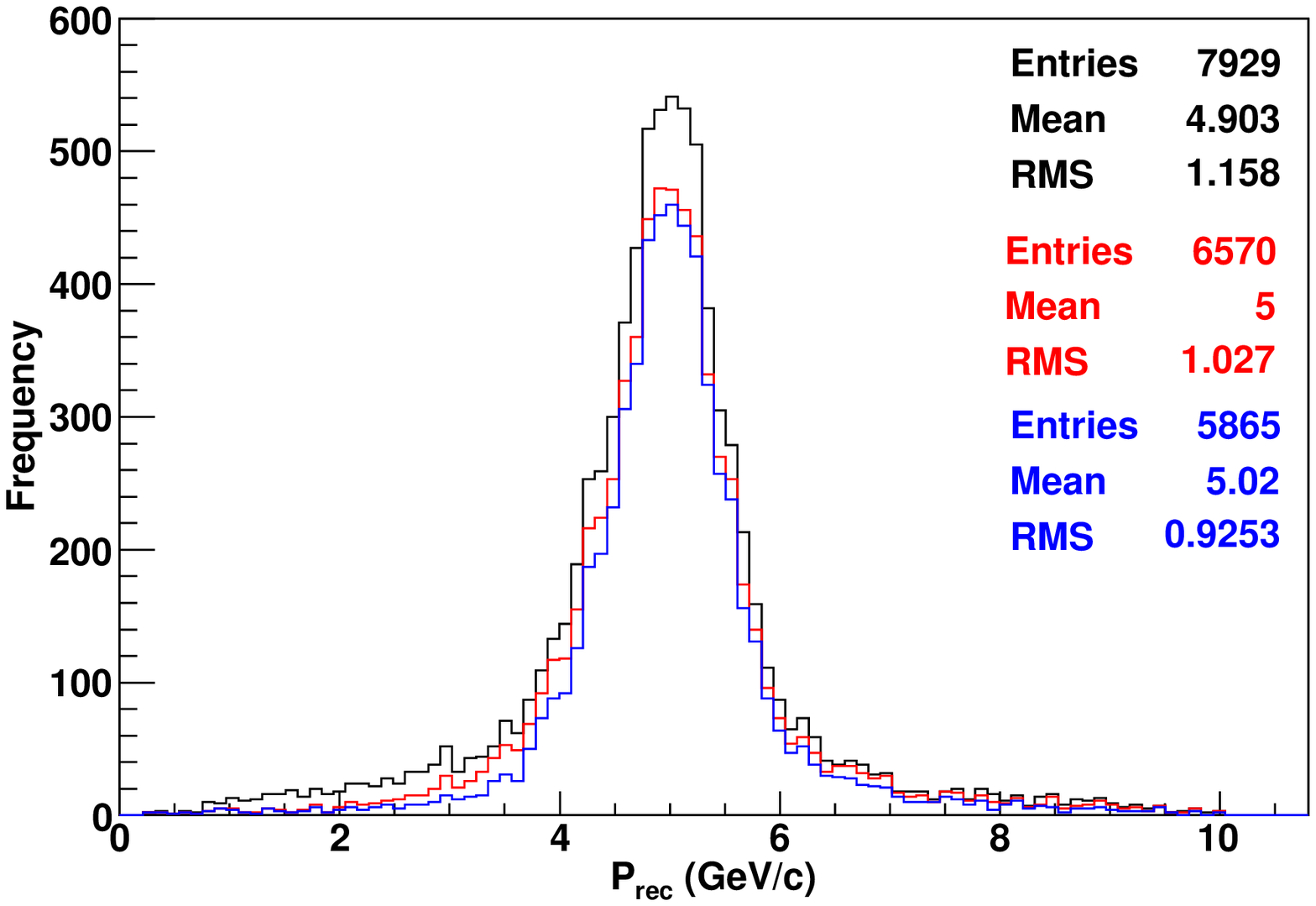}
\caption{The figures show the reconstructed momenta $P_{\rm rec}$ using the selection criteria $N_{hits}>n_0$ for partially contained events in
the side regions 9 (left) and 10 (right) for ($P_{\rm in}$, $\cos\theta$) =
(5 GeV/c, 0.65) with $n_0 = 15$.
Fully contained events have no $N_{hits}$ constraint. In each figure,
the black curve is without constraints on $N_{hits}$, red is with
$N_{hits}/\cos\theta > n_0$ and blue is for $N_{hits} > n_0$.}
\label{fig:fixedcuts-side}
\end{figure}

Similarly, the effect of the selection criteria on the reconstruction in
the side regions is shown in Fig.~\ref{fig:fixedcuts-side}. Here the
constraint on the partially contained events is not as strongly marked
as in the peripheral region: while there is certainly a decrease in the
RMS width of the distribution and in the number of selected events when
the constraint is applied, a larger fraction of events are lost due to
the constraint, with a number of ``good'' events being lost from
the peak of the distribution as well, unlike in the peripheral region.

The final choice of selection criteria will be guided by the physics
study. In case the requirement is good momentum resolution, then the
choice $n_0=20$ may be appropriate (that is, either $N_{hits}>20$ or
$N_{hits}/\cos\theta>20$). However, since the shape of the distribution is
already reasonable for $n_0=15$, this choice may be used when the focus
is not so much on precision reconstruction but on higher event
reconstruction rates. In the rest of this paper, we shall apply the
constraint $N_{hits}/\cos\theta > 15$ as being appropriate and sufficient.
This choice also improves the reconstruction efficiency of large angle
(small $\cos\theta$) events whose tracks naturally contain fewer hits
and are harder to reconstruct.

In the next section, we present the results on muon resolution and
efficiencies in the peripheral and side region using these selection
criteria.

\section{Muon Response in the Peripheral and Side Regions}

\subsection{Momentum Reconstruction Efficiency}

\begin{figure}[htp]
\renewcommand{\figurename}{Fig.}
  \centering
\includegraphics[width=0.48\textwidth]{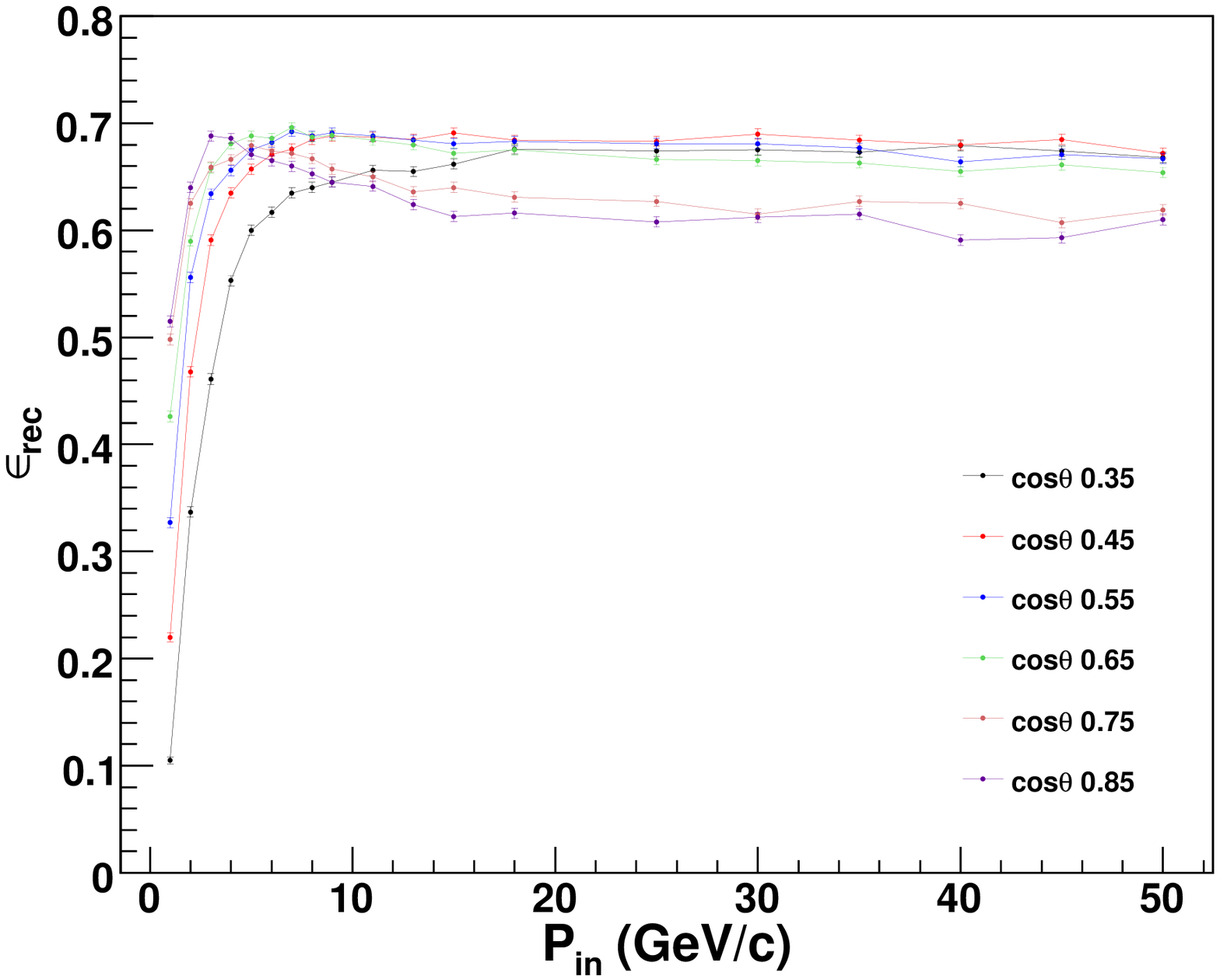}
\includegraphics[width=0.48\textwidth]{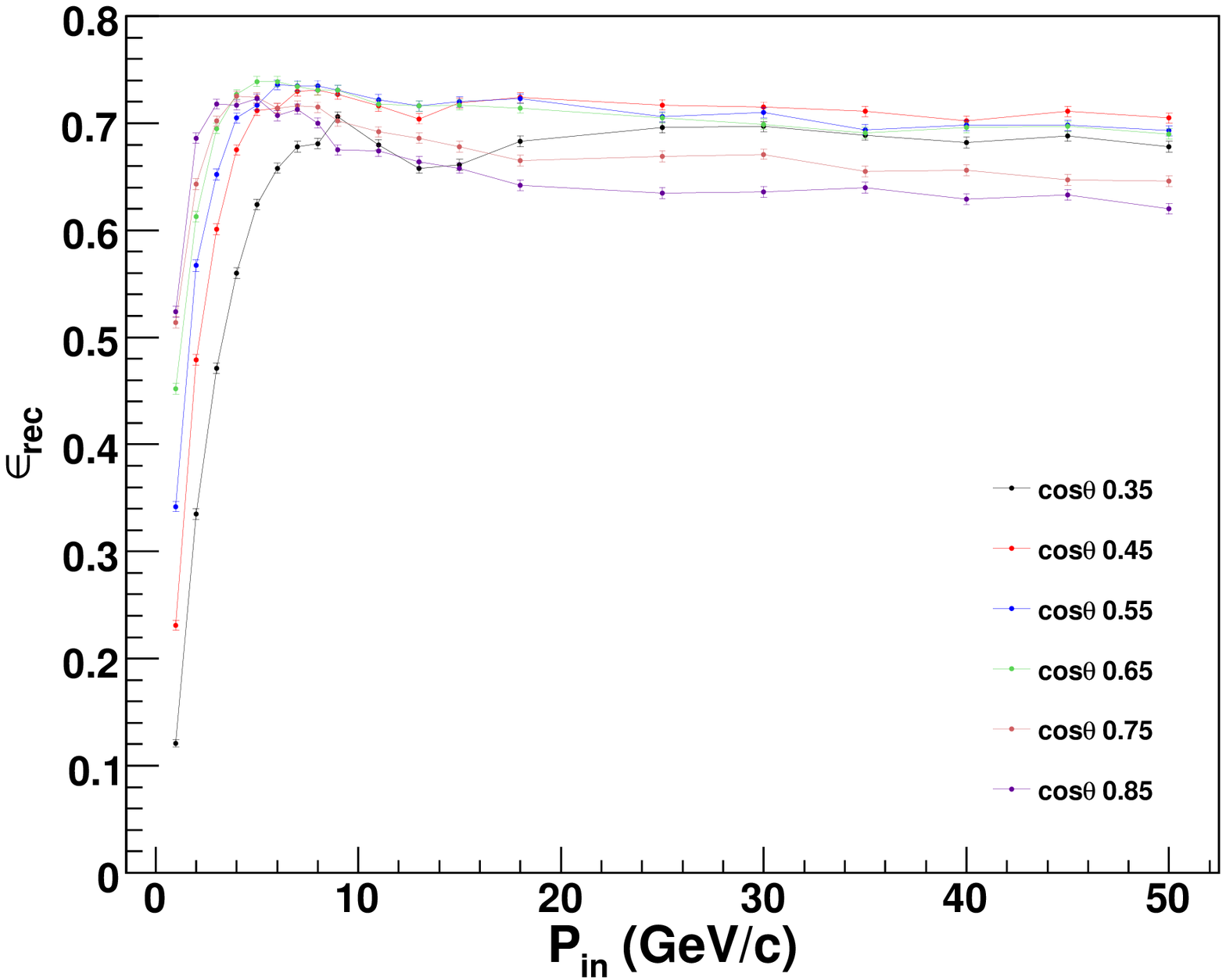}
\caption{Reconstruction efficiency averaged over all $\phi$ bins as a
function of the input momentum $P_{in}$ (GeV/c) for different zenith
angles $\cos\theta$ in the peripheral (left) and side 9 (right) regions. For a
discussion of the selection criteria see the text.}
\label{fig:recoeff-avrg}
\end{figure}

The reconstruction efficiency is defined as the ratio
of the number of reconstructed events $n_{\rm rec}$ (irrespective of
charge) to the total number of events, $N_{total}$. We have
\begin{eqnarray}
\epsilon_{\rm rec} & = & \frac{n_{\rm rec} }{N_{\rm
total}}~, \\ \nonumber
\hbox{with error  } \delta \epsilon_{\rm rec} & = &
\sqrt{\epsilon_{\rm rec}(1-\epsilon_{\rm rec})/N_{\rm total}}~~.
\end{eqnarray} 
Fig.~\ref{fig:recoeff-avrg} shows the reconstruction efficiency averaged
over $\phi$ as a function of input momentum for different $\cos\theta$
values in the peripheral and side regions.

The reconstruction efficiency increases for all angles, from $P_{in}
= 1$ GeV/c since the number of hits increases as the particle crosses
more layers. Since there are fewer hits for more slant-angled
muons, the efficiency at a given momentum is better for larger values
of $\cos\theta$. Also, the reconstruction efficiency is very similar
for all the peripheral and side regions.

In all cases, the slight worsening of the efficiency for $\cos\theta
= 0.85$ at higher momenta is spurious and is due to the selection
criterion that the event should reconstruct exactly one track. At large
angles, it is more likely that two portions of a track on either side
of a dead space such as a support structure are reconstructed as two
separate tracks. Efforts are on to retrieve such events by improving
the reconstruction code \cite{Kolahal}. When these tracks are correctly
reconstructed, the efficiency is expected to saturate rather than fall
off at these momentum values. Again, such tracks are not expected to be
troublesome in genuine neutrino events, as discussed earlier.

\subsection{Relative Charge Identification Efficiency}

The charge identification (cid) of the particle is critical in many studies since it distinguishes events initiated by neutrinos and anti-neutrinos;
these have different matter effects as they propagate through the
Earth and hence give the required sensitivity to the neutrino mass
hierarchy. The charge of the particle is determined from the direction
of curvature of the track in the magnetic field. Relative charge
identification efficiency is defined as the ratio of the number of events
with correct charge identification, $n_{\rm cid}$, to the total number of
reconstructed events, $n_{\rm rec}$, i.e., 
\begin{eqnarray} \epsilon_{\rm
cid} & = & \frac{n_{\rm cid} } {n_{\rm rec}}~,
\end{eqnarray} 
\hbox{where
the errors in $n_{\rm cid}$ and $n_{\rm rec}$ are correlated so that
the error in the ratio is calculated as:} 
\begin{eqnarray}
\delta \epsilon_{\rm cid} & = &
\sqrt{\epsilon_{\rm
cid}(1-\epsilon_{\rm
cid})/n_{\rm rec}}~. \nonumber
\end{eqnarray}

\begin{figure}[htp]
\renewcommand{\figurename}{Fig.}
  \centering
\includegraphics[width=0.48\textwidth]{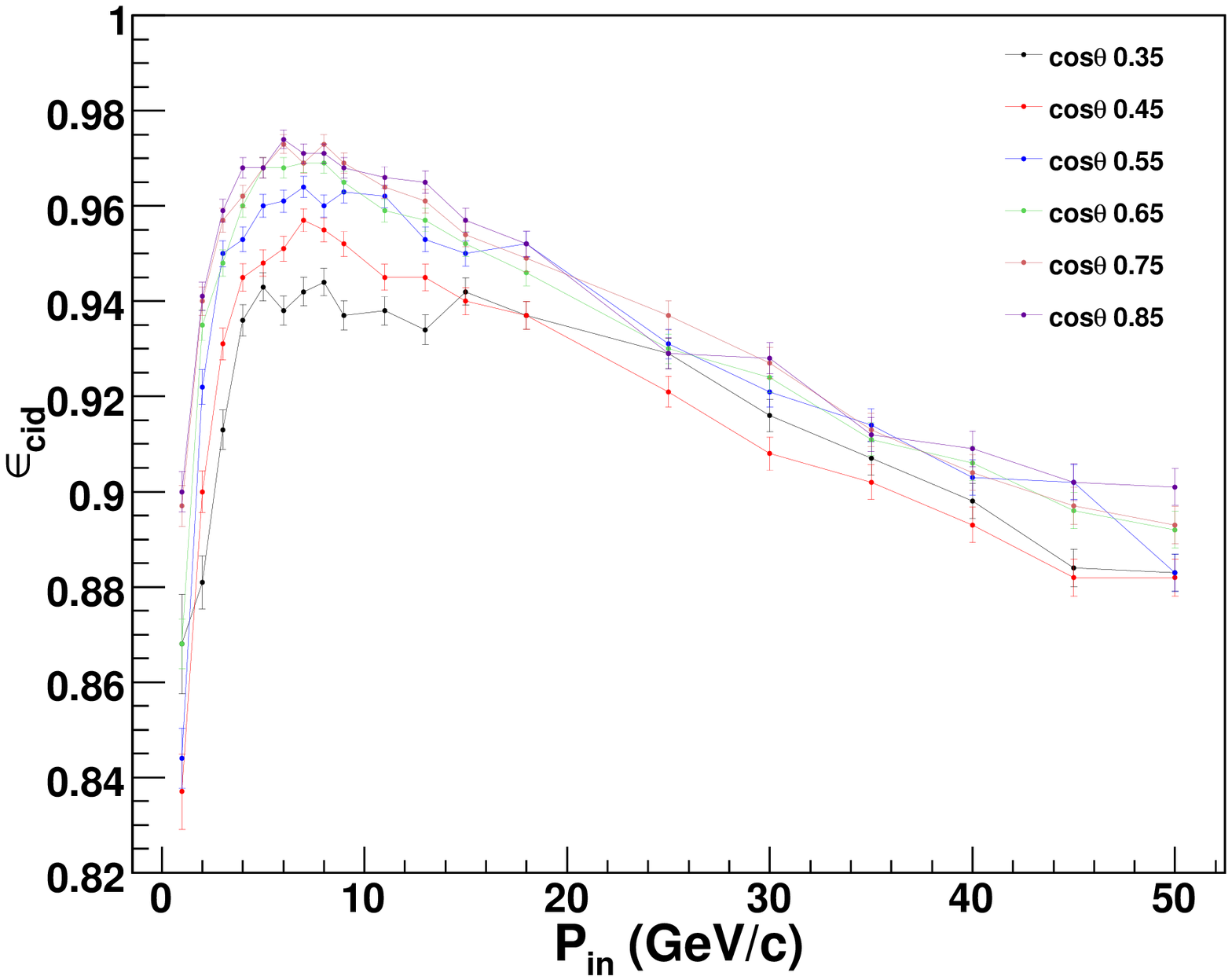}
\includegraphics[width=0.48\textwidth]{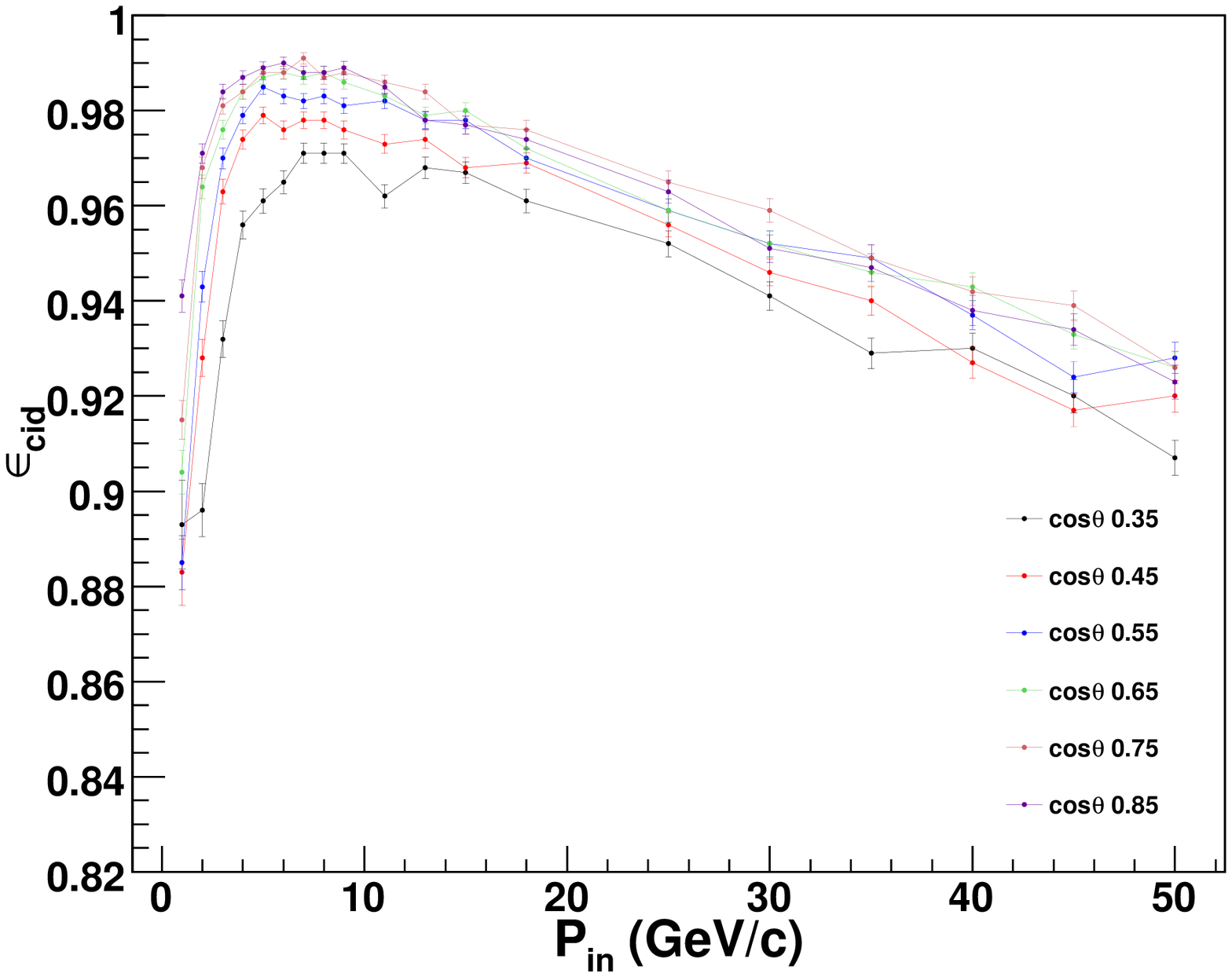}
\caption{Charge identification (cid) efficiency averaged over all $\phi$ bins as a function of the input momentum $P_{in}$ (GeV/c) for different zenith
angles $\cos\theta$ in the peripheral (left) and side 9 (right) regions. For a
discussion of the selection criteria see the text.}
\label{fig:cideff-avrg}
\end{figure}

Fig.~\ref{fig:cideff-avrg} shows the relative charge identification
efficiency as a function of input momentum for different $\cos\theta$
values in the peripheral and side region 9. (Similar results apply for
side region 10). The muon undergoes multiple scattering while propagating
in the detector; for small momentum, since the number of layers traversed
is small, this may lead to an incorrectly reconstructed direction
of bending, resulting in the wrong charge identification. Hence the
charge identification efficiency is relatively poor at lower energies
but as the energy increases cid efficiency also improves. At very high
input momenta, bending due to the magnetic field is less. For partially
contained events, only the initial relatively straight portion of the track
is contained within the detector; this leads to large momentum
uncertainty as well as mis-identification of charge. Overall the
relative charge identification efficiency is marginally smaller than in
the central region because of the smaller magnetic field.

\subsection {Direction (up/down) Reconstruction}

\begin{figure}[htp]
\renewcommand{\figurename}{Fig.}
  \centering
\includegraphics[width=0.48\textwidth]{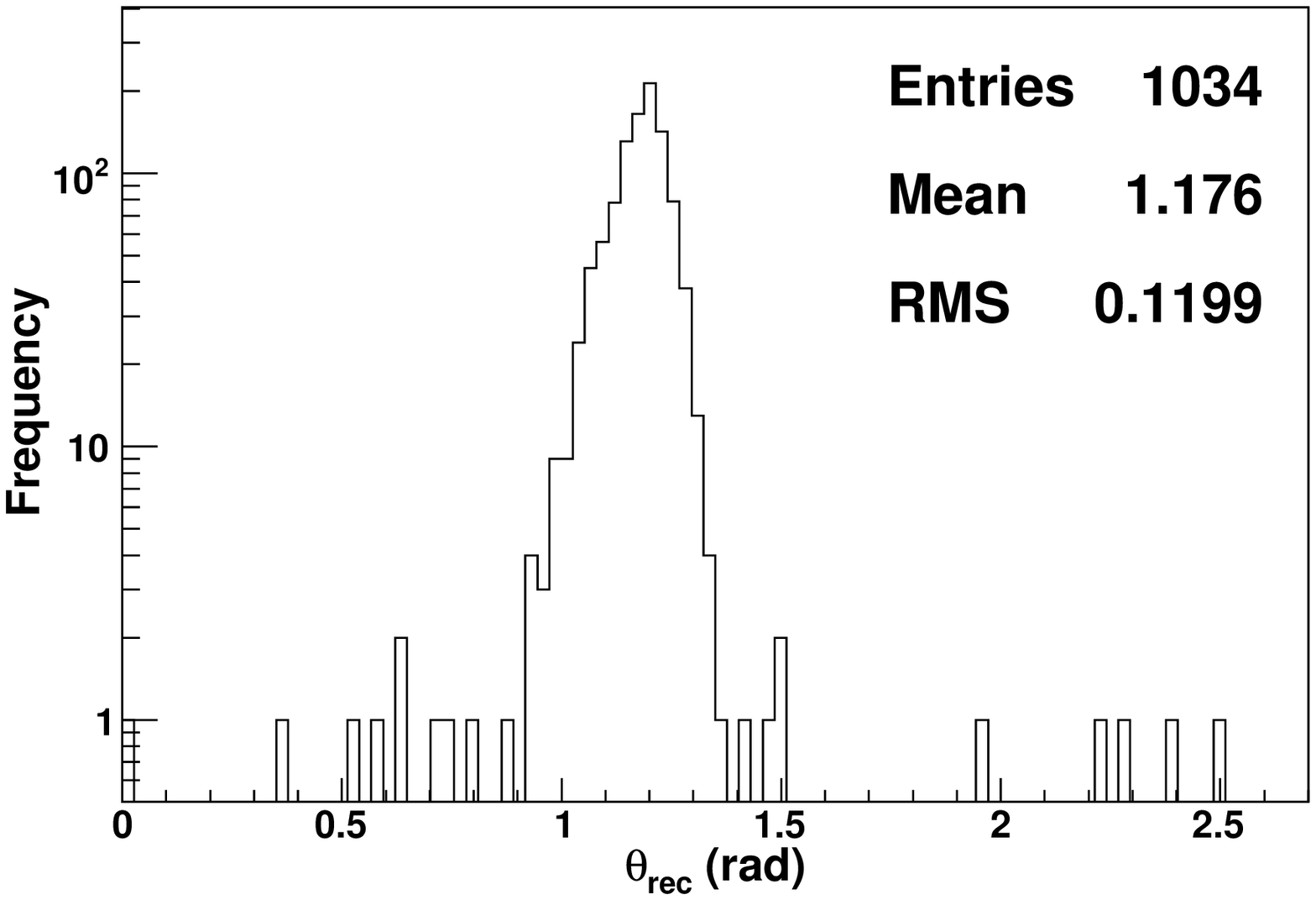}
\includegraphics[width=0.48\textwidth]{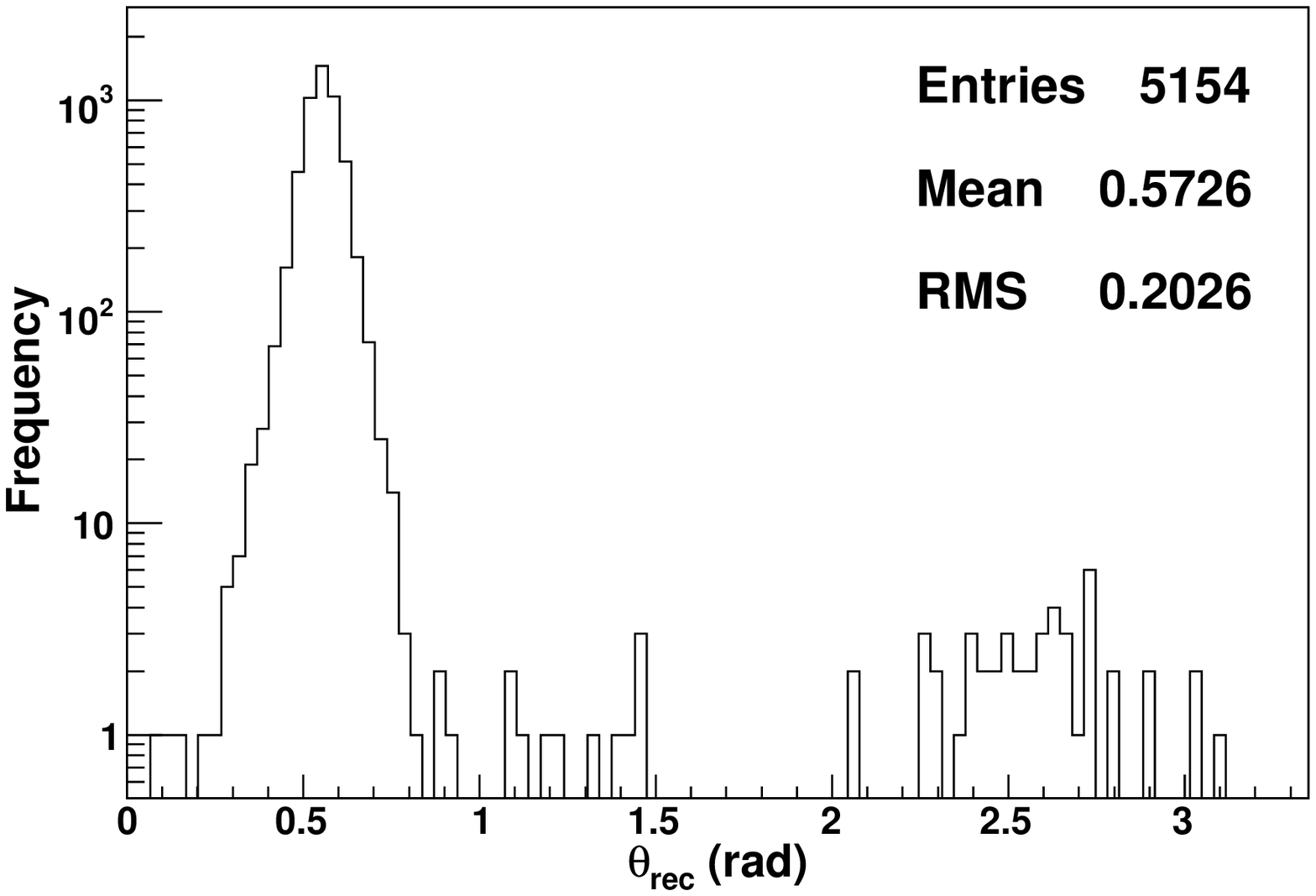}
\caption{Reconstructed zenith angle distributions for $P_{in}$ = 1 GeV/c at $\cos\theta$ = 0.35 (left) and 0.85 (right) respectively, in the peripheral region. Note that the y-axis scales are different for the two plots.}
\label{fig:direction_per}
\end{figure}

\begin{figure}[htp]
\renewcommand{\figurename}{Fig.}
  \centering
\includegraphics[width=0.48\textwidth]{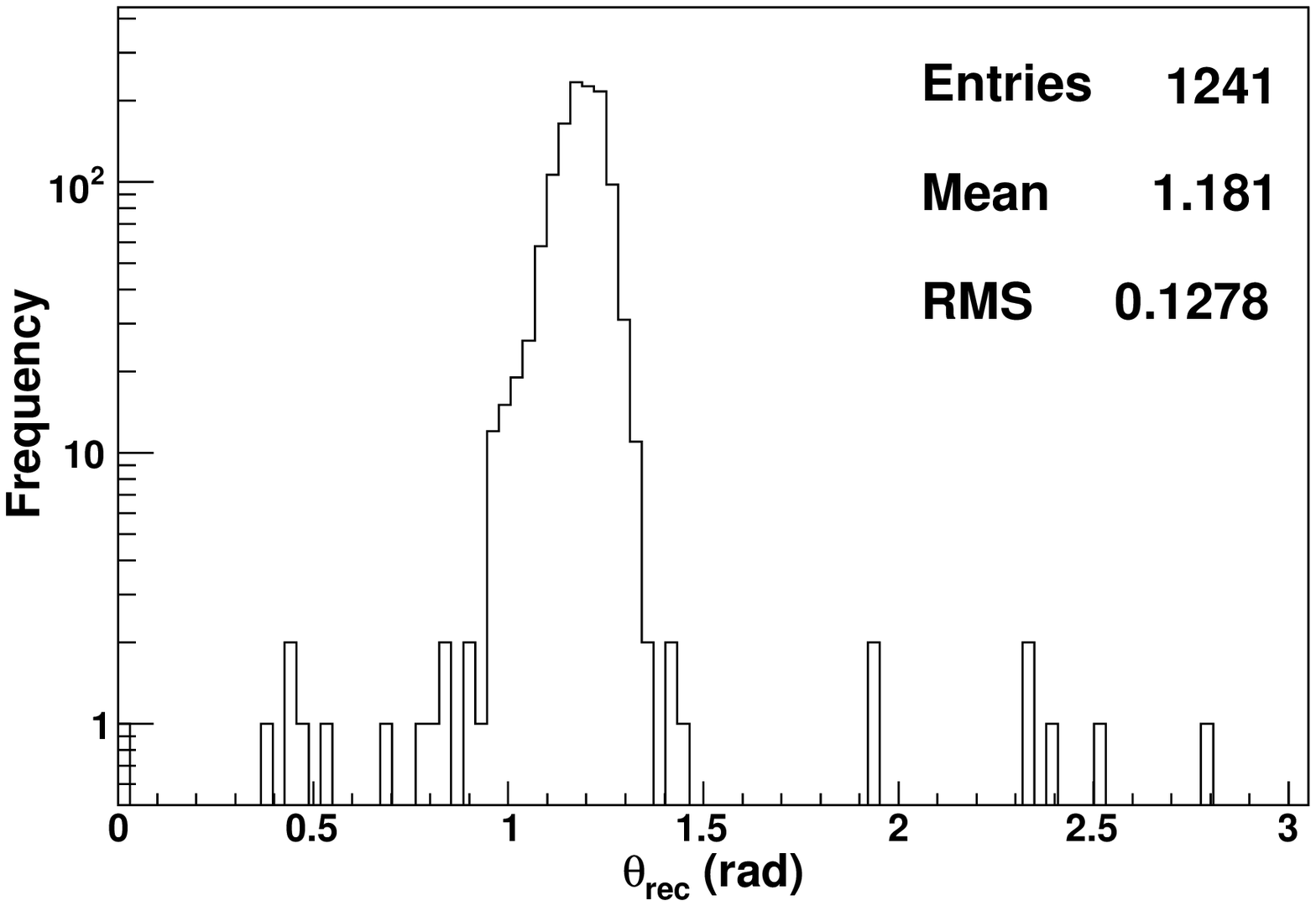}
\includegraphics[width=0.48\textwidth]{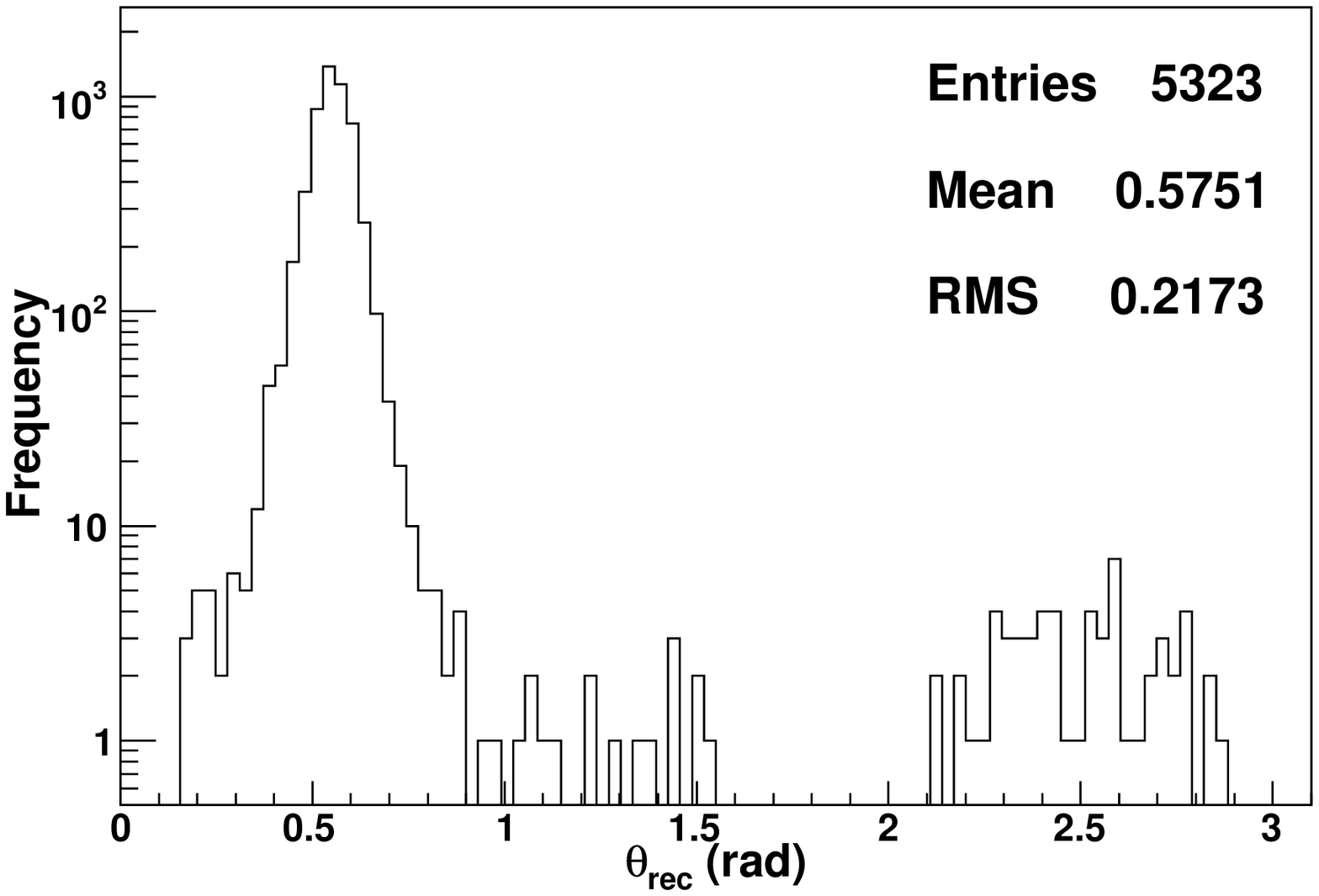}
\caption{Reconstructed zenith angle distributions for $P_{in}$ = 1 GeV/c at $\cos\theta$ = 0.35 (left) and 0.85 (right) respectively, in the side region 9. Note that the y-axis scales are different for the two plots.}
\label{fig:direction_side9}
\end{figure}

The reconstructed zenith angle distributions for $P_{in}$ = 1 GeV/c
at $\cos\theta$ = 0.35 and $\cos\theta$ = 0.85 in the peripheral
and side region 9 are shown in Figs.~\ref{fig:direction_per} and
\ref{fig:direction_side9} respectively.

From Fig.~\ref{fig:direction_per}, it is noticed that there are few
events reconstructed in the downward direction (wrong direction) with $\theta_{rec} > \pi/2$. For $P_{in}$ = 1 GeV/c with $\cos\theta$ = 0.35 (0.85), this fraction is about 0.48 (0.89)\% and it drops to a negligible value at higher energies for all $\cos\theta$.  Similar results are obtained for the side region 9
as can be seen from Fig.~\ref{fig:direction_side9}. This small fraction
also contributes to wrong cid since the relative bending in the magnetic
field is measured w.r.t the muon momentum direction.

The direction determination depends on the time resolution while the
charge identification depends also on the strength of the magnetic
field. A 1 GeV/c muon with $\cos\theta \sim 1$ traverses about 12 layers;
this corresponds to a time difference between first and last hit of about
4 ns. Since the RPCs have a time resolution of 1 ns, this explains why
the fraction of muons whose direction is wrongly determined is small.

\subsection{Zenith Angle Resolution}

\begin{figure}[htp]
\renewcommand{\figurename}{Fig.}
\begin{center}
\includegraphics[width=0.48\textwidth,height=0.33\textwidth]{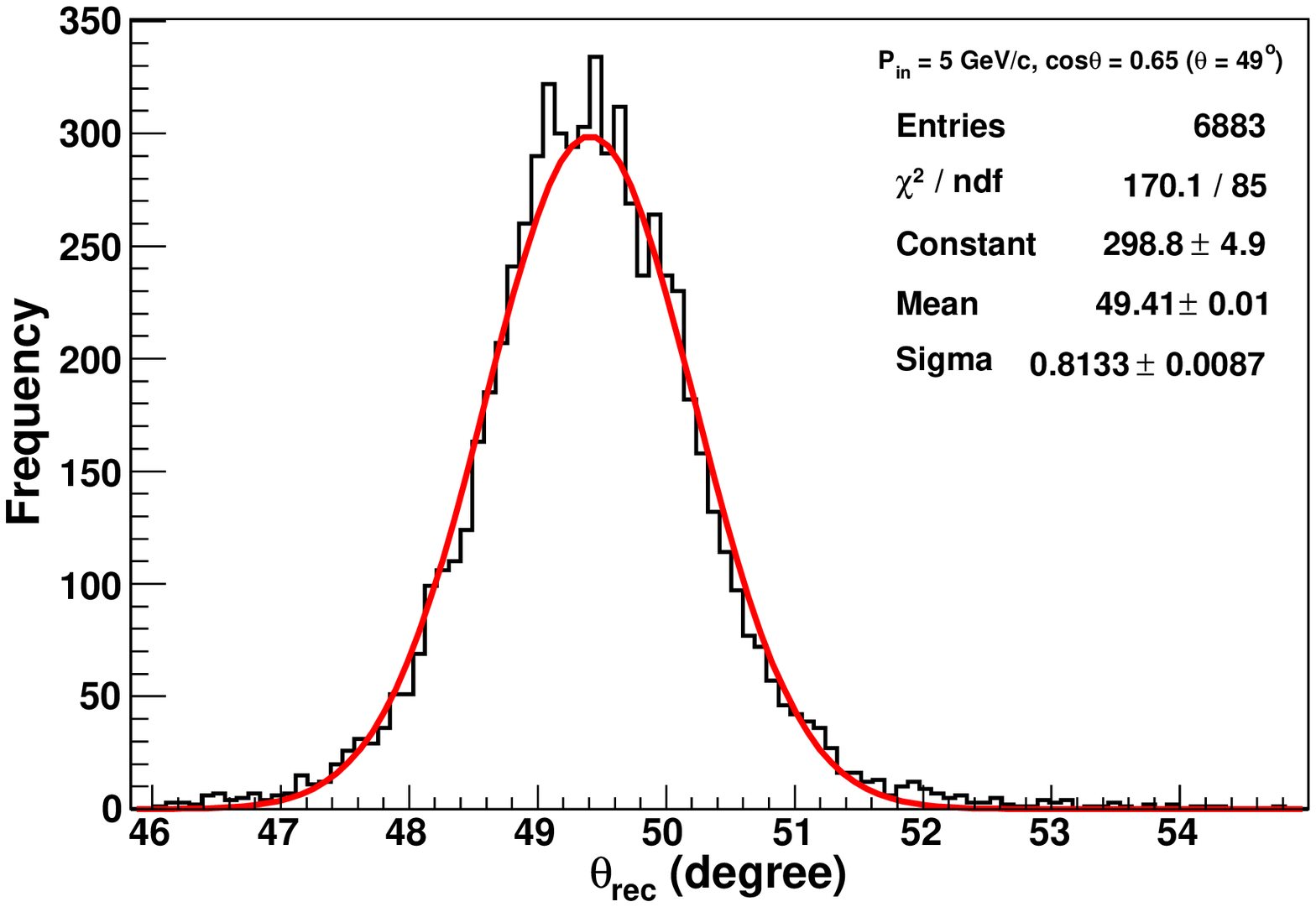}
\includegraphics[width=0.48\textwidth,height=0.33\textwidth]{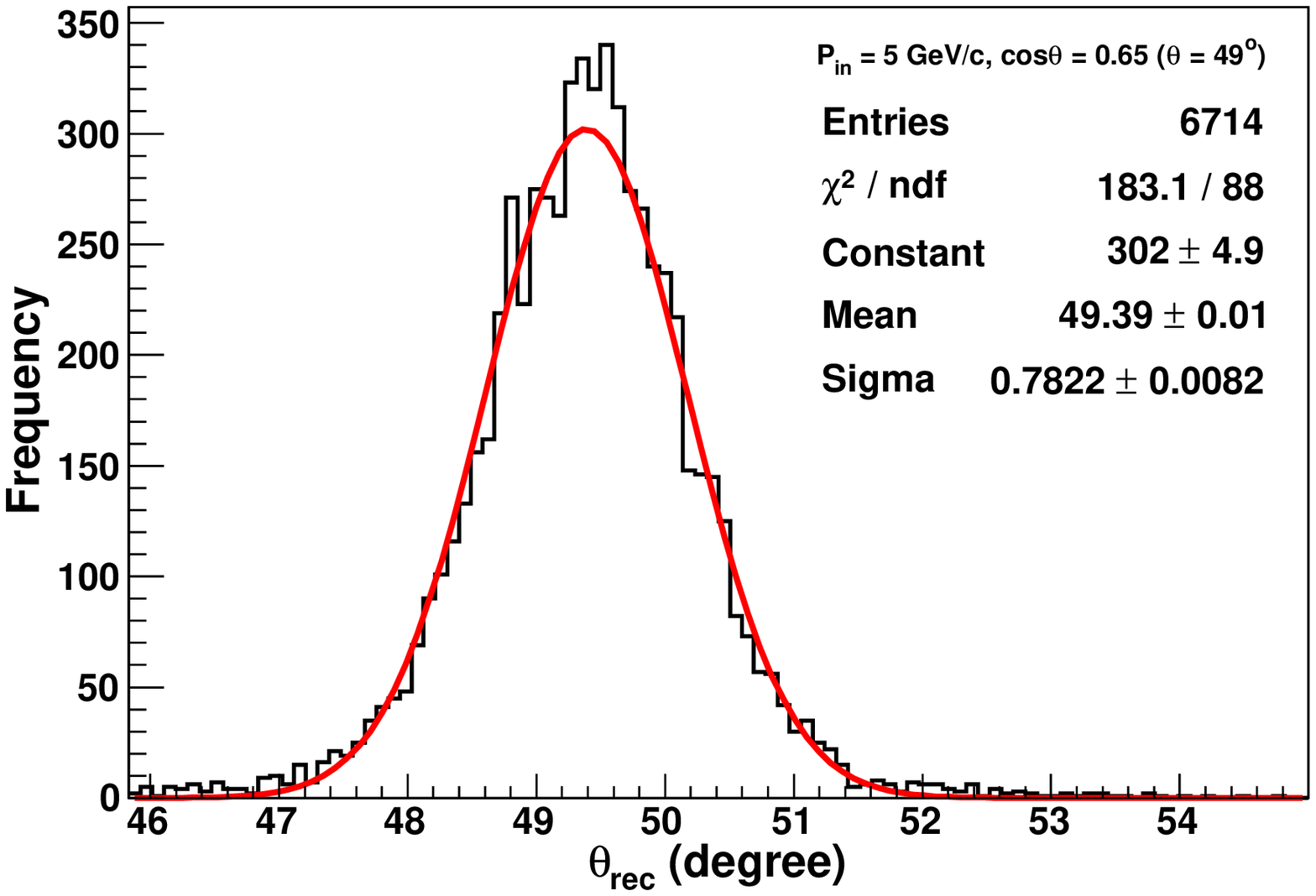}
\end{center}
\caption{Reconstructed distribution $\theta_{rec}$ for input $(P_{\rm
in}, \cos\theta) = (5\hbox{ GeV/c}, 0.65)$ in the peripheral region (left)
and side region 9 (right). The selection criteria are the same as before.}
\label{fig:theta_histo}
\end{figure}

Those events which are successfully reconstructed (for all $\phi$) are
analysed for their zenith angle resolution. The events distribution as
a function of the reconstructed zenith angle $\theta_{rec}$ is shown in
Fig.~\ref{fig:theta_histo} for a sample input $(P_{\rm in}, \cos\theta) =
(5\hbox{ GeV/c}, 0.65)$ for the peripheral and side region 9 respectively.

\begin{figure}[htp]
\renewcommand{\figurename}{Fig.}
\begin{center}
\includegraphics[width=0.48\textwidth,height=0.33\textwidth]{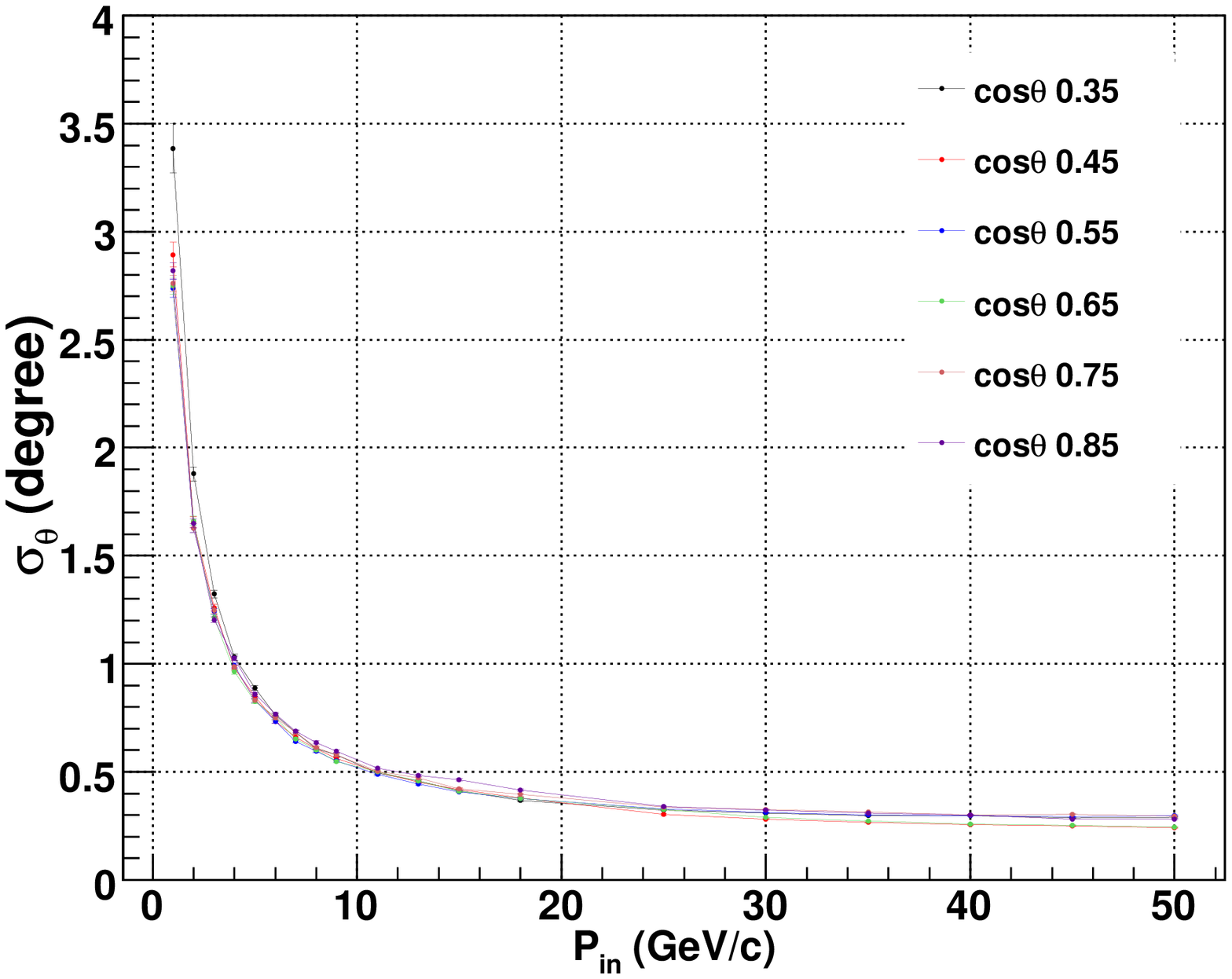}
\includegraphics[width=0.48\textwidth,height=0.33\textwidth]{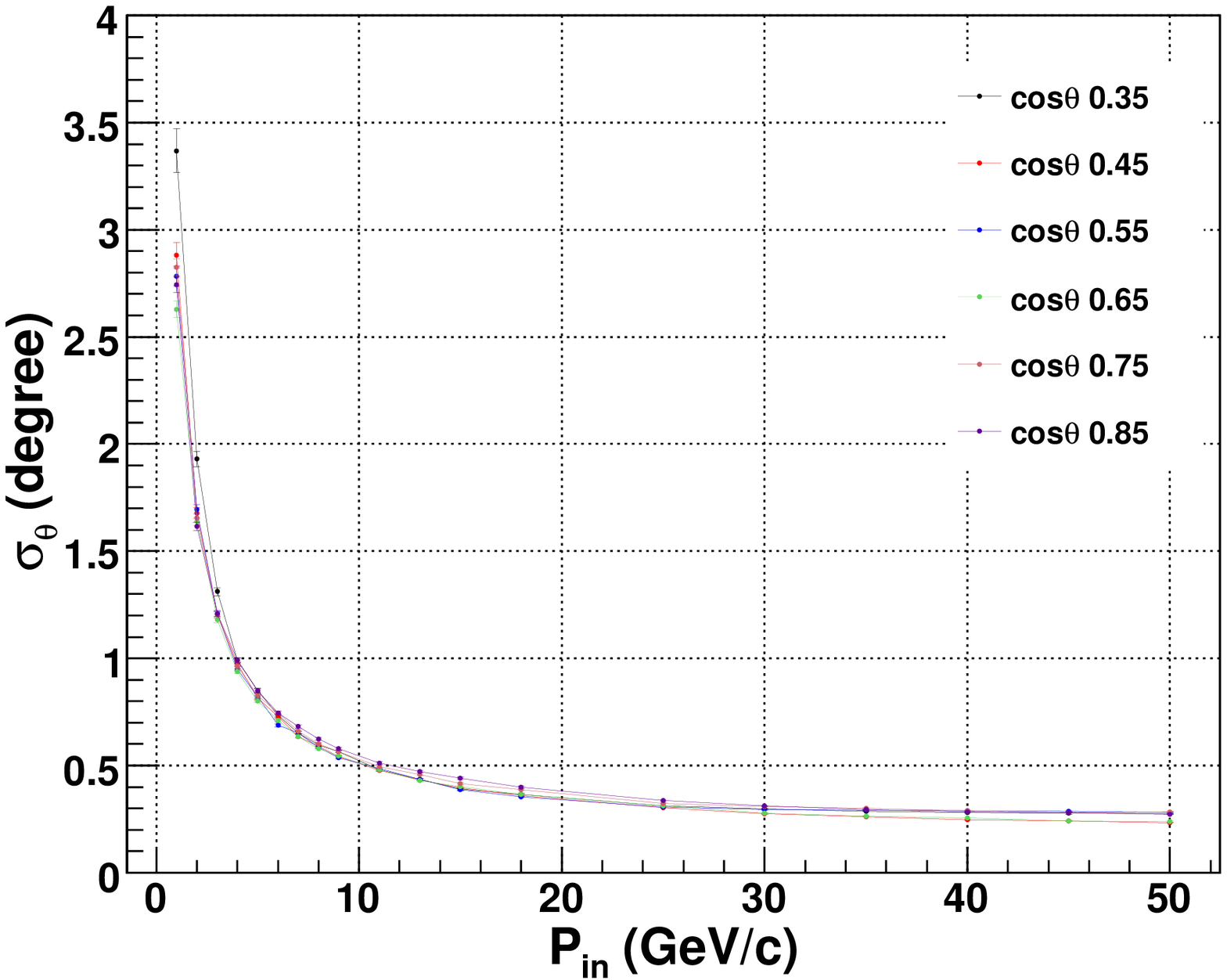}
\end{center}
\caption{Resolution, $\sigma_\theta$, of reconstructed angle
$\theta_{rec}$ as a function of the input momentum $P_{\rm in}$ (GeV/c)
for different values of input $\cos\theta$ in the peripheral region (left)
and side region 9 (right). The selection criteria are the same as before.}
\label{fig:theta}
\end{figure}

The angular resolution is good in both the regions and is in fact better
than about a degree for input momentum greater than a few GeV, as seen
in Fig.~\ref{fig:theta}, with the resolution being marginally better in
the side region. Similar results are obtained in side region 10 as well.
In addition, the fraction of events reconstructed in the wrong direction
(wrong quadrant of $\cos\theta$) is negligibly small, being less than
0.5\% for $P_{\rm in} \ge 2$ GeV/c.

\subsection{Muon Momentum Response}

\begin{figure}[bhp]
\renewcommand{\figurename}{Fig.}
  \centering
\includegraphics[width=0.44\textwidth]{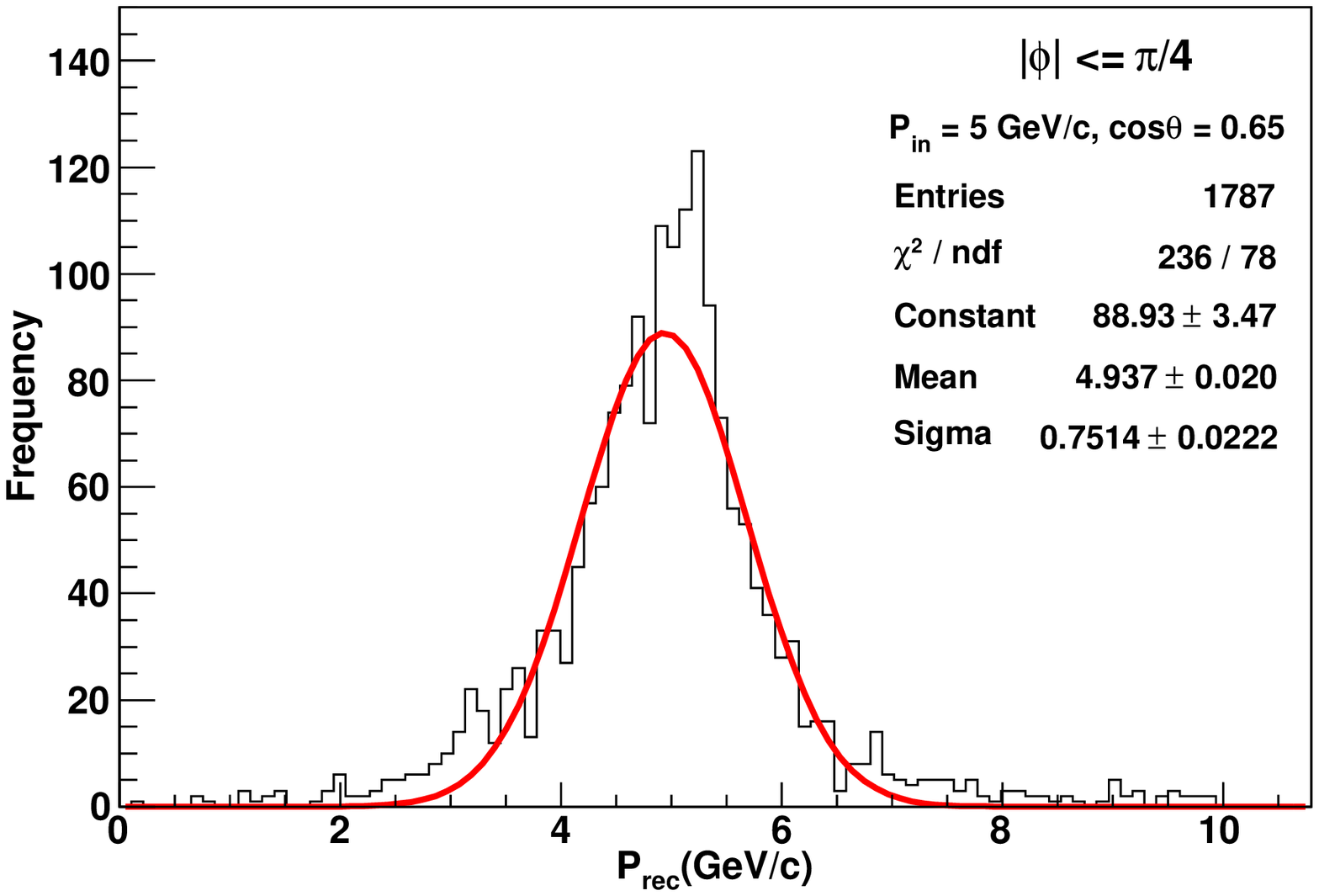}
\includegraphics[width=0.44\textwidth]{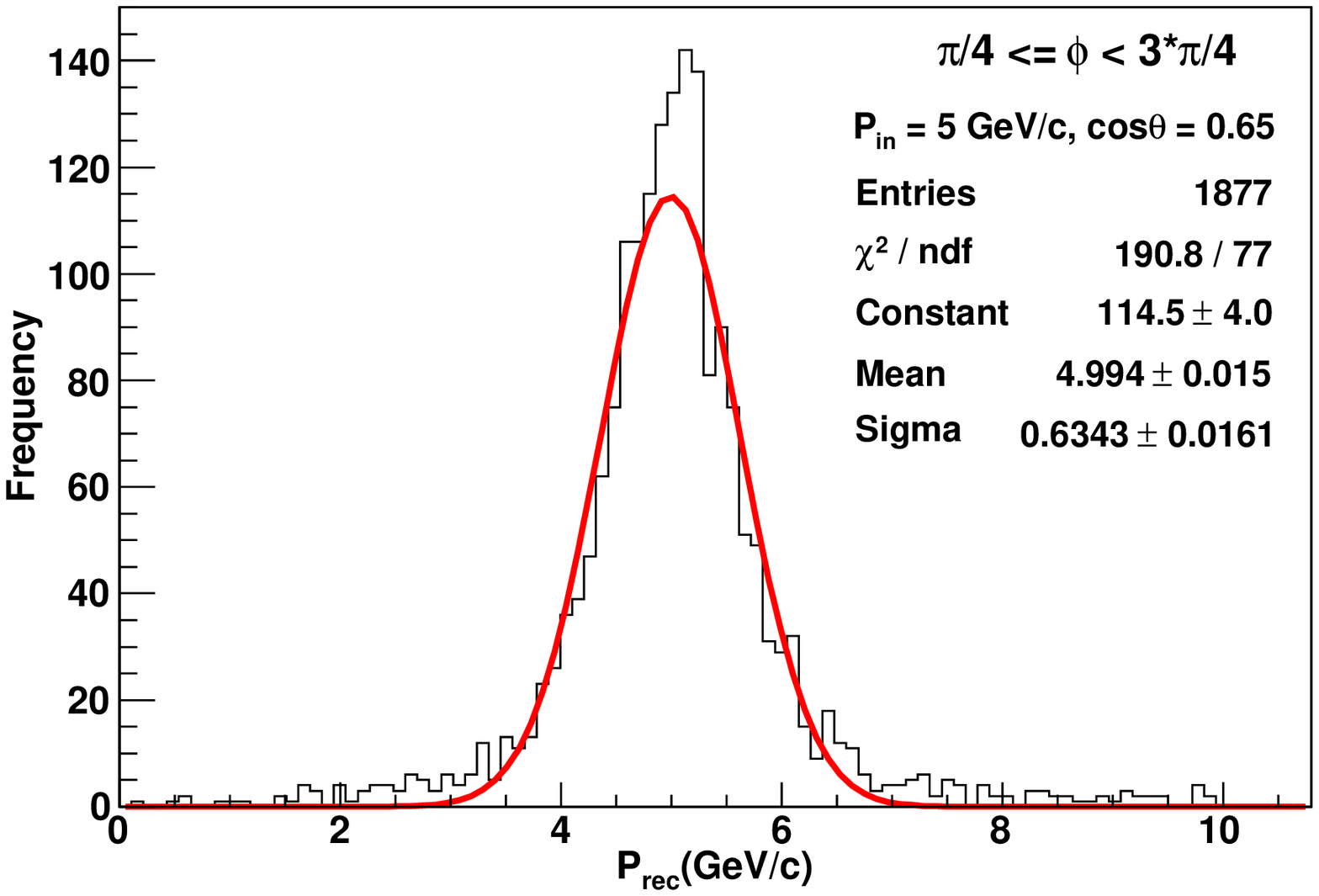}
\includegraphics[width=0.44\textwidth]{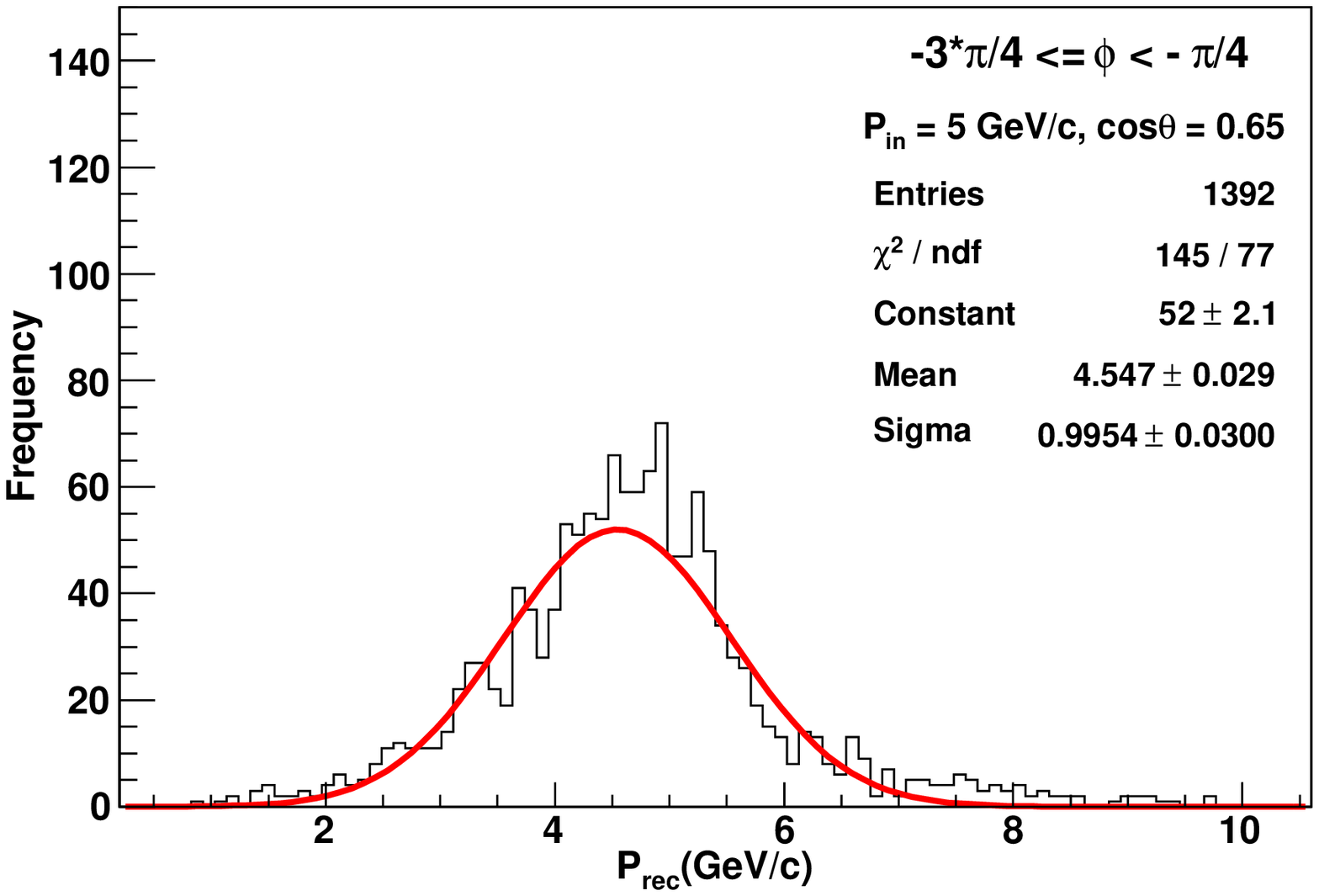}
\includegraphics[width=0.44\textwidth]{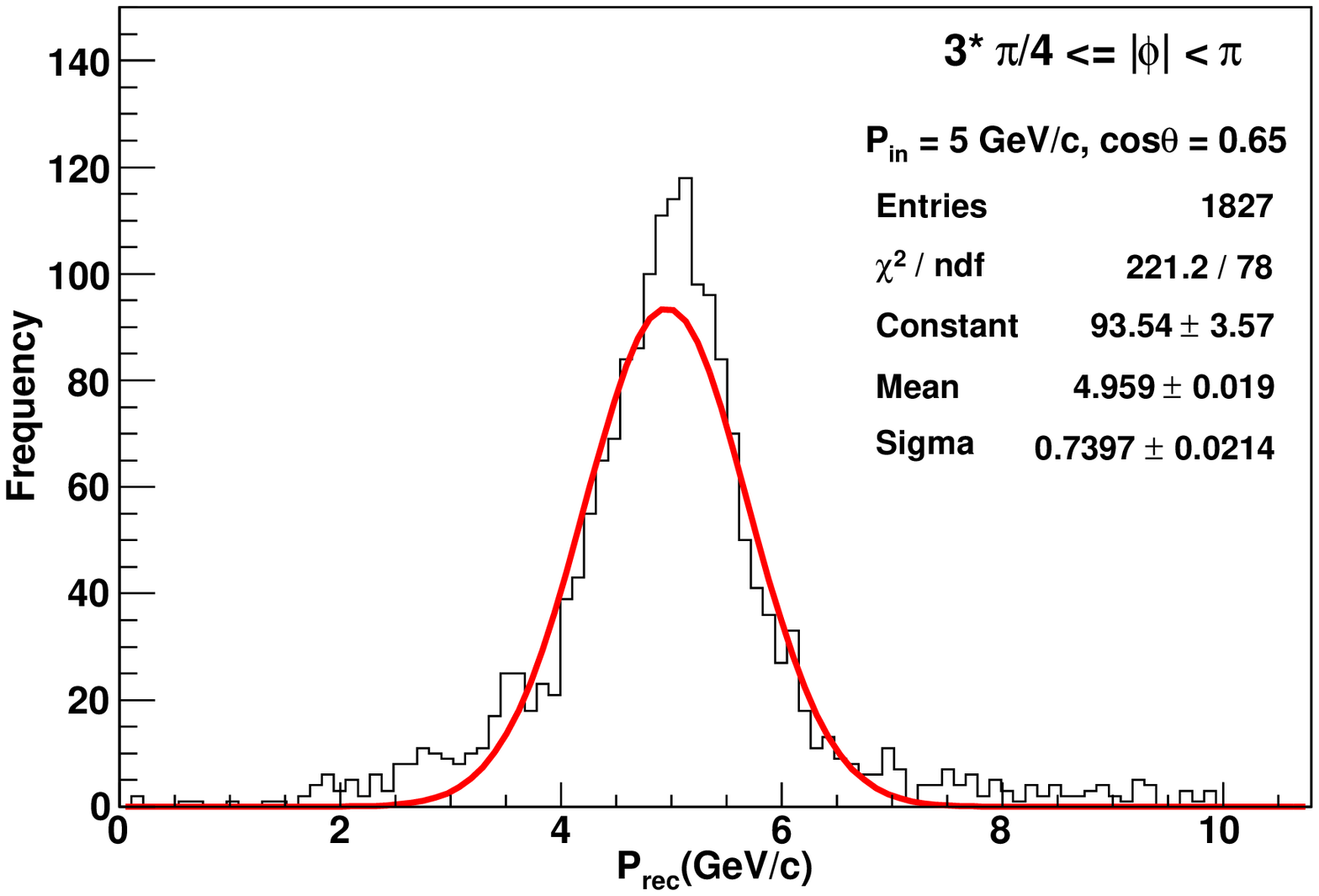}
\caption{Gaussian fits to reconstructed momentum distributions $P_{rec}$
(GeV/c) for muons with fixed energy $(P_{in}, \cos\theta) = (5 \hbox{
GeV/c}, 0.65)$ in four different bins of azimuthal angle in the
peripheral region. See text for details on the bins and the selection
criteria used.}
\label{fig:f5_65_4phi_nhit15_per}
\end{figure}

While the cid efficiency and zenith angle resolution are insensitive to
the azimuthal angle $\phi$, due to the reasons given above, we analyse
the muon momentum response in different $\phi$ bins. The response is shown in
Fig.~\ref{fig:f5_65_4phi_nhit15_per} for the peripheral region with the
constraint $N_{hits}/\cos\theta > 15$ being applied as usual to the partially
contained events, for sample input values of $(P_{in}, \cos\theta) =
(5 \hbox{ GeV/c}, 0.65)$.

The histograms in $P_{rec}$ have been fitted with Gaussian
functions. The width of each distribution of the four sets differs while the mean remains similar. As expected, $\phi$ bin III
(with most muons exiting the detector from the side) has the smallest number of reconstructed events and the worst resolution.
Bin II has the best resolution, while bins I and IV have a similar
response.
 This is in contrast to the response in the central region
\cite{central} where the reconstruction efficiencies were roughly equal
in all $\phi$ bins.

\begin{figure}[htp]
\renewcommand{\figurename}{Fig.}
 \centering
\includegraphics[width=0.44\textwidth]{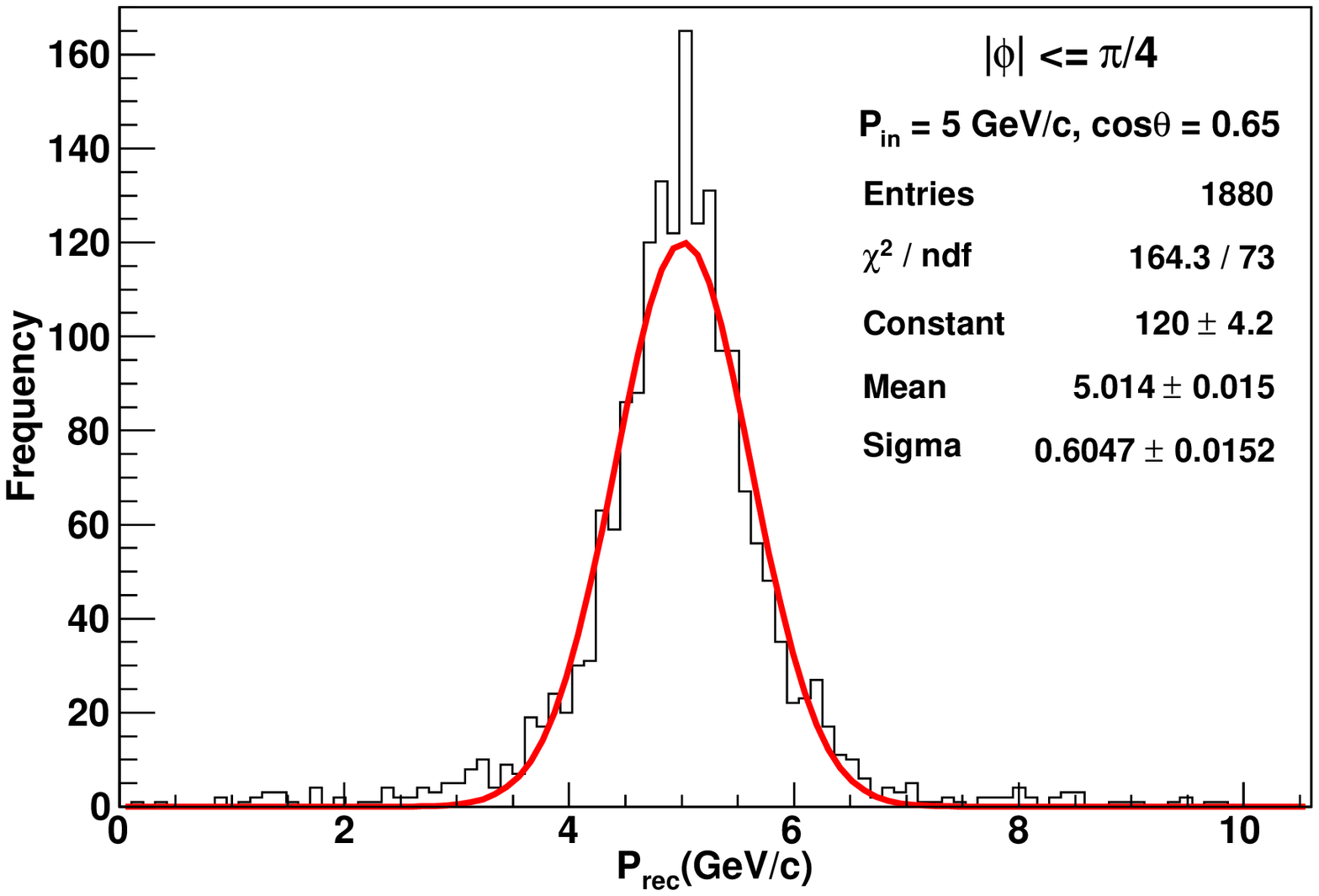}
\includegraphics[width=0.44\textwidth]{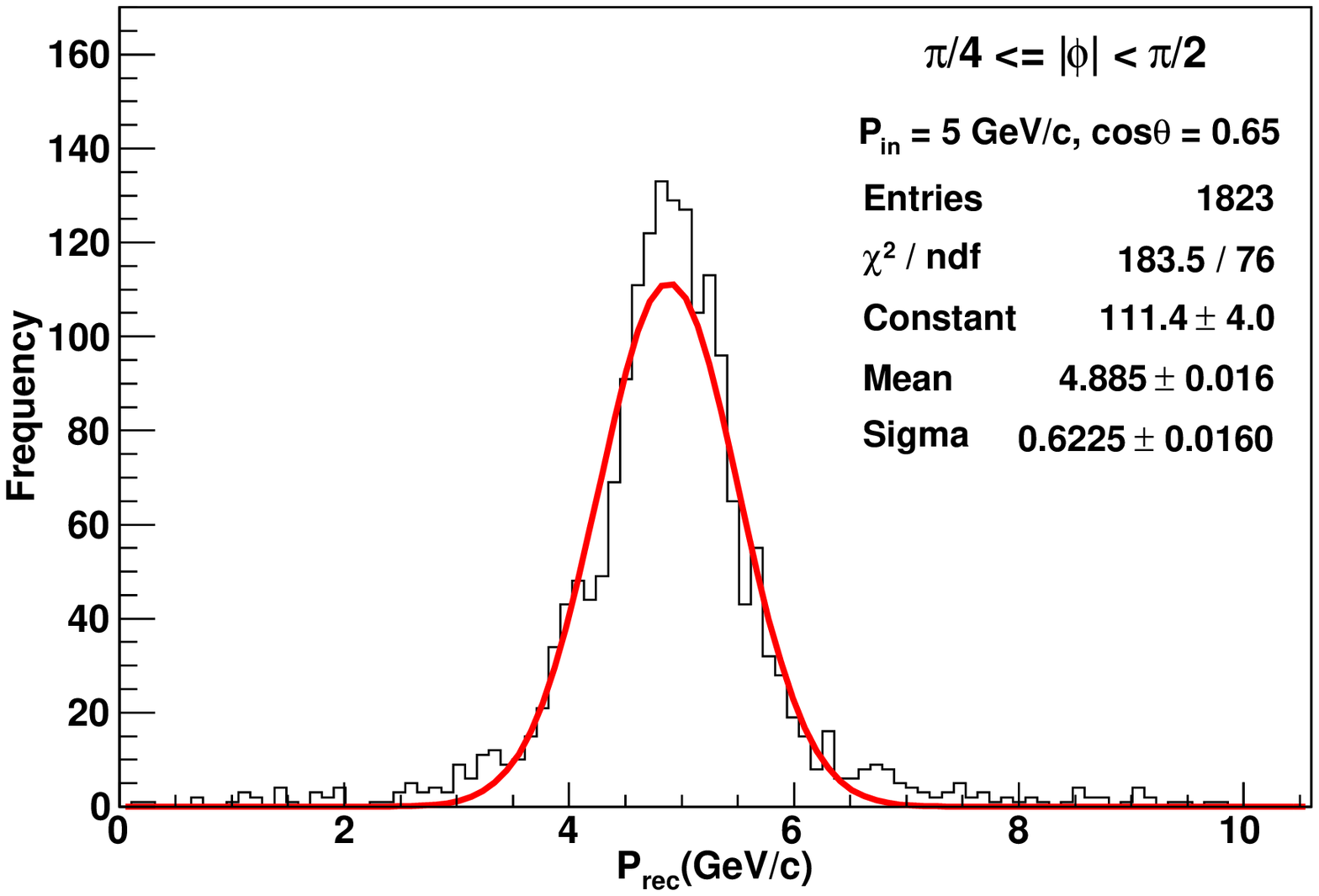}
\includegraphics[width=0.44\textwidth]{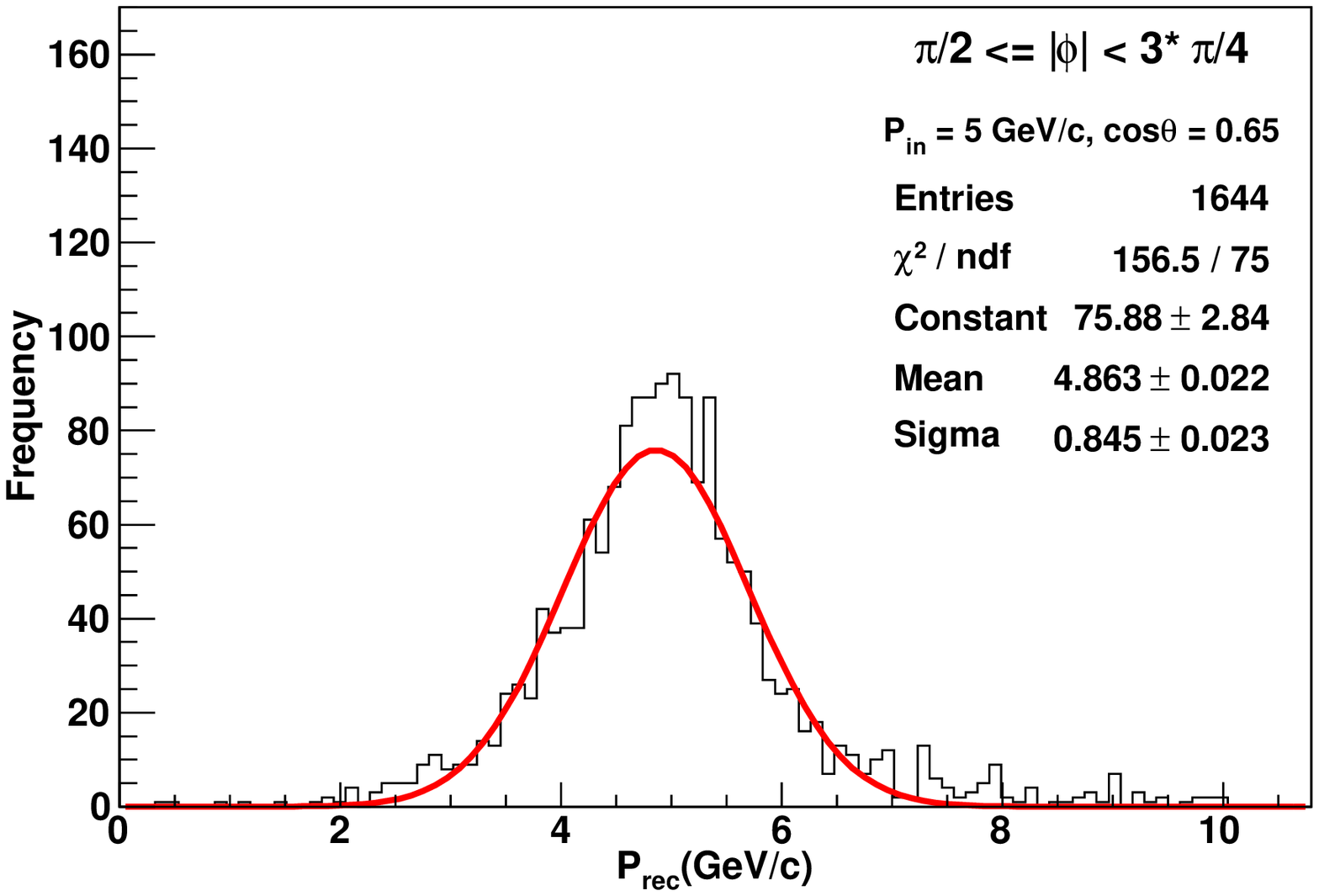}
\includegraphics[width=0.44\textwidth]{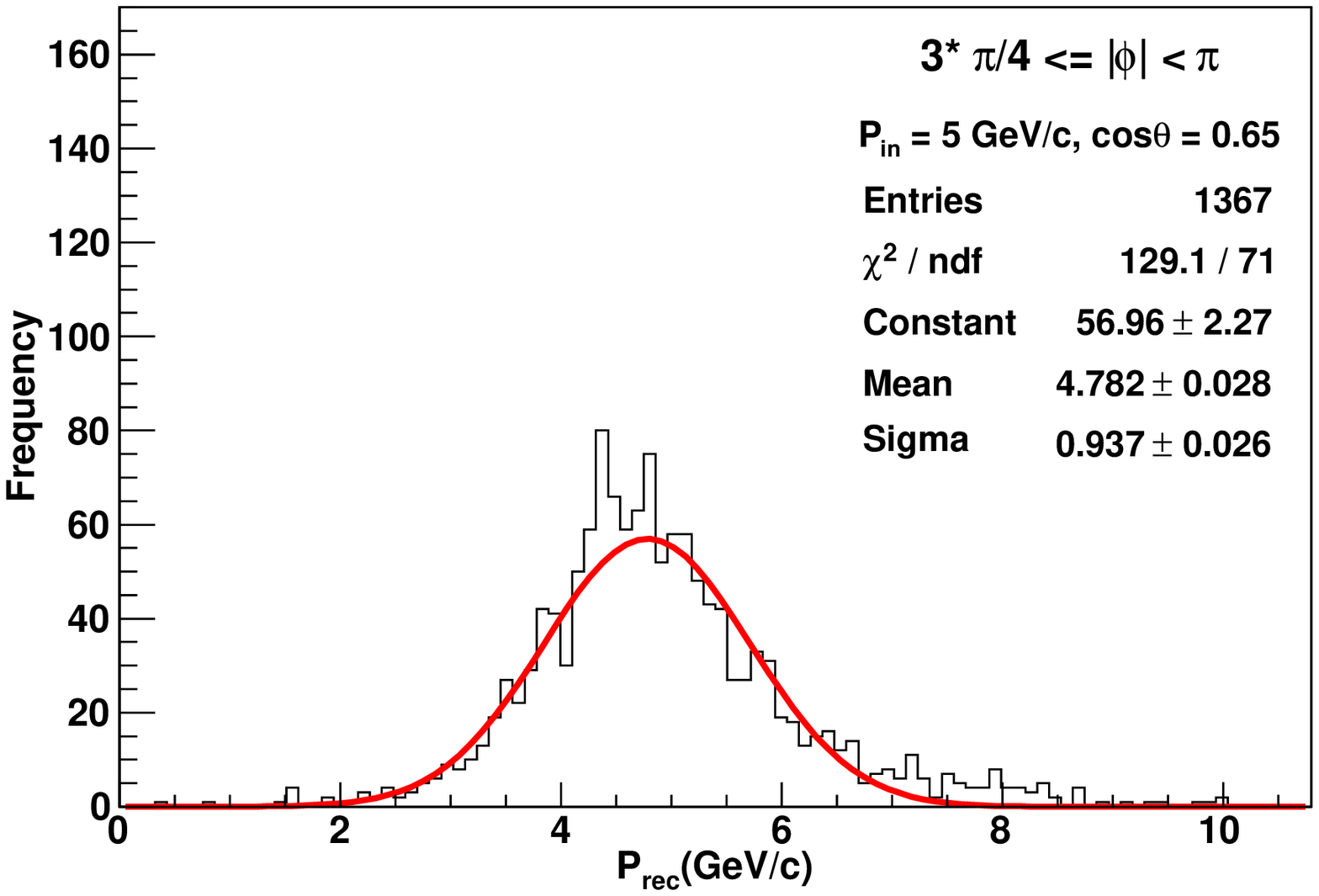}
\caption{Gaussian fits to reconstructed momentum distributions $P_{rec}$
(GeV/c) for muons with fixed energy $(P_{in}, \cos\theta) = (5 \hbox{
GeV/c}, 0.65)$ in four different bins of azimuthal angle in the
side region 9. See text for details on the bins and the selection
criteria used.}
\label{fig:side9-f5_65_4phi}
\end{figure}

Similar histograms are shown in Fig.~\ref{fig:side9-f5_65_4phi} for side
region 9. As discussed earlier, $\phi$ bin IV has both the worst
reconstruction and the worst resolution, while bin I has the best ones. Unlike
the peripheral case where the bins I and IV had similar response, here
bins II and III are not similar because the side region is not symmetric
between these two bins: muons in bin III are more prone to exit the
detector and hence the detector response is worse in both efficiency and quality of reconstruction.

The results in region 10 are similar to region 9 with interchange of
bins I and IV, and bins II and III as can be easily understood from
Fig.~\ref{fig:map}. However, overall the quality of reconstruction is
better in region 10 by about 15\% due to the nature of the forces in this
region as discussed earlier; see Fig.~\ref{fig:map}. We shall show results
for side region 9 everywhere, as being the more conservative result and
simply remark on similarities/differences to be expected in region 10.

\begin{figure}[tbp]
\renewcommand{\figurename}{Fig.}
  \centering
\includegraphics[width=0.48\textwidth]{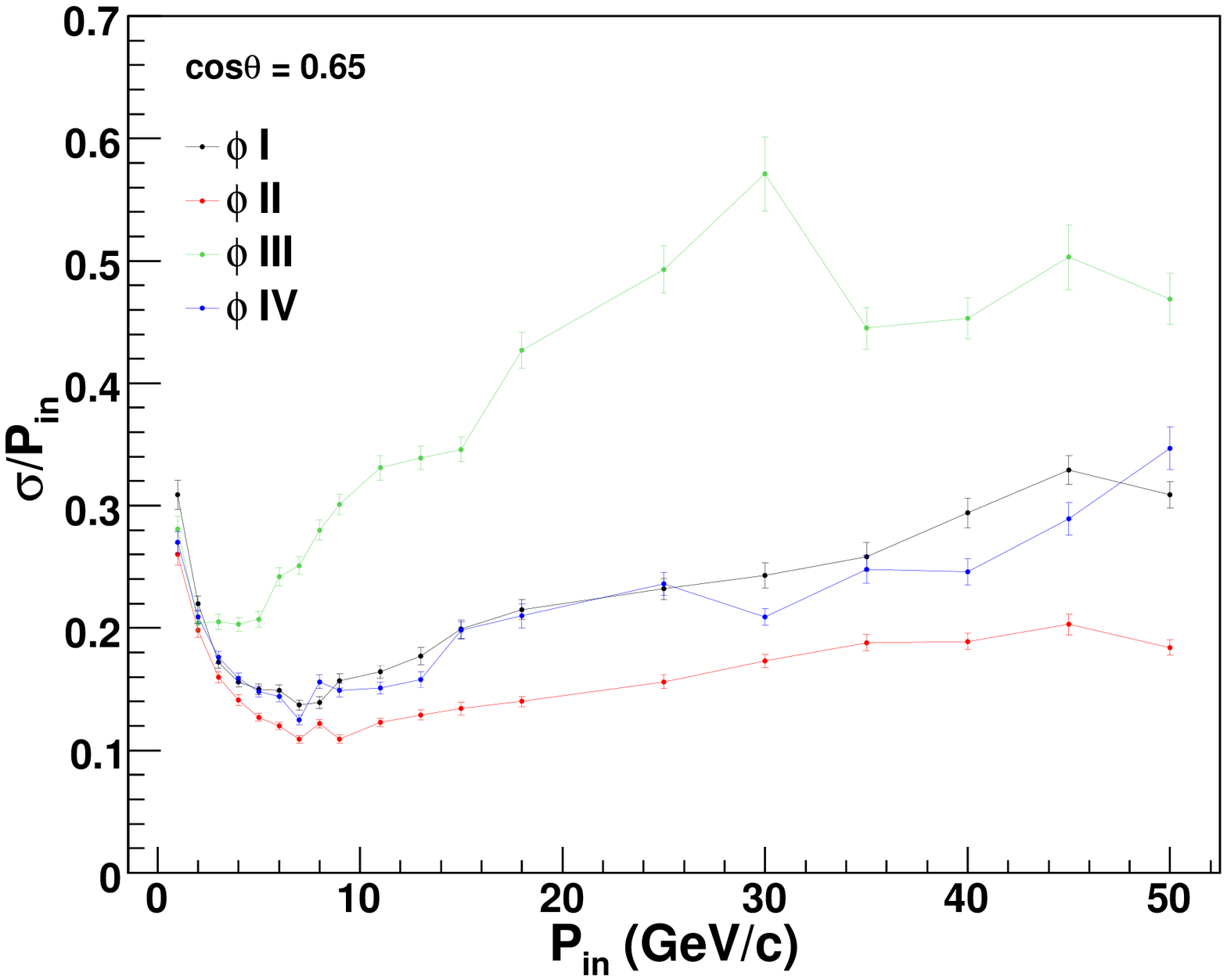}
\includegraphics[width=0.48\textwidth]{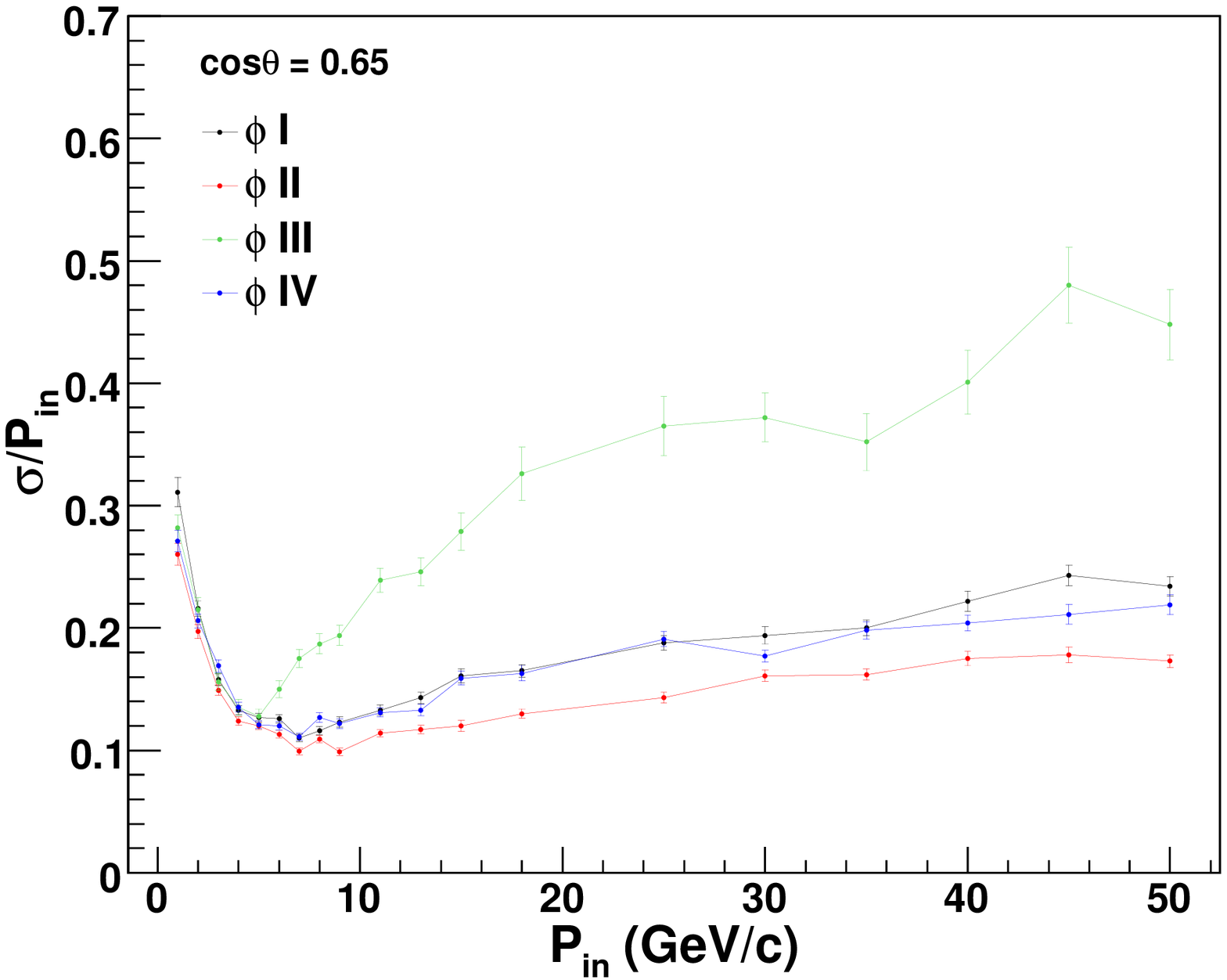}

\caption{Muon resolution in the peripheral region as a function of input
momentum $P_{\rm in}$ (GeV/c) for $\cos\theta$ = 0.65
in different bins of $\phi$ with $N_{hits}/\cos\theta > n_0$ cut, where $n_0$ = 15 (20) left (right).}
\label{fig:resol-nhits15-20}
\end{figure}

Fig.~\ref{fig:resol-nhits15-20} shows the momentum resolution as a function of $P_{in}$ in the peripheral region for the four $\phi$ bins using the selection criteria $N_{hits}/\cos\theta > n_{0}$ with $n_0$ = 15, 20, for $\cos\theta$ = 0.65.

\subsection{Momentum Resolution as a Function of $(\theta, \phi)$}

\begin{figure}[htp]
\renewcommand{\figurename}{Fig.}
  \centering
\includegraphics[width=0.44\textwidth,height=0.27\textwidth]{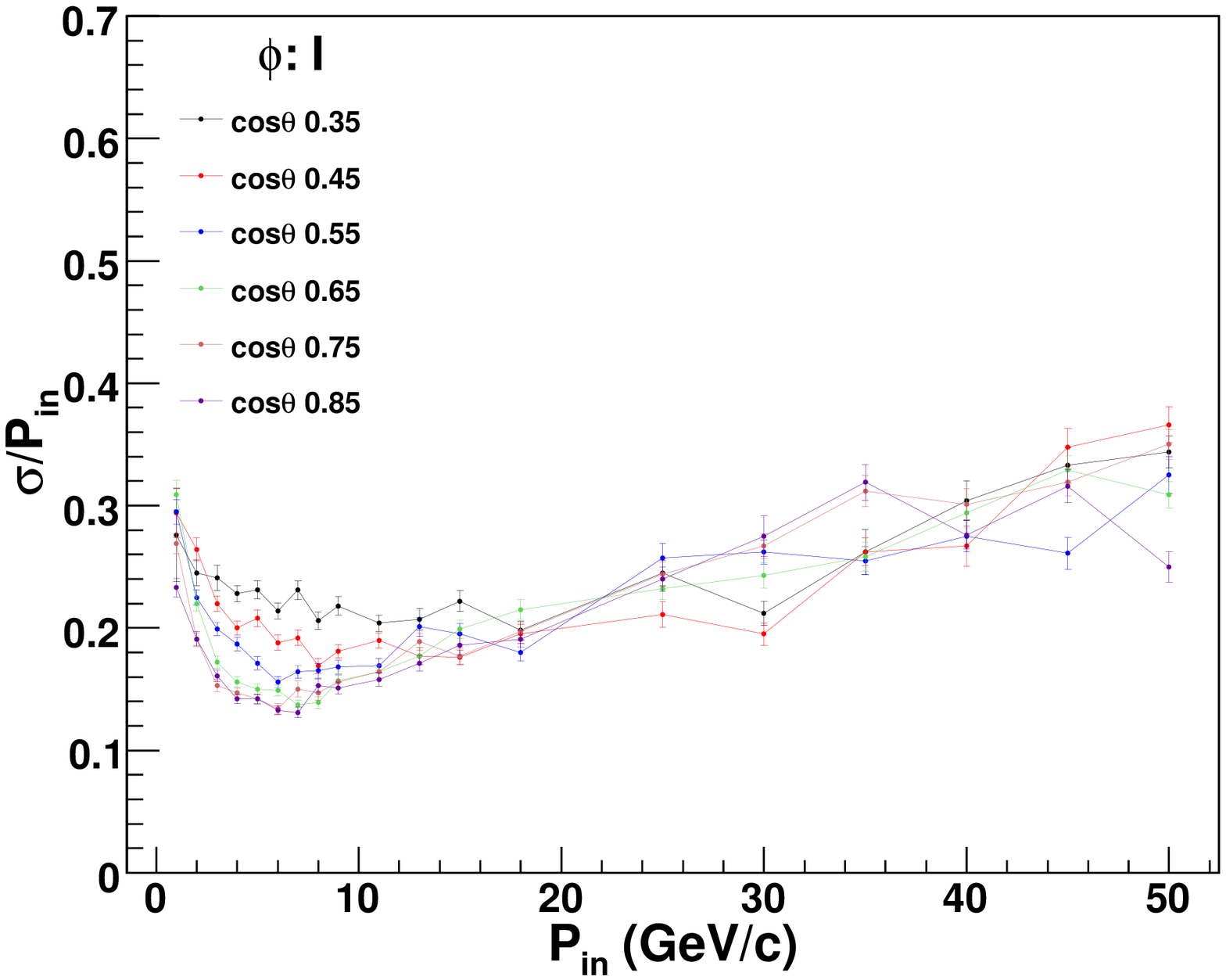}
\includegraphics[width=0.44\textwidth,height=0.27\textwidth]{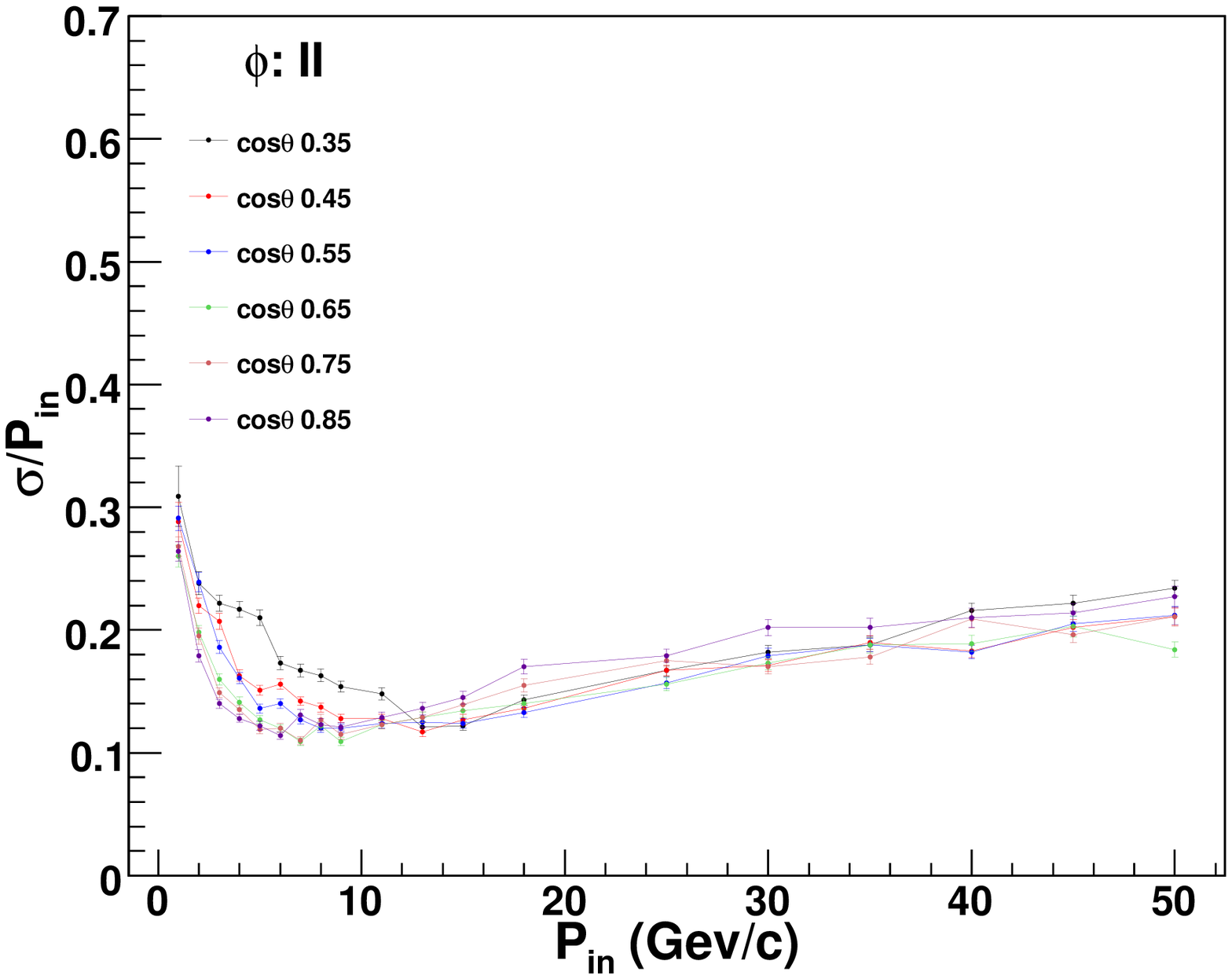}
\includegraphics[width=0.44\textwidth,height=0.27\textwidth]{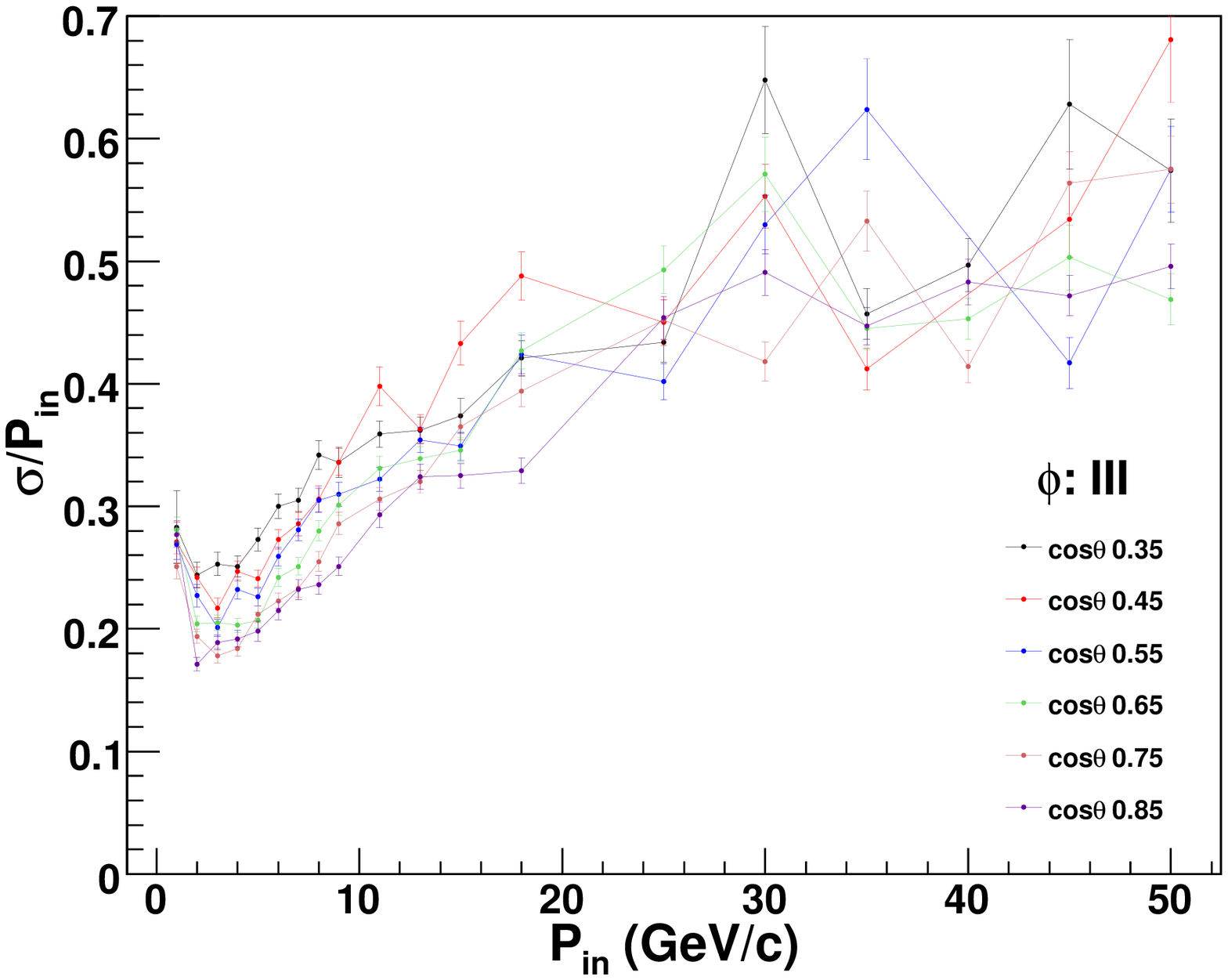}
\includegraphics[width=0.44\textwidth,height=0.27\textwidth]{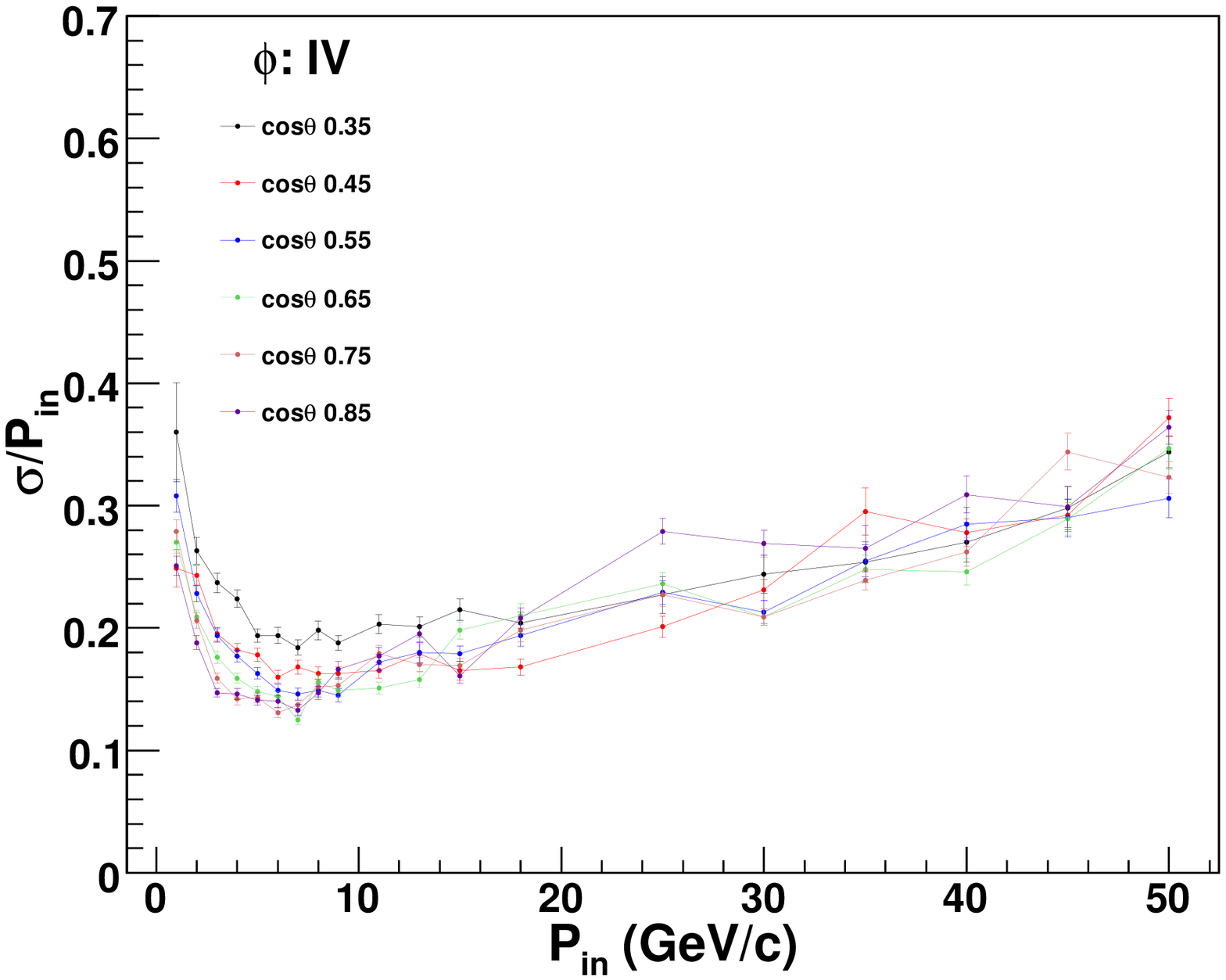}
\caption{Muon resolution in the peripheral region as a function of
input momentum $P_{\rm in}$ (GeV/c) for different values of $\cos\theta$
in different bins of $\phi$.}
\label{fig:per-resol-reg}
\end{figure}

Gaussian fits to the reconstructed momentum distribution in these
regions give the reconstructed mean and RMS width $\sigma$. The momentum resolution ($R$) is
defined from these fits as,
\begin{eqnarray}
R & = & \sigma /P_{\rm in}, \\ \nonumber
\hbox{with error  } \delta R & = & \delta\sigma/P_{\rm in}~.
\end{eqnarray}
Fig.~\ref{fig:per-resol-reg} shows the variation of resolution as a
function of $P_{in}$ from 1 to 50 GeV/c for different values of $\cos\theta$ from 0.35 to 0.85 in the different $\phi$ bins of the peripheral region. In all bins, the momentum resolution improves with the increase of energy
upto about $P_{in} \sim 6$ GeV/c as the number of hits increases, but
worsens at higher momenta since the particle then begins to exit the
detector. This effect is considerable in the $\phi$ bin III which
therefore has the worst resolution while $\phi$ bin II has the best
resolution, as expected from the earlier discussions. In general, the
resolution improves for more vertical angles (larger $\cos\theta$)
as the number of hits in a track increases.

\begin{figure}[tbp]
\renewcommand{\figurename}{Fig.}
  \centering
\includegraphics[width=0.44\textwidth,height=0.27\textwidth]{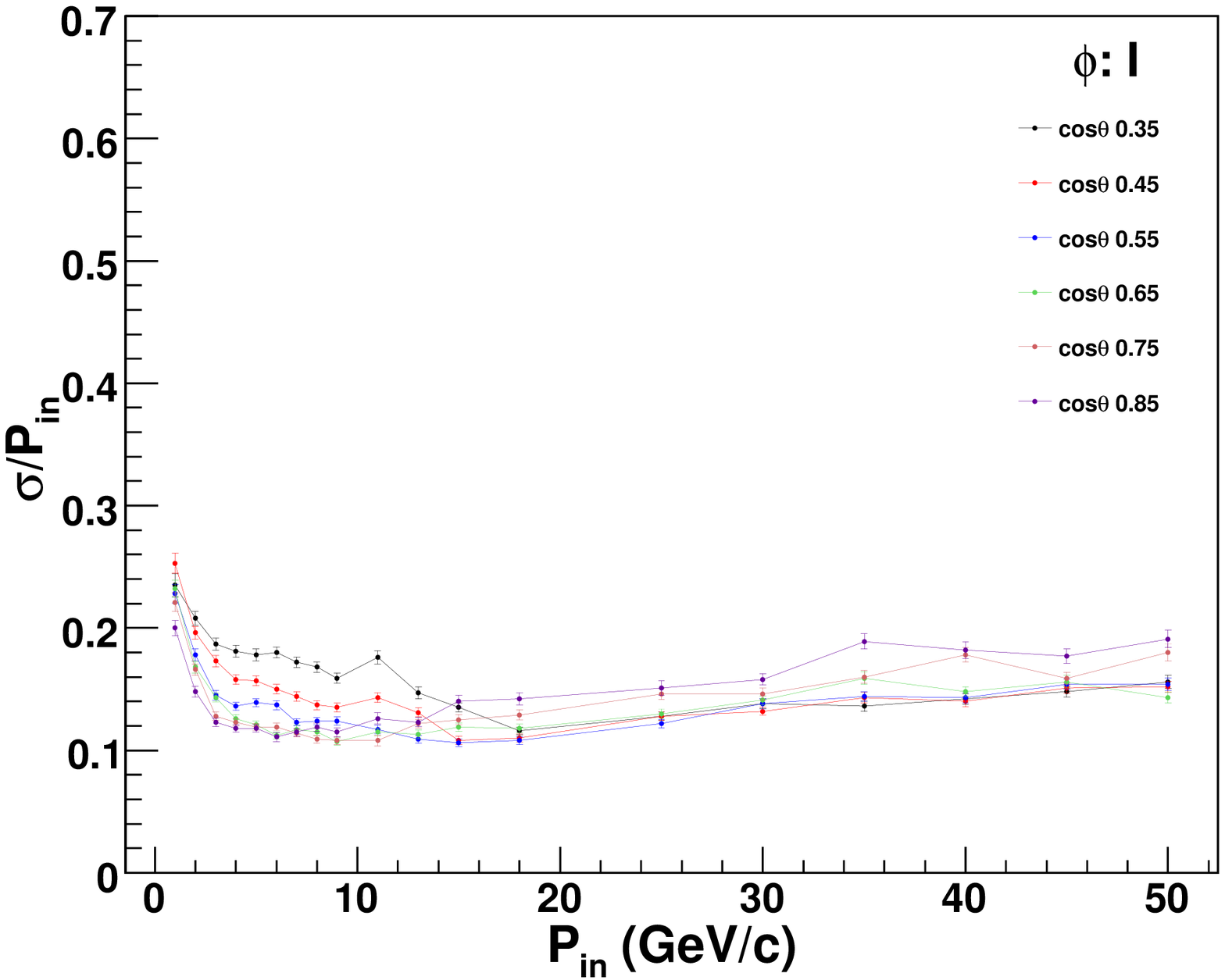}
\includegraphics[width=0.44\textwidth,height=0.27\textwidth]{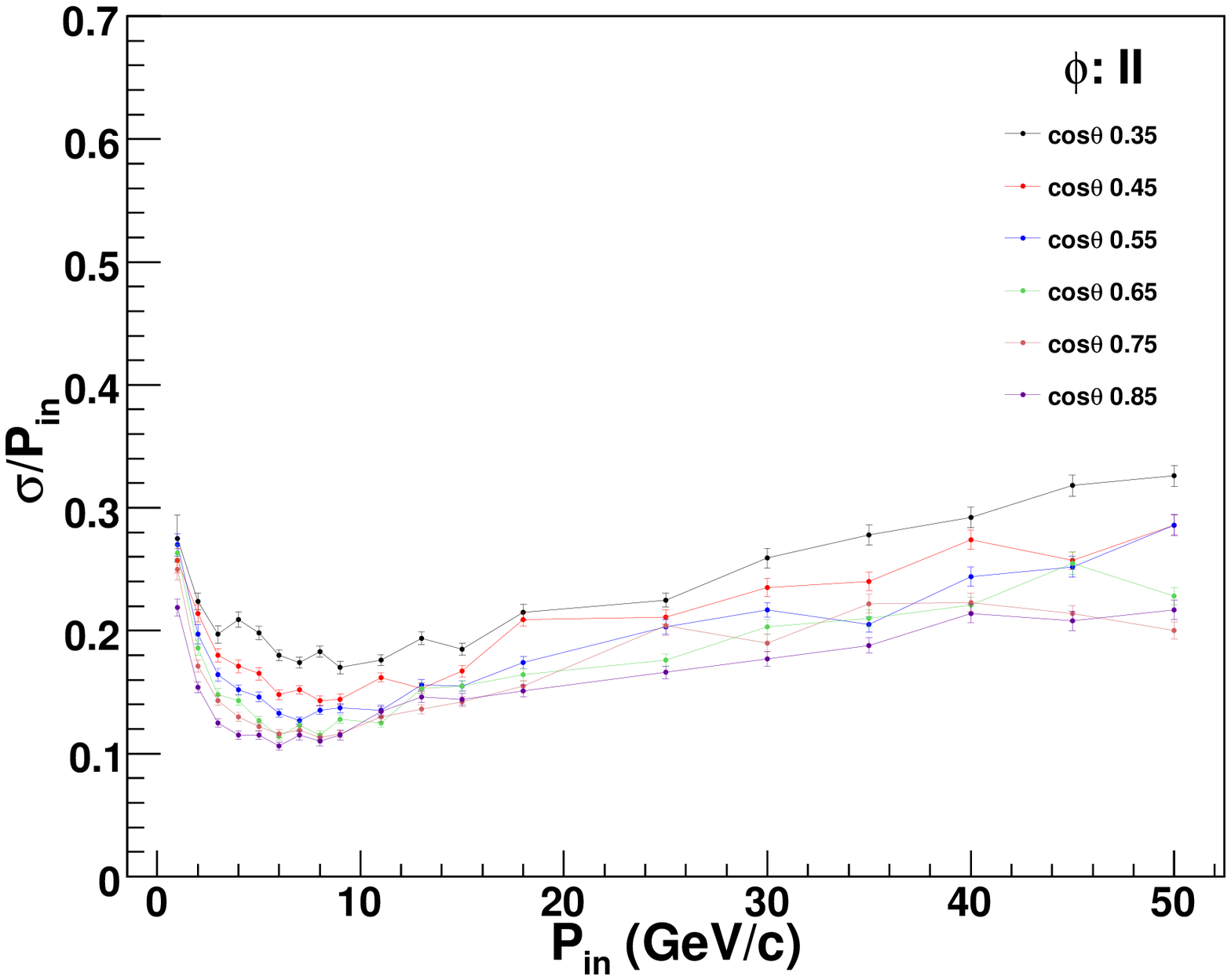}
\includegraphics[width=0.44\textwidth,height=0.27\textwidth]{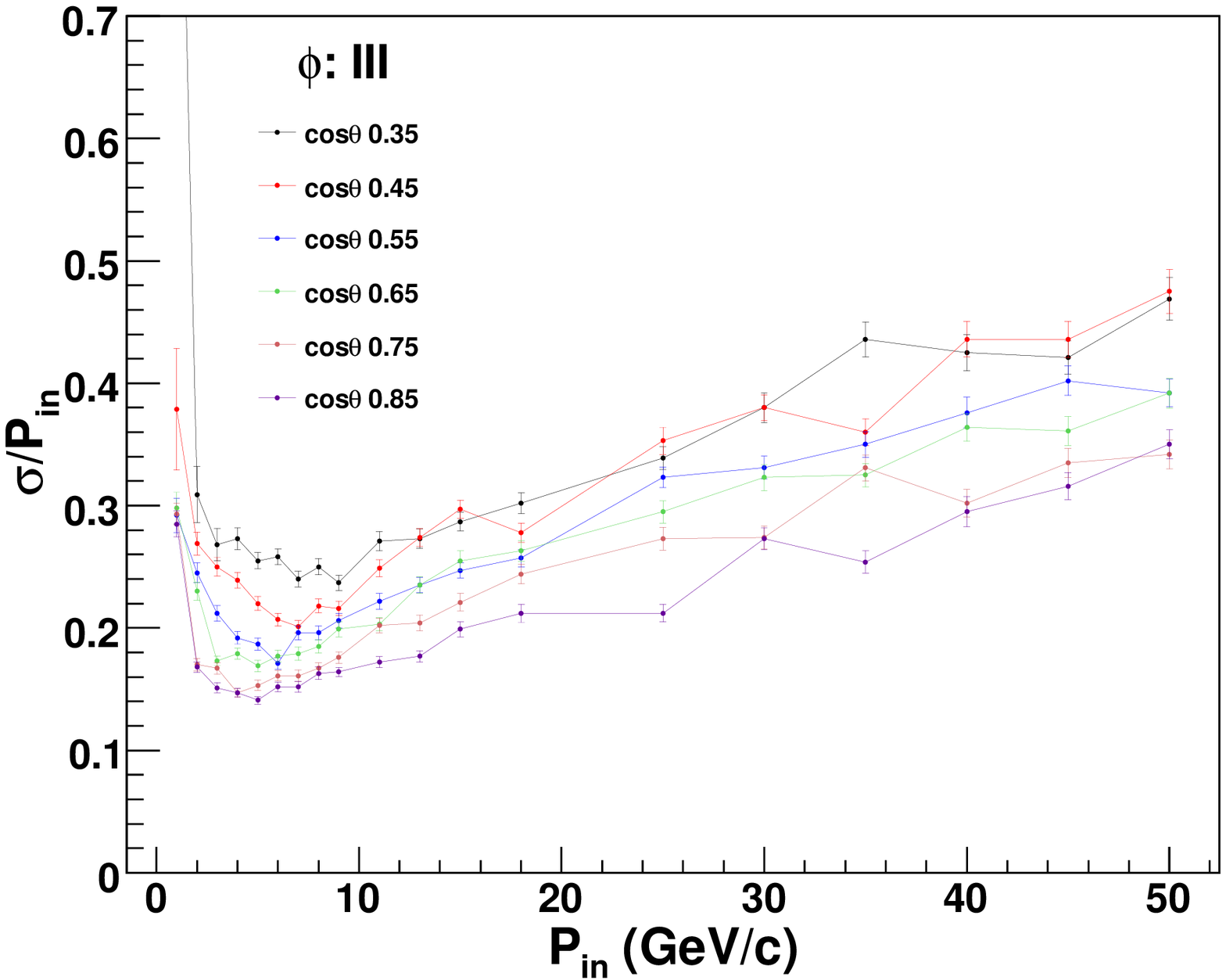}
\includegraphics[width=0.44\textwidth,height=0.27\textwidth]{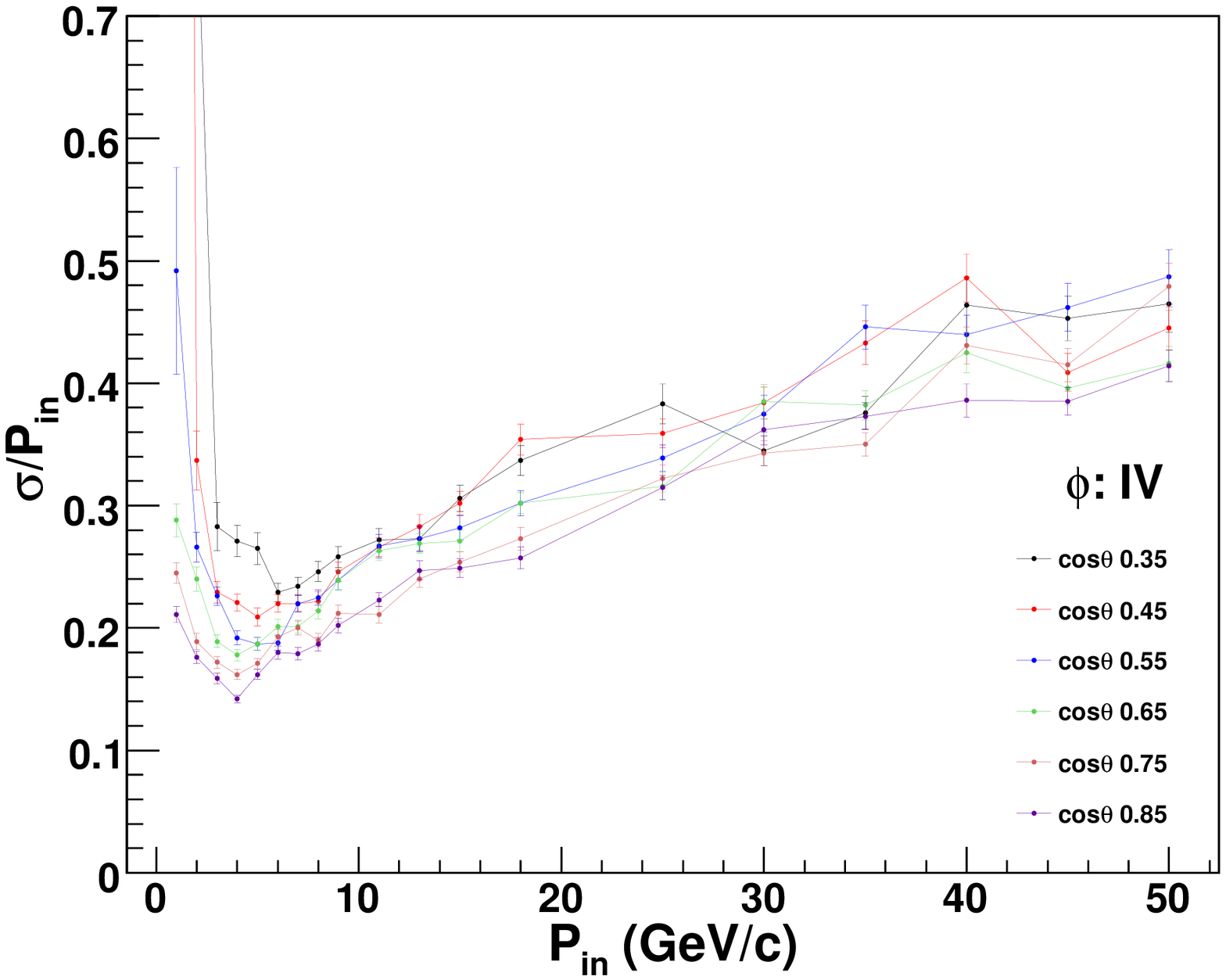}
\caption{Muon resolution in the side region 9 as a function of input
momentum $P_{\rm in}$ (GeV/c) for different values of $\cos\theta$ in different bins of $\phi$.}
\label{fig:side9-resol-reg}
\end{figure}

Fig.~\ref{fig:side9-resol-reg} shows similar results for the side region
9. Again, it is observed that for all the angles and energies, $\phi$
bin I has the best response while the resolutions worsens in bins
III and IV. Results in region 10 are similar to those in region 9 with
interchange of response in $\phi$ bins (I, IV) and (II, III), but with a
few percent better resolution in all cases.

The resolution for a given $P_{in}$ is marginally better in the side
region than in the peripheral region due to the somewhat larger and
uniform magnetic field. A detailed comparison of the response in different
regions will be presented in the next section.

\section{Comparison of Muon Response in Different Regions of ICAL}

We compare the muon response in the peripheral and side regions with
that in the central region as presented in Ref.~\cite{central}. 

\begin{figure}[htp]
\renewcommand{\figurename}{Fig.}
  \centering
\includegraphics[width=0.48\textwidth]{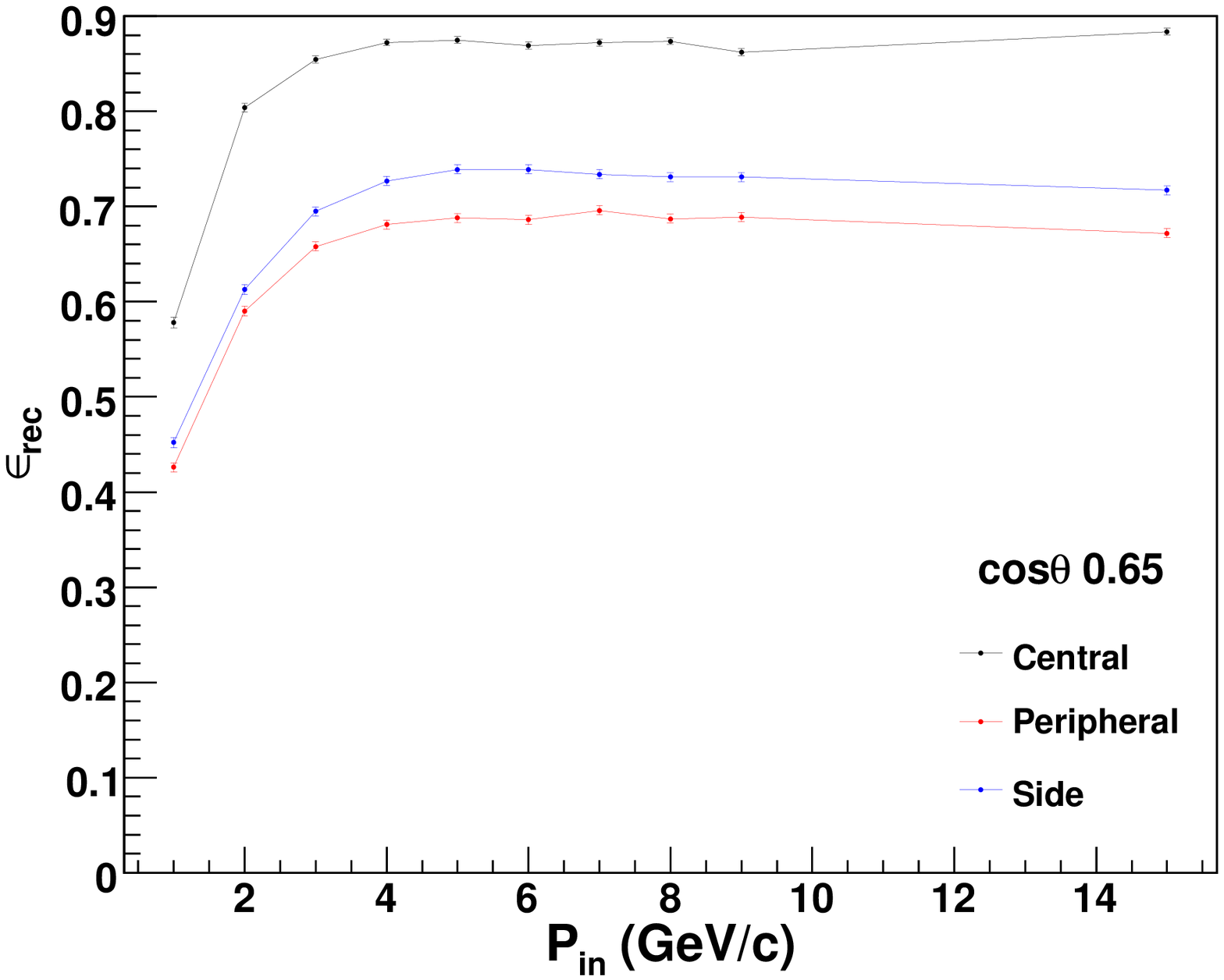}
\includegraphics[width=0.48\textwidth]{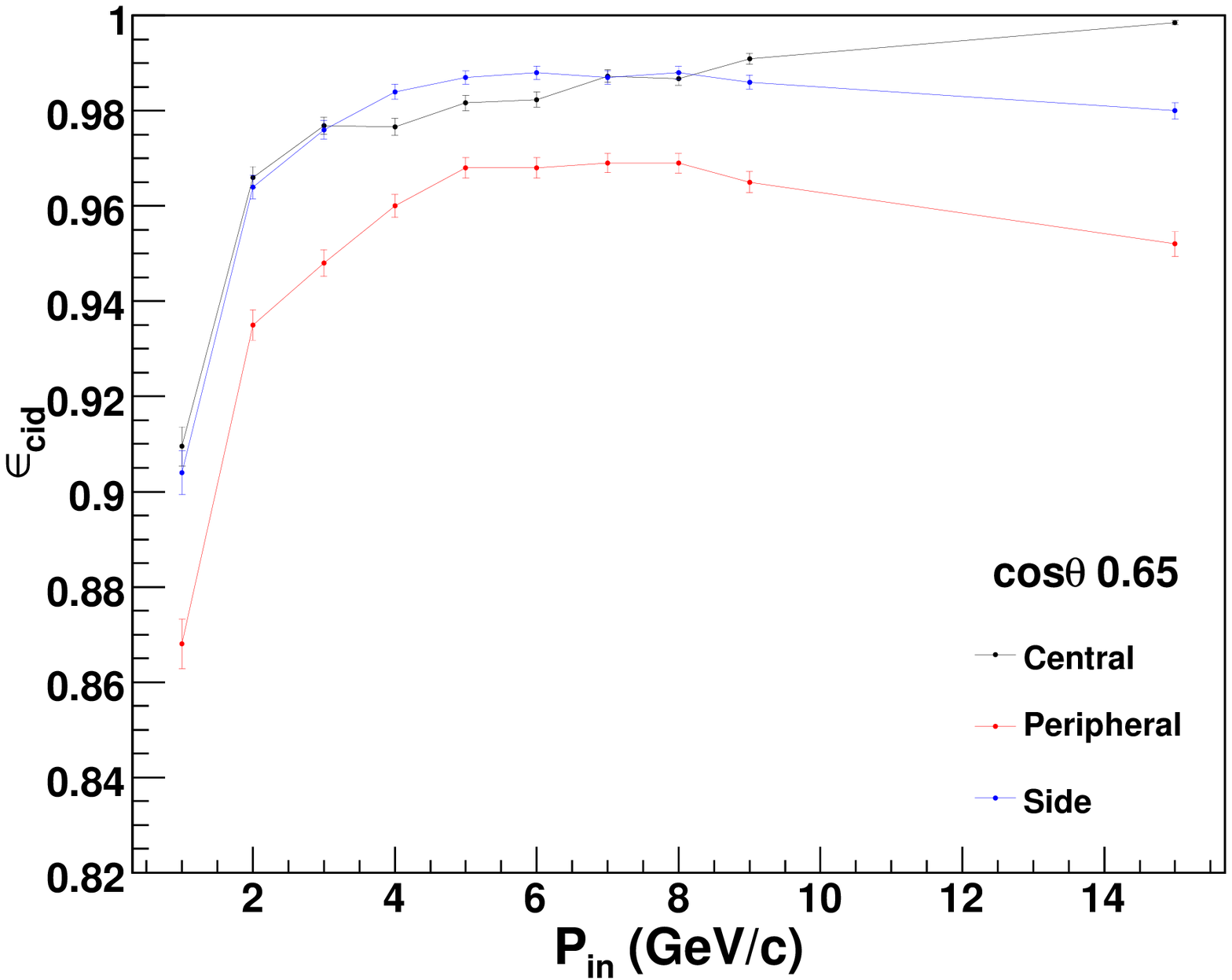}
\caption{Comparison of reconstruction (left) and cid efficiency (right) of
central, peripheral and side regions as a function of $P_{in}$ (GeV/c) at
$\cos\theta = 0.65$. Note that the y-axis scales are different for the two plots.}
\label{fig:comp-eff}
\end{figure}

For all choices of selection criteria, the reconstruction and cid
efficiencies in the central region are better than either the peripheral
or side region as shown in Fig.~\ref{fig:comp-eff} for
$\cos\theta=0.65$; however, for input momenta upto $P_{in} \sim 8$
GeV/c, the central and side region cid efficiencies are comparable. Note
that applying more stringent selection criteria in order to improve
the momentum resolution in the peripheral and side regions (and hence
overall resolution of the detector) will further worsen the reconstruction
efficiencies in these regions.

In addition, the angular resolutions are very similar between the
peripheral and side regions, as can be seen from Fig.~\ref{fig:theta} and
are in fact similar to those obtained earlier in the central region
\cite{central}. 

\begin{figure}[htp]
\renewcommand{\figurename}{Fig.}
~\hspace{-0.5cm} \includegraphics[width=0.35\textwidth,
height=0.33\textwidth]{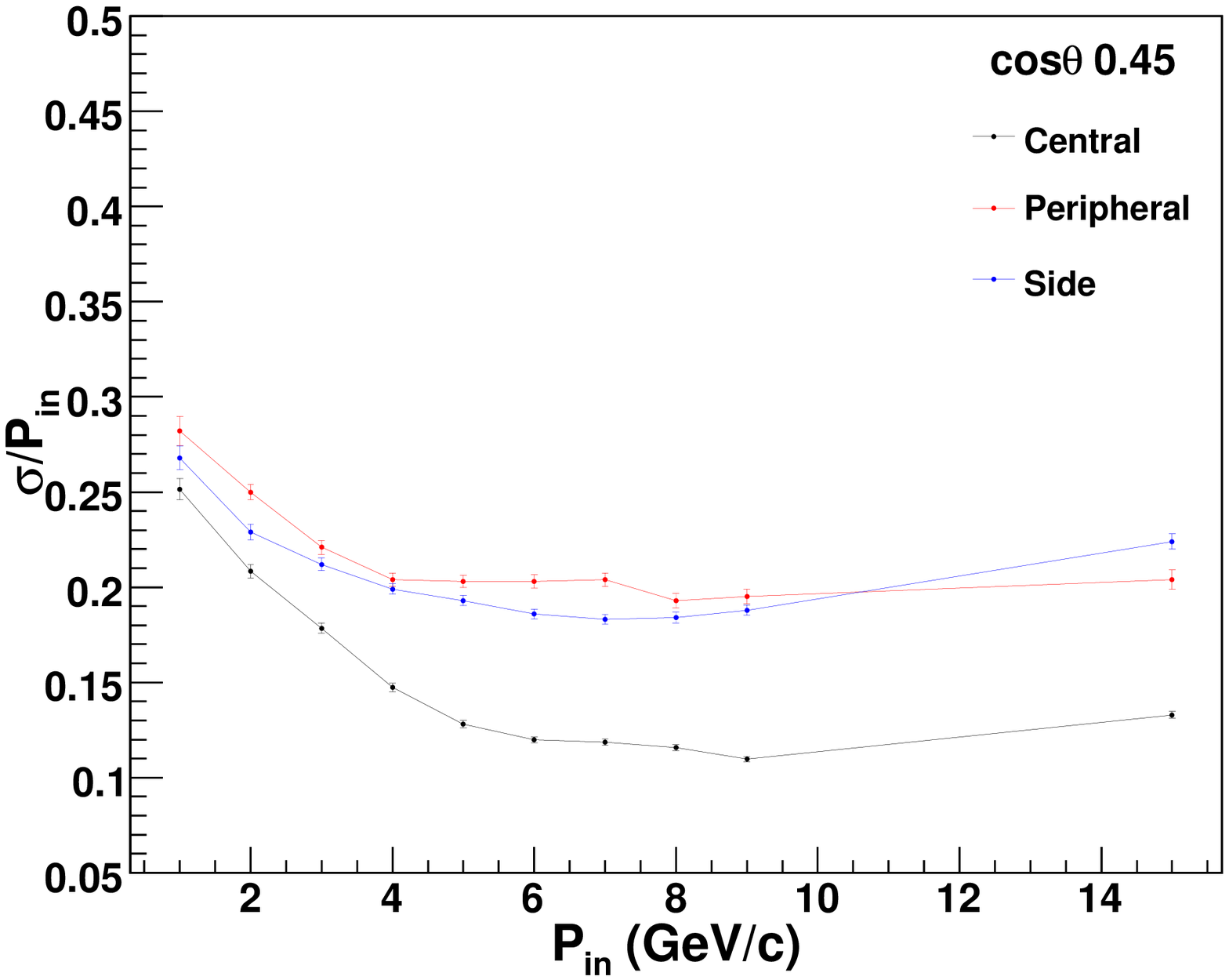}
\hspace{-0.5cm} 
\includegraphics[width=0.35\textwidth,
height=0.33\textwidth]{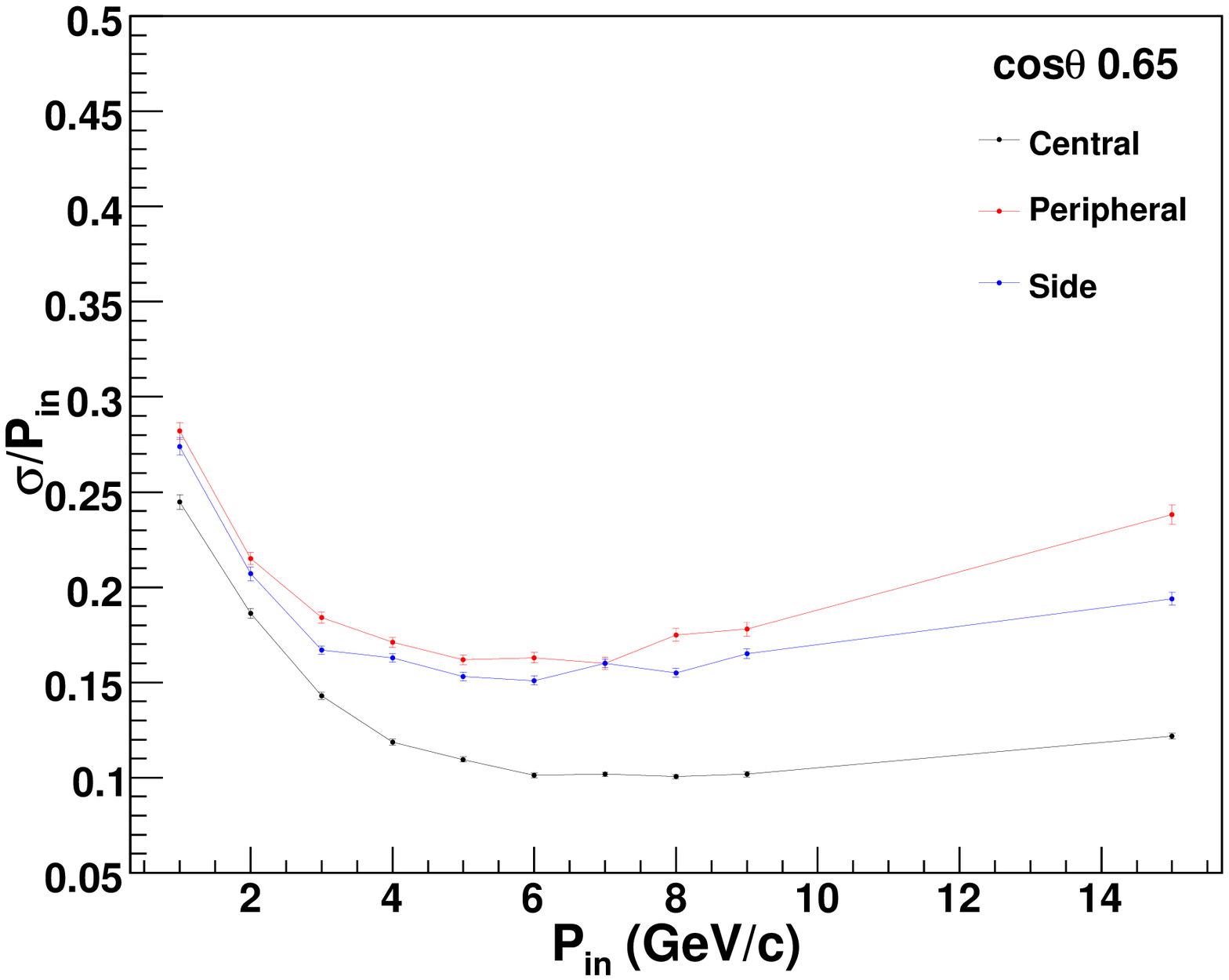}
\hspace{-0.5cm} 
\includegraphics[width=0.35\textwidth, height=0.33\textwidth]{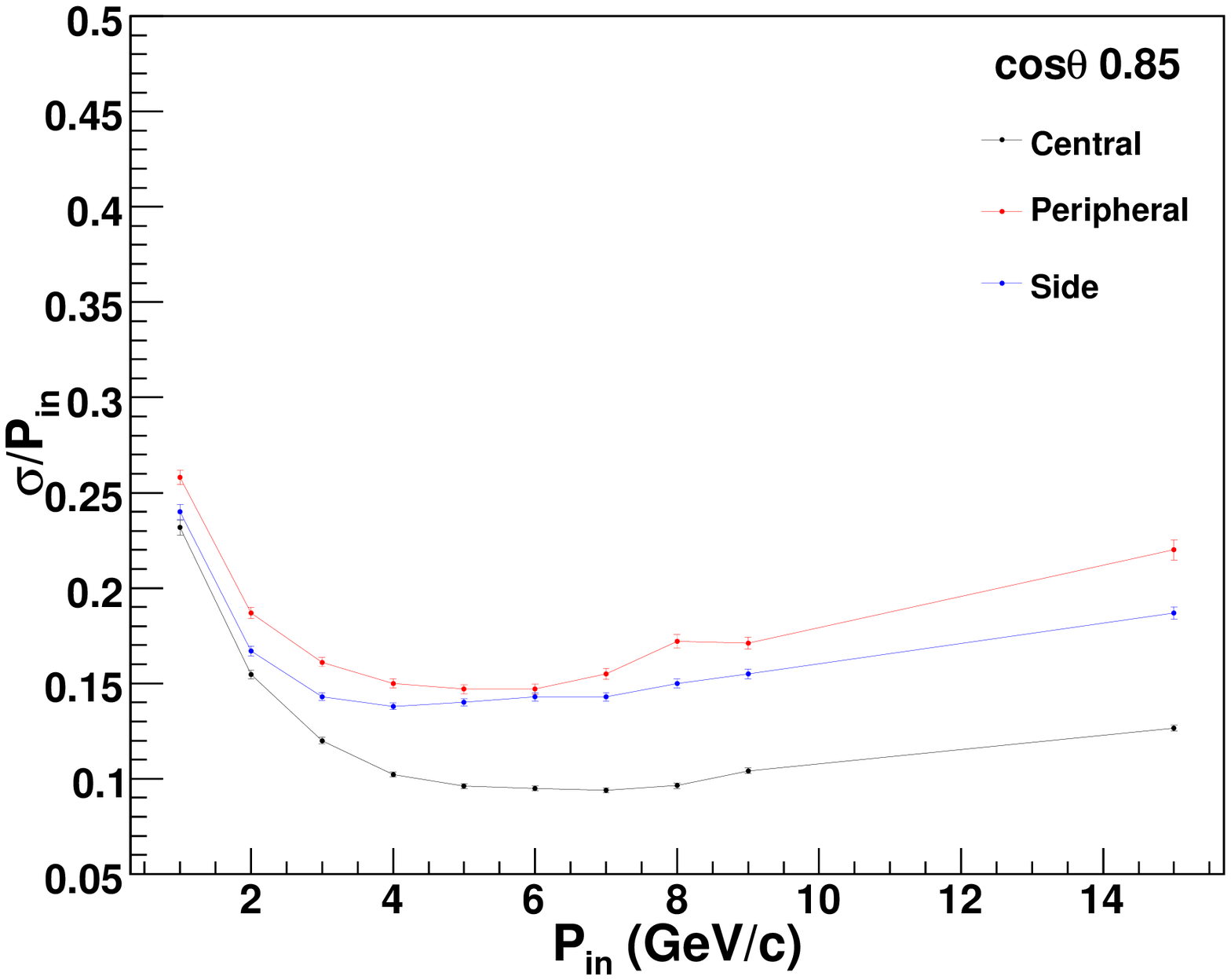}
\caption{Comparison of resolutions in peripheral and side region 9 as a
function of the input momentum $P_{in}$ (GeV/c) along with
earlier results in the central region \cite{central} for different
values of $\cos\theta = 0.45, 0.65, 0.85$.}
\label{fig:comp}
\end{figure}

The comparison of the $\phi$-averaged peripheral and side region
momentum resolutions as a function of input momentum $P_{in}$ from
1 to 15 GeV/c is shown in Fig.~\ref{fig:comp} for $\cos\theta = 0.45,
0.65, 0.85$. We have also shown the $\phi$-averaged central region
results \cite{central} in the same plots. The criterion of a
single reconstructed track only was also applied to the central region, but no constraint was placed on $N_{hits}$. While the side region resolutions
are only marginally better than those in the peripheral region, the
central region gives the best resolution, as expected. However, we
note that the results are $\phi$ averaged and so the resolutions can
be much improved in the peripheral and side regions depending on the
$\phi$ bin chosen. The peripheral and side region resolutions can be
improved by changing the selection criteria at the cost of reconstruction
efficiency. The resolutions in all regions are comparable at low momenta,
$P_{\rm in} \le 3$ GeV/c, since almost all tracks are fully contained
in this case.

~\hspace{-0.5cm} 

\section{Discussions and Conclusion}

The goal of the proposed ICAL detector is to study neutrino oscillations
using atmospheric neutrinos. It is more sensitive to muons and hence
the physics will focus on charged current scattering of $\nu_\mu$
($\overline{\nu}_\mu$) in the detector. Hence a simulations study of
the response of ICAL to muons is crucial. The ICAL geometry was simulated
using GEANT4 software and the detector response was studied for muons with momenta from 1 to 50 GeV/c, polar angle $\cos\theta \ge 0.35$ and
smeared over all azimuthal angles, $-\pi \le \phi \le \pi$. In the current
study, muons were generated in the peripheral and side region of the
ICAL detector where the magnetic field is non-uniform in both magnitude
and direction and where edge effects are important. The study showed
that a crucial selection criterion on the number of hits
$N_{hits}/\cos\theta > n_0$ for partially contained tracks was necessary to achieve good detection efficiency. The magnetic field
and the detector geometry break the azimuthal symmetry; hence the muon
response was analysed in different $\phi$ bins.

Results using $N_{hits}/\cos\theta > 15$ show that the best momentum
resolutions of about 10--15\% are obtained in bin II ($\pi/4 \le
\phi < 3\pi/4$) at input momenta of $P_{\rm in} \ge 4$ GeV/c in the
peripheral region and in bins I and II ($\vert \phi \vert \le \pi/4$
and $\pi/4 < \vert \phi \vert \le \pi/2$) in Side region 9 (see
Figs.~\ref{fig:map} and \ref{fig:phi_choice} for definitions of these
regions).

Also, $\phi$-averaged results are obtained with $N_{hits}/\cos\theta > 15$
for the reconstruction efficiency, charge identification efficiency
and momentum resolution as shown in Figs.~\ref{fig:comp-eff} and
\ref{fig:comp} for the peripheral and side regions of the ICAL detector
in comparison with earlier results in the central region \cite{central}.

A reconstruction efficiency of about 60--70\% and a correct charge
identification of about 97\% of the reconstructed muons was obtained for
$P_{\rm in} \ge 4$ GeV/c and this decreased to about 90\% for higher
momenta $P_{\rm in} \sim 50$ GeV/c in both regions. Average (over
$\phi$) resolutions obtained are between 15--25\% over $P_{\rm in} =
1$--15 GeV/c in the peripheral region and marginally better in the side
region, with the central region response being the best. Note
that these responses are relevant for studies such as precision
measurement of neutrino oscillation parameters or the mass hierarchy
determination with ICAL. For the case of physics studies such as rock
muons or cosmic ray muons, the response in only certain $\phi$ bins are
relevant since the muons in these cases are always entering the detector
from outside; for this reason, the performance will be better than the
averages shown here.

In contrast, good angular resolution of better than a degree for $P_{\rm
in} \ge 4$ GeV/c is obtained in the peripheral and side regions, which
is comparable to that in the central region.

The simulations indicate that the detector has a good response to muons, with reconstruction of momentum with 15--24\% resolution, direction reconstruction of about a degree for muon energies greater than 4 GeV and charge identification of about 97\%. While fully contained events are reconstructed with the same
efficiency as in the central region, only those partially contained ones which have at least $N_{hits}/\cos\theta > n_0$ in their tracks, $n_0 \sim 15$, are well reconstructed in the simulations. This implies
a loss of reconstruction efficiency due to this criterion. However, the number of events reconstructed in these regions, which is expected to be about 50\% from naive considerations of detector geometry, is about 60--70\%, due to the effect of the magnetic field, which increases the recontruction efficiency in the peripheral region.

\paragraph{Acknowledgements}: We thank Naba K Mondal for suggestions
and support during this work. We also thank the INO simulations group
for their comments and suggestions on the results; Gobinda Majumder
and Asmita Redij for code-related discussions; and Shiba Behera for
discussions on the magnetic field map. R. Kanishka acknowledges UGC/DST
(Govt. of India) for financial support.

\end{document}